# Messung der Membrankapazität mit Hilfe der Saugelektrodentechnik an Kolonkryptzellen der Ratte

Diplomarbeit

vorgelegt von

Christian Schill

aus Freiburg i. Br.

im August 1998

Fakultät für Physik der Albert-Ludwigs-Universität Freiburg i. Br.

angefertigt im

Physiologischen Institut der Albert-Ludwigs-Universität Freiburg i. Br.

**Leiter der Arbeit:** Prof. Dr. R. Greger / Prof. Dr. K. Königsmann

# Inhaltsverzeichnis

# 1. Einleitung

## 1.1 Die Zelle

Eine tierische Zelle (Abb.1) besteht aus dem Zellkern und dem ihn umgebenden Zytoplasma mit den Zellorganellen. Das Zytoplasma wird von dem Zytoskelett, einem Netzwerk aus feinen Proteinfäden (Aktin- und Tubulin-Filamenten) durchzogen, das die Form der Zelle stabilisiert.

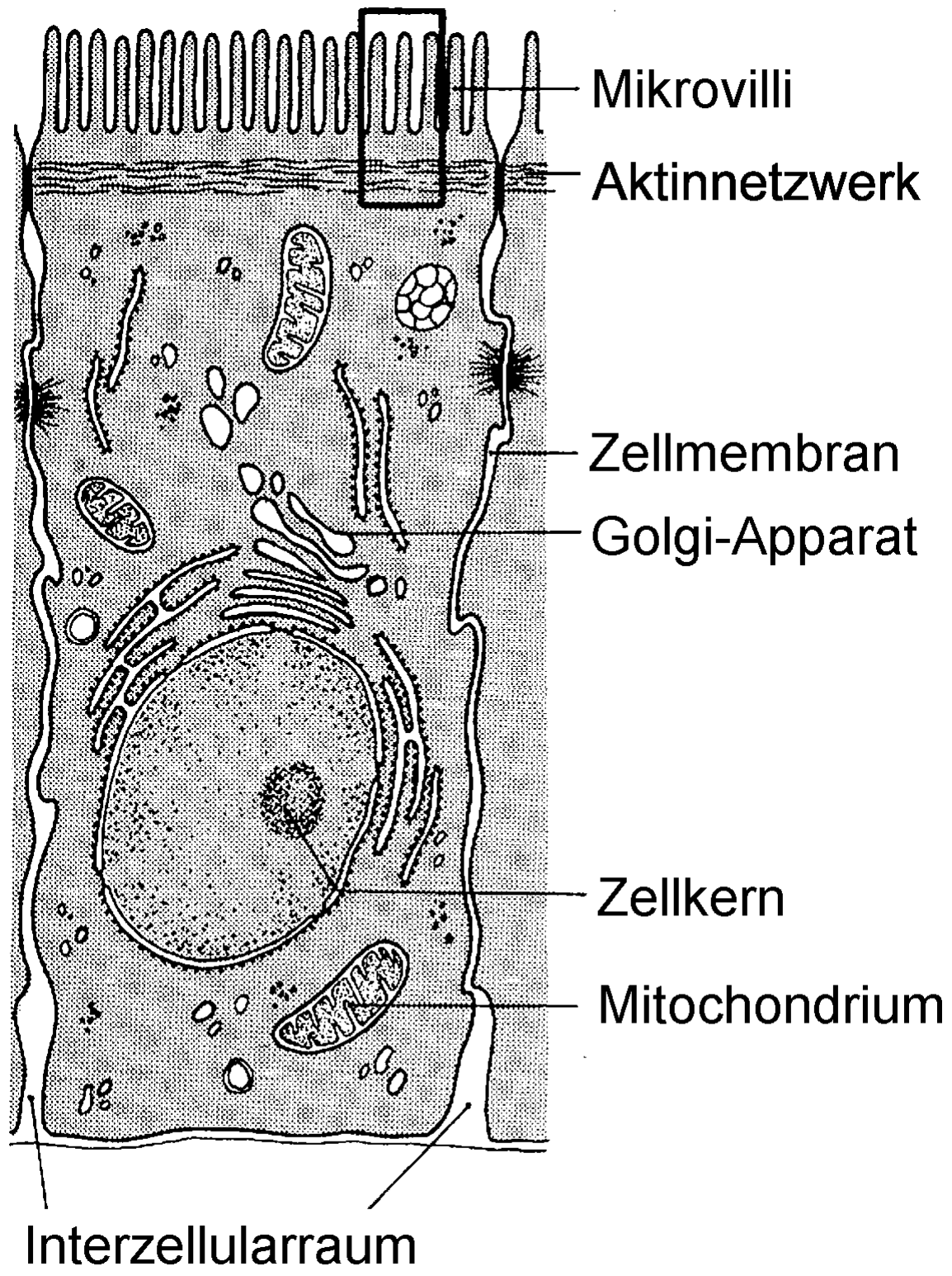

Abb.1: *Schematischer Aufbau einer Zelle am Beispiel einer Darmepithelzelle (nach [92]).*

Die äußere Begrenzung der Zelle stellt die Zellmembran (Abb.2) dar. Diese besteht aus einer Doppelschicht von Lipidmolekülen, ist ein elektrischer Isolator und bildet eine Diffusionsbarriere für wasserlösliche Substanzen. In die Membran sind verschiedene Proteine eingelagert, die für die elektrischen Eigenschaften der Zelle verantwortlich sind und als Ionenkanäle, -pumpen und -transporter den Ionentransport durch die Zellmembran ermöglichen.

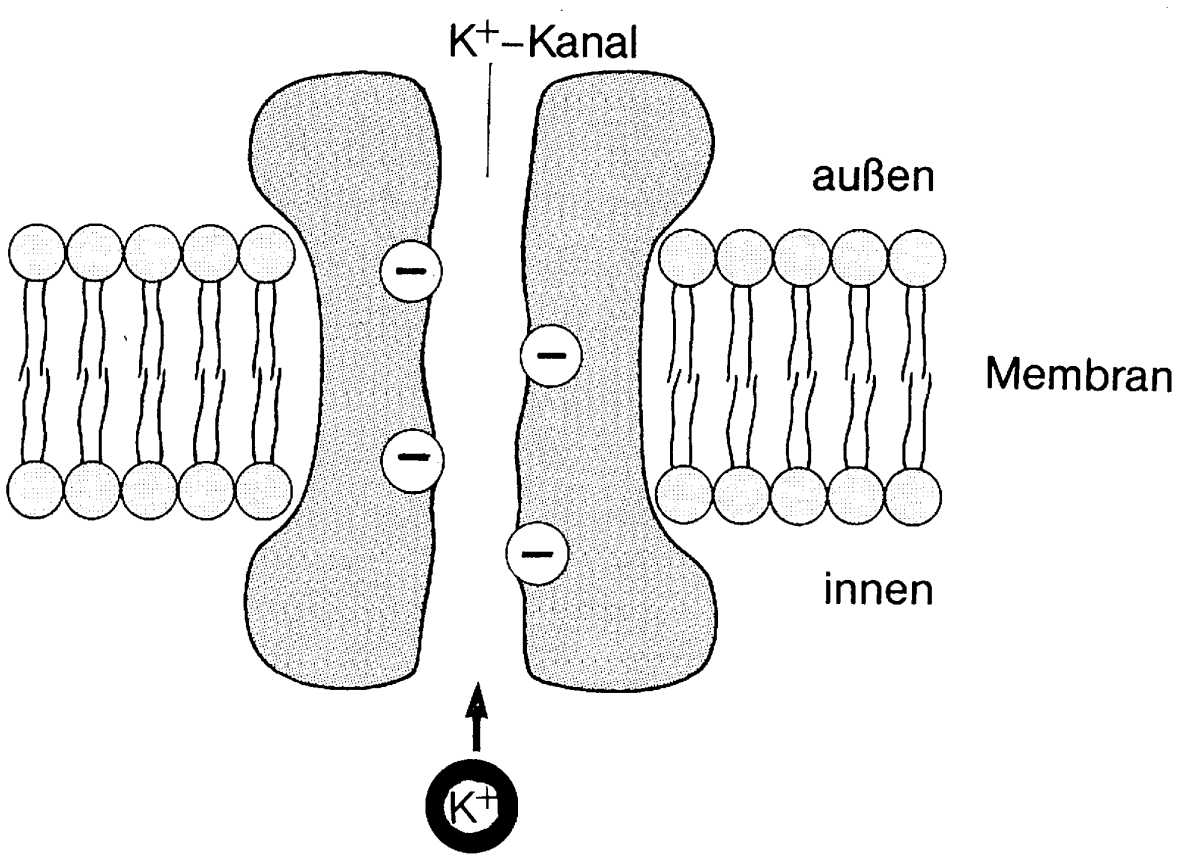

Abb.2: *Schematischer Aufbau einer Zellmembran mit einem eingelagerten Ionenkanal (nach [93]).*

Ionenkanäle erlauben selektiv den Durchtritt einer oder mehrerer Ionenarten durch die Zellmembran. Sie können durch Konformationsänderung des Proteins zwei diskrete Zustände, „Offen“ und „Geschlossen“, einnehmen. Man unterscheidet Kanäle für $Na^+$, $K^+$, $Ca^{2+}$ und $Cl^-$-Ionen sowie nichtselektive Kationenkanäle.

Voraussetzung für den Ionentransport durch Kanäle ist ein elektrochemischer Gradient über die Zellmembran. Er kann sowohl aktiv durch Ionenpumpen als auch sekundär aktiv durch Transporter erzeugt werden.

Ionenpumpen binden ein oder mehrere Ionen auf einer Seite der Zellmembran und transportieren sie aktiv auf die andere Seite der Membran. Dieser Transport wird durch Spaltung der energiereichen chemischen Verbindung Adenosintriphosphat in der Zelle angetrieben und kann auch gegen einen Konzentrationsgradienten erfolgen.

Transporter binden selektiv mehrere Ionen in einem stöchiometrisch festen Verhältnis und transportieren sie durch die Zellmembran. Sie nutzen den von den Ionenpumpen aufgebauten Konzentrationsgradienten eines Ions als Energiequelle für den Transport anderer Ionenarten. Der wichtigste Ionentransporter in Kolonzellen ist der $Na^+2Cl^-K^+$-Kotransporter, der gleichzeitig ein Natrium-, zwei Chlorid- und ein Kaliumion in die Zelle transportiert.

## 1.2 Das Membranpotential

Die ionale Zusammensetzung der extrazellulären Lösung unterscheidet sich von der des Zytoplasmas. Daher gibt es für jedes Ion einen Konzentrationsgradienten zwischen Extra- und Intrazellulärraum. Besitzt die Zellmembran eingelagerte Ionenkanäle, die für ein spezifisches Ion permeabel sind, so gibt es eine chemische Triebkraft für die Diffusion durch die Membran. Ihr entgegen wirkt das elektrische Feld, das durch die Diffusion der geladenen Ionen entsteht. Es bewirkt eine Potentialdifferenz zwischen Intra- und Extrazellulärraum, das Membranpotential.

Die Nernst-Gleichung gibt die Potentialdifferenz zwischen Intra- und Extrazellulärraum an, bei der die chemische Triebkraft in die eine und die elektrische Kraft auf ein einzelnes Ion in die andere Richtung gerade im Gleichgewicht zueinander stehen:

$$E = \frac{RT}{zF} \ln\left(\frac{c_a}{c_i}\right) \qquad (1)$$

*Nernst-Gleichung:* *E: Nernstpotential,*
*R: allgemeine Gaskonstante,*
*T: absolute Temperatur,*
*z: Ladung des Ions in Einheiten der Elementarladung,*
*F: Faraday-Konstante,*
*$c_i$: Ionenkonzentration in der Zelle,*
*$c_a$: Ionenkonzentration außerhalb der Zelle.*

In Tab.1 sind als Beispiel die Nernstpotentiale der wichtigsten Ionen an der Kolonepithelzelle zusammengestellt:

| Ion | $K^+$ | $Cl^-$ | $Na^+$ | $Ca^{2+}$ |
|---|---|---|---|---|
| Nernst-Potential E | ca. -90 mV | ca. -30 mV | ca. +60 mV | ca. +120 mV |

**Tab.1:** *Nernstpotentiale der wichtigsten Ionen an der Kolonepithelzelle (nach [31,32,39,61]).*

Da die Zellmembran für unterschiedliche Ionen permeabel ist und für jede Ionenart ein anderer Konzentrationsgradient zwischen Extra- und Intrazellulärraum herrscht, setzt sich das resultierende Potential der Gesamtzelle, das sogenannte *Membranpotential*, aus den Nernst-Potentialen mehrerer Ionen zusammen. Die Goldman-Gleichung beschreibt das Membranpotential der Zelle als Summe der Nernstpoten-tiale der einzelnen Ionen, gewichtet mit der fraktionellen Leitfähigkeit für jedes Ion [93].

$$E_m = \sum_{alle\ Ionen} f_i E_i$$
$$E_m = f_{K^+} E_{K^+} + f_{Na^+} E_{Na^+} + f_{Cl^-} E_{Cl^-} + f_{Ca^{2+}} E_{Ca^{2+}} \quad (2)$$

*Goldman-Gleichung:* *$E_m$: Membranpotential,*
*$f_i$ : fraktionelle Leitfähigkeit für das jeweilige Ion i,*
*$E_i$ : Nernst-Potential für das jeweilige Ion i.*

Öffnen oder schließen sich bei der Stimulation der Zelle für eine Ionenart spezifische Ionenkanäle, so ändert sich die fraktionelle Leitfähigkeit für dieses Ion. Dies hat zur Folge, daß sich auch das Membranpotential der Zelle verschiebt. Eine Erhöhung des Membranpotentials wird als Hyperpolarisation, eine Erniedrigung als Depolarisation bezeichnet.

## 1.3 Die Membranleitfähigkeit und Membrankapazität

Die Lipidschicht der Zellmembran stellt einen sehr guten elektrischen Isolator dar. Daher wird die elektrische Leitfähigkeit einer Zellmembran hauptsächlich von den eingelagerten Ionenkanälen bestimmt. Ionenpumpen und Ionentransporter tragen nur einen kleinen Anteil zur Gesamtleitfähigkeit bei. Die Leitfähigkeit der Zellmembran $G_m$ wird in der Physiologie als Kehrwert ihres Ohmschen Widerstandes $R_m$ definiert und üblicherweise in der Einheit Siemens ($1\ S = \frac{1}{\Omega}$) angegeben. Sie liegt bei den meisten Zellen zwischen 1 und 100 Nanosiemens (nS).

Eine Zelle kann durch die Leitfähigkeit ihrer Zellmembran und zusätzlich durch ihre elektrische Kapazität charakterisiert werden: Die elektrisch isolierende Doppellipidschicht der Zellmembran bildet ein Dielektrikum, das die zwei elektrisch leitenden Lösungen im Zellinneren und außerhalb der Zelle trennt. Vergleicht man die Zelle mit einem annähernd kugelförmigen Kondensator, so stellen das Zellinnere und die extrazelluläre Lösung die beiden Kondensatorpole und die Zellmembran die Schicht des Dielektrikums dar. Die Kapazität der Zelle läßt sich daher aus ihrer Membranfläche *A*, der Dicke *d*, der Dielektrizitätskonstanten der Doppellipidschicht $\varepsilon_r$ und der Dielektrizitätskonstanten des Vakuums $\varepsilon_0$ berechnen:

$$C_m = \varepsilon_0 \varepsilon_r \frac{A}{d} \tag{3}$$

Da die Dicke und die Dielektrizitätskonstante einer Lipidmembran durch ihren molekularen Aufbau feststehen, sind $\varepsilon_r$ und *d* für alle Zellmembranen in guter Näherung konstant. Die Kapazität $C_m$ einer Zelle ist daher proportional zu ihrer Oberfläche *A*:

$$C_m = \alpha \cdot A \qquad \alpha = \textit{Proportionalitätskonstante} \tag{4}$$

Die Proportionalitätskonstante $\alpha$ liegt für die bisher untersuchten Zellen in der Größenordnung von 1 $\mu F/cm^2$ [76,82].

Die Messung der Zellkapazität hat eine besondere Bedeutung bei der Untersuchung von Transportprozessen in der Zelle gewonnen [82]. Aus licht- und elektronenmikroskopischen Untersuchungen ist bekannt, daß der Transport von Stoffen aus der Zelle, die sogenannte Exozytose, nach folgendem Mechanismus abläuft: Die Stoffe, die aus der Zelle transportiert werden sollen, werden im Zellinnern in kleine, membranumschlossene Bläschen, sogenannte Vesikel (Abb.3), eingeschlossen. Wird die Zelle zur Exozytose stimuliert, so transportiert das Zytoskelett die Vesikel an die Zelloberfläche, wo sie mit der Zellmembran verschmelzen. Dabei öffnet sich die Innenseite des Vesikels nach außen, sein Inhalt wird freigesetzt [93].

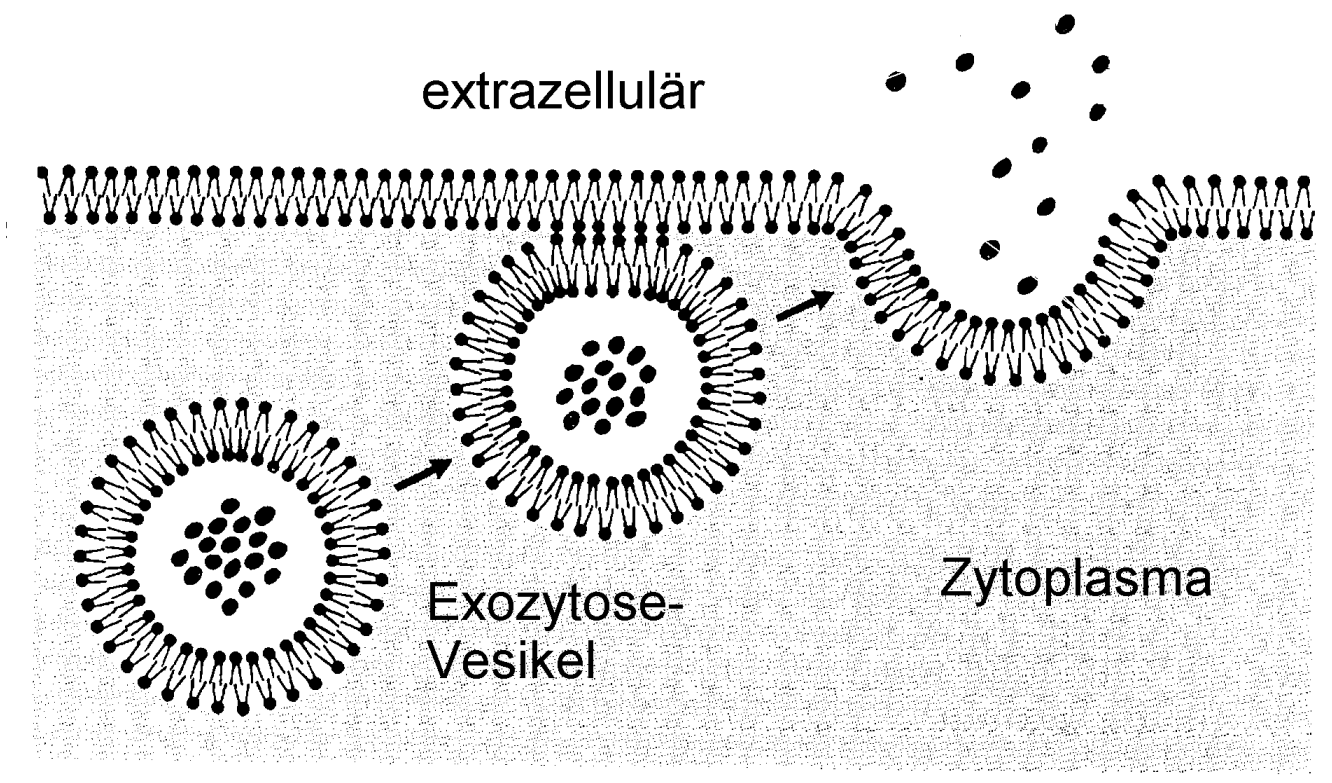

Abb.3: *Schematische Darstellung einer Exozytose. Die Exozytosevesikel verschmelzen mit der Zellmembran, ihr Inhalt wird nach außen abgegeben (nach [93]).*

Dieser Vorgang hat auch Auswirkungen auf die Zellmembrankapazität: Das Vesikel, das ganz im Zellinnern liegt, trägt nicht zur Kapazität der Zellmembran bei, da es von allen Seiten vom Zytoplasma umschlossen ist, d.h. nur zu einem Pol des Kondensators Verbindung hat. Sobald sich aber beim Verschmelzen mit der Zellmembran eine Pore nach außen bildet, hat das Vesikel Verbindung zu beiden Kondensatorpolen: Die Kapazität des Vesikels addiert sich zur Kapazität der übrigen Zellmembran [4,67]. Der plötzliche Kapazitätsanstieg, der durch das Verschmelzen eines einzelnen Vesikels mit der Zellmembran entsteht, konnte an speziellen Zellen des Immunsystems mit sehr großen Exozytosevesikeln, den sogenannten Mastzellen [4,21,64,110], sowie an chromaffinen Zellen der Nebenniere [2,76] beobachtet werden. Bei den meisten anderen zur Exozytose fähigen Zellen sind die Vesikel jedoch so klein, daß nur ein quasi-kontinuierlicher Kapazitätsanstieg erkennbar ist [82,90].

## 1.4 Die Rolle der Exozytose bei der Regulation von Ionenleitfähigkeiten der Zelle

Die Regulation von Ionenkanälen einer Zelle kann prinzipiell auf zwei verschiedenen Wegen erfolgen:

- Bereits in der Zellmembran vorhandene Ionenkanäle können durch den Einfluß intrazellulärer Botenstoffe, sogenannter *second messenger* wie z.B. Calcium-Ionen ($Ca^{2+}$) oder cyclischem Adenosinmonophosphat (cAMP), geöffnet oder geschlossen werden.

- Neue Ionenkanäle werden auf ein intrazelluläres Signal hin in die Membran durch Exozytose eingebaut und später wieder durch Endozytose in die Zelle aufgenommen. Dieser „membrane traffic“ kann durch *second messenger* wie z.B. cAMP beeinflußt werden [8,44].

In der Literatur wird eine Beteiligung der Exozytose an der Regulation von Ionenkanälen in unterschiedlichen epithelialen Geweben diskutiert: bei Pankreas-Azinus-Zellen [69-71,90], respiratorischen Epithelzellen [48], Epithelzellen des proximalen Tubulus der Niere [94,98] und Epithelzellen des Dickdarms (Kolons) [24,35].

Bei vorangegangenen Untersuchungen im Physiologischen Institut der Universität Freiburg an $HT_{29}$-Zellen, einer Kulturzellinie aus dem Kolonepithel, ergaben sich mehrere Hinweise auf eine Beteiligung der Exozytose an der Regulation der Ionenkanäle dieser Zellen [35]: Nach der Stimulation der Zellen mit Agonisten, die die *second messenger* cAMP und $Ca^{2+}$ freisetzen, konnte ein Anstieg der Membrankapazität beobachtet werden. Inhibitoren der Exozytose, wie *Cytochalasin B* oder *Clostridien-Toxin C2,* konnten diesen Aktivierungs-Mechanismus unterbrechen.

In der vorliegenden Arbeit sollte daher erstmals die Rolle der Exozytose bei der Aktivierung von Ionenkanälen an einem nativen Epithelgewebe des Verdauungstraktes, den Kryptbasiszellen aus dem Kolon der Ratte untersucht werden.

Der Regulationsmechanismus des Elektrolyttransports im Kolonepithel ist in vielen Einzelheiten noch nicht geklärt und wird intensiv untersucht [18,34,36,106]. Im folgenden Abschnitt soll deshalb die Morphologie und Funktion des Kolonepithels zunächst eingehend erläutert werden.

## 1.5 Morphologie und Funktion des Kolonepithels

Das Kolon (Dickdarm) ist wie der gesamte Verdauungstrakt aus vier aneinandergrenzenden Schichten aufgebaut: Die Schleimhaut (*Tunica mucosa*) kleidet die Innenseite des Darmes aus und ist für den Transport von Elektrolyten und Wasser zuständig. Daran angrenzend finden wir eine Bindegewebsschicht (*Tela submucosa*), die für die mechanische Verbindung zur außenliegenden Muskelschicht (*Tunica muscularis*) sorgt. Die äußere Begrenzung des Kolons bildet wiederum eine Bindegewebshaut (*Tunica serosa*).

Die Oberfläche des Dickdarms wird durch zahlreiche sackartige Einstülpungen, die etwa einen halben Millimeter tiefen Kolonkrypten vergrößert. Die Krypten sind von einem einschichtigen, hochprismatischen Epithel ausgekleidet. Sie besitzen einen Bürstensaum (*Mikrovilli*) zur Vergrößerung der Resorptionsfläche.

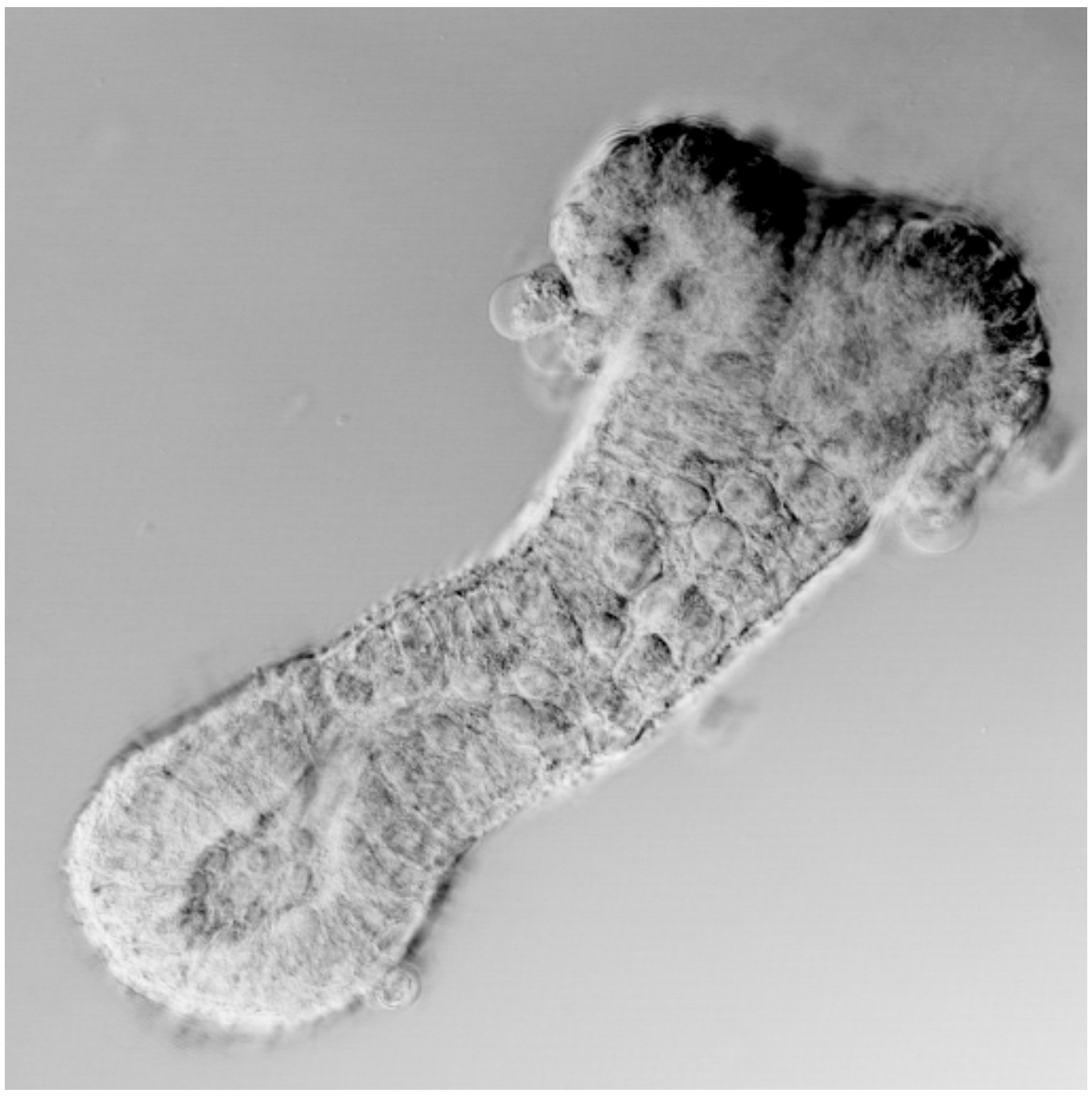

Abb.4: *Photographie einer isolierten Kolonkrypte. Nach [36].*

Man unterscheidet zwischen der Kryptbasis und der Kryptoberfläche. In der Kryptbasis liegen Stammzellen des Epithels, die sich ständig teilen und neue Epithelzellen bilden. Innerhalb einiger Tage wandern die Epithelzellen zur Kryptoberfläche, wo sie schließlich abgestoßen werden.

Die Hauptaufgabe des Dickdarmes ist die lebenswichtige Rückresorption von Wasser und Salzen, die mit den Verdauungssäften in den Darm gelangen [75]. Aus einem Liter Wasser und etwa 5-10 Gramm Kochsalz, die täglich beim Menschen aus dem Dünn- in den Dickdarm gelangen, bleiben nach der Dickdarmpassage nur noch 100 Milliliter Wasser mit etwa 20-100 Milligramm gelöstem Kochsalz zurück [27]. Andererseits kann der Dickdarm auch Flüssigkeit und Salze sezernieren [27]. Während an der Kryptoberfläche hauptsächlich Kochsalz-Resorption stattfindet, sind die Zellen der Kryptbasis ausschließlich zur Sekretion fähig [18,58]. Das Ausmaß der Sekretion wird durch die vagale Innervation und verschiedene Gewebsmediatoren (z.B. ATP, Prostaglandin $E_2$) gesteuert. Normalerweise überwiegt die Elektrolyt-Resorption im Kolon. Findet - beispielsweise durch das bakterielle Choleratoxin - eine übermäßige Sekretion von Elektrolyten und Wasser statt [109], so spricht man von sekretorischer Diarrhöe. Sie kann zu lebensbedrohlichen Elektrolyt- und Wasserverlusten führen.

## 1.6 Die Kochsalzsekretion in der Kolonkrypte

Nach Untersuchungen der NaCl-Sekretion an der Rektaldrüse des Dornhaies (*Squalus acanthias*) wurde 1984 von R. Greger ein Modell für die Elektrolytsekretion in Epithelien (Abb.5) vorgestellt [33]. Seine Gültigkeit konnte inzwischen auch an vielen anderen Geweben und insbesondere an der Kolonkrypte bestätigt werden [19,37,38,62,108].

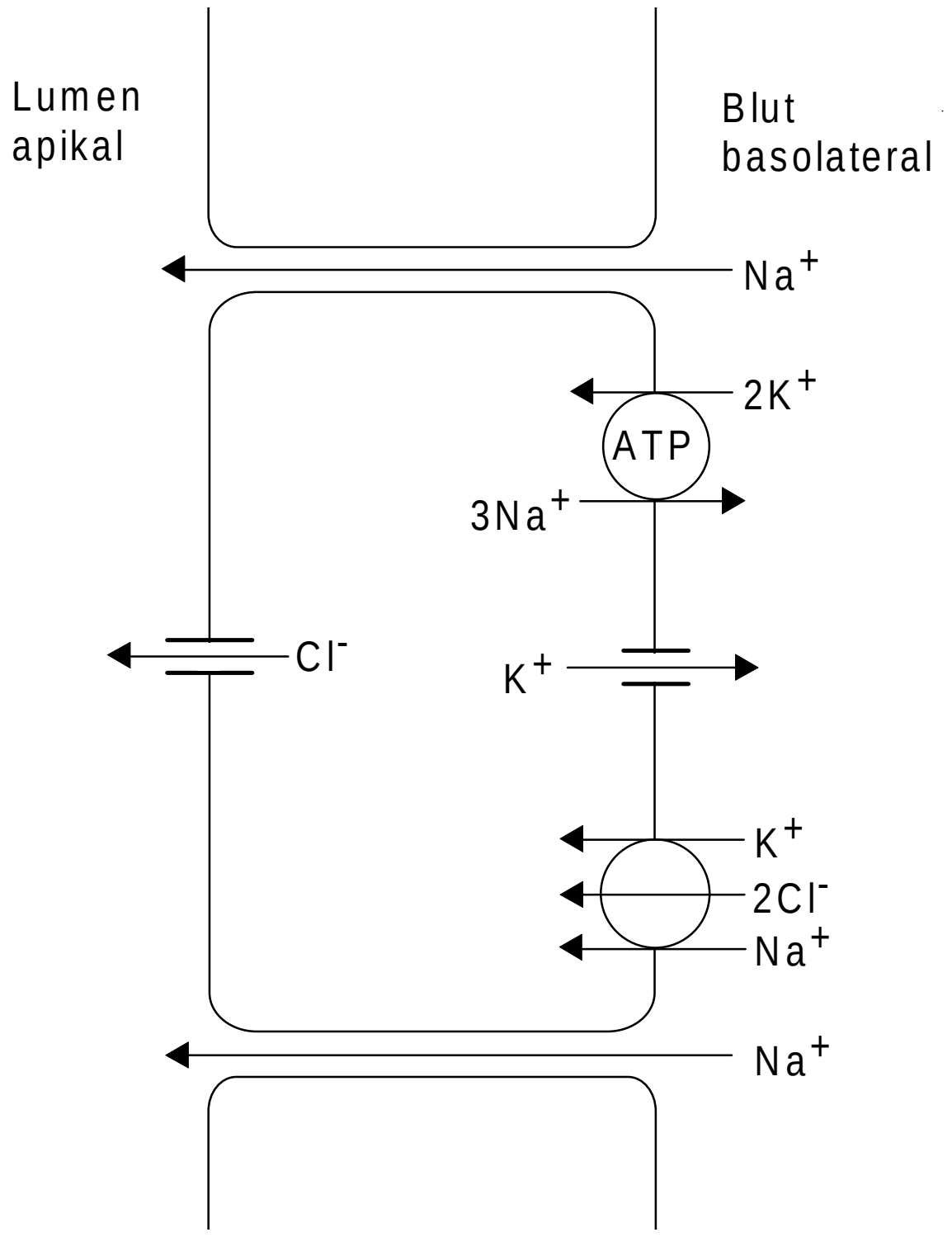

Abb.5: *Modell einer NaCl sezernierenden Epithelzelle. Nach [36].*

Wie in Abbildung 5 schematisch dargestellt, sind an der Sekretion, d.h. dem Transport von $Na^+$ und $Cl^-$ -Ionen in das Darmlumen, verschiedene Ionenkanäle, Ionentransporter und Ionenpumpen (ATPasen) beteiligt. Der Transport eines $Cl^-$-Ions von der Blut- auf die Lumen-Seite erfolgt transzellulär: Zwei $Cl^-$-Ionen werden durch den $Na^+2Cl^-K^+$-Kotransporter in die Kolonzelle aufgenommen. Das gleichzeitig transportierte $K^+$-Ion rezirkuliert durch basale $K^+$-Kanäle zurück auf die basolaterale Seite. Die Energie für den $Cl^-$-Transport liefert der $Na^+$-Gradient über die Zellmembran [33,41]. Er wird durch die $Na^+K^+$-ATPase aufrecht erhalten, die unter Spaltung eines energiereichen ATP-Moleküls 3 $Na^+$ heraus und 2 $K^+$ in die Zelle transportiert.

Durch luminale $Cl^-$-Kanäle gelangt das $Cl^-$-Ion schließlich in den Darm. Das $Na^+$-Ion folgt ihm passiv parazellulär nach, getrieben durch ein lumennegativen elektrischen

Potential-Gradienten. Das lumennegative Potential wird durch die basolaterale $K^+$ und die luminale $Cl^-$-Leitfähigkeit verursacht: Während sich über der basolateralen Membran mit hauptsächlicher $K^+$-Leitfähigkeit das Nernst-Potential von $K^+$, -90 mV ausbildet, liegt an der luminalen Membran nur das niedrigere $Cl^-$-Potential von -30 mV an. Insgesamt resultiert also eine lumennegative Spannung von maximal (-90 mV -(-30 mV)) = -60 mV.

Die tatsächliche Höhe des lumennegativen elektrischen Gradienten stellt ein Maß für die Transportaktivität eines Epithels dar. Er kann mit Hilfe einer sogenannten Ussing-Kammer gemessen werden.

Die NaCl-Sekretion über das Kolonepithel erzeugt außerdem gleichzeitig einen osmotischen Gradienten für Wasser, das von der basolateralen auf die luminale Epithelseite strömt.

## 1.7 Die Regulation der Sekretion

Die Regulation der Sekretion geschieht in der Kolonkrypte über verschiedene Gewebshormone und -mediatoren. Allen Regulationswegen gemeinsam ist die Übertragung eines äußeren Signals eines extrazellulären Botenstoffs (Hormons) in ein Signal in der Zelle. Der grundsätzliche Ablauf dieser Signal-Transduktionskaskade ist für alle Hormone sehr ähnlich. Der Botenstoff bindet an der Außenseite der Zellmembran an einen Rezeptor, der in die Membran eingelagert ist. Er transportiert das Signal durch die Zellmembran ins Zellinnere, wo es über mehrere Zwischenschritte zur Freisetzung eines intrazellulären Botenstoffs, des sogenannten *second messenger* kommt. Der *second messenger* seinerseits kann auf die Ionenkanäle der Zelle Einfluß nehmen und z.B. ihre Offen- und Geschlossenwahrscheinlichkeit verändern.

Im Kolonepithel sind die drei „second messenger" zyklisches Adenosinmonophosphat (cAMP), cyclisches Guanosinmonophosphat (cGMP) und Calcium-Ionen ($Ca^{2+}$) an der Regulation der Sekretion beteiligt [36]:

Prostaglandin $E_2$, Sekretin oder VIP binden auf der extrazellulären Seite an membranständige Rezeptoren. Über G-Proteine wird die Adenylatcyclase aktiviert, die unter ATP-Verbrauch cAMP herstellt. cAMP bindet an die regulatorische Untereinheit der Proteinkinase A. Dadurch wird die katalytische Untereinheit von der regula-torischen abgespalten und aktiviert [72]. Die aktivierte katalytische Untereinheit der Proteinkinase A phosphoryliert bestimmte Proteine. Dadurch wird z.B. im Kolon basolateral eine Kalium-Leitfähigkeit [66,106] sowie luminal eine Chlorid-Leitfähigkeit [3,57,59] aktiviert. Beide Effekte steigern die Sekretion.

Der Botenstoff $Ca^{2+}$ reguliert viele zelluläre Prozesse. Die $Ca^{2+}$-Aktivität einer ruhenden Zelle beträgt nur etwa $10^{-7}$ Mol/Liter, während die Calciumkonzentration im Extrazellulärraum etwa 1 mmol/l beträgt. Ein Calcium-Anstieg in der Zelle kann einerseits durch Freisetzung von Calcium-Ionen aus intrazellulären Calcium-Speichern und andererseits durch $Ca^{2+}$-Einstrom hervorgerufen werden. Als Auslöser für die $Ca^{2+}$-Freisetzung wirken die beiden Substanzen Inositoltrisphosphat ($IP_3$) und Diacylglycerol (DaG), die beide von dem Enzym Phospholipase C durch Spaltung des Membranlipids Phosphatidyl-Inositol gebildet werden [54]. An der Kolonzelle führt eine $[Ca^{2+}]_i$-Erhöhung zum Öffnen basolateraler $K^+$-Kanäle [7,37,38,79] und wahrscheinlich ebenfalls luminaler $Cl^-$-Kanäle [3,33,60]. Die Sekretion wird damit gesteigert.

Neben der Regulation der an der Sekretion beteiligten Ionenkanäle ist auch die Regulation des $Na^+2Cl^-K^+$-Kotransporters zur Kontrolle der Sekretion nötig. Neue Untersuchungen an der Rektaldrüse des Dornhaies (*Squalus acanthies*) haben gezeigt, daß sowohl ein Absinken der intrazellulären $Cl^-$-Konzentration als auch ein Anstieg des cAMP die Transportrate dieses Proteins erhöht [105].

## 1.8 Aufgabenstellung

In der vorliegenden Arbeit sollte die Rolle der Exozytose bei der Aktivierung von Ionenkanälen an der Kolonkrypte untersucht werden. Dazu sollten zunächst die Änderungen der Zellmembrankapazität, die ein Maß für die Exozytoseaktivität darstellen, bei der Stimulation der Zellen zur Kochsalzsekretion bestimmt werden. Daher war die Aufgabe der vorliegenden Arbeit:

- Kapazitätsmessungen an Kryptbasiszellen aus nativem Gewebe der Ratte mit Hilfe der Saugelektrodentechnik durchzuführen und die Auswirkungen verschiedener Agonisten, die die NaCl-Sekretion erhöhen, auf die Membrankapazität sowie die Membranleitfähigkeit und das Membranpotential der Kolonzelle zu untersuchen.

- Hierzu sollten zunächst die Meßgenauigkeit und die Meßfehler der bereits etablierten Zweifrequenz-Kapazitätsmeßapparatur bestimmt und Verbesserungen der Meßtechnik vorgenommen werden.

Im zweiten Teil sollte die inhibitorische Wirkung verschiedener Toxine auf das Zytoskelett der Kolonkrypte ausgenutzt werden, um eine eventuell auftretende Exozytose zu unterbinden. Die Aufgabe bestand darin:

- Veränderungen in der Aktivierbarkeit der Ionenkanäle und der Membrankapazität der Kolonepithelzelle nach der Hemmung der Exozytose mit den Toxinen A und B aus *Clostridium difficile*, *Cytochalasin B, Phalloidin* und *Colchicin* zu untersuchen.

# 2. Methoden

## 2.1 Methoden zur Kapazitätsmessung an Zellen

### 2.1.1 Erste Messungen der Membrankapazität

Voraussetzung für jede Messung der Membrankapazität ist ein elektrischer Zugang zum Zytoplasma, über den Spannungen an das Zellinnere angelegt und Ströme über die Zellmembran gemessen werden können. Die ersten Kapazitätsmessungen an lebenden Zellen wurden in den Jahren 1978/79 von L. Jaffee et al. und J. Gillespie durchgeführt [26,50]. Sie nutzten in die Zelle eingestochene Glas-Mikroelektroden als elektrische Verbindung zum Zytoplasma. Kleine Kapazitätsänderungen, wie sie z.B. bei Exozytoseereignissen auftreten, lagen jedoch unterhalb des Auflösungsvermögens dieser Meßmethode [89]. Die Meßgenauigkeit wurde im wesentlichen durch den hohen Ohmschen Widerstand der intrazellulären Ableitelektrode eingeschränkt, die sehr dünn sein mußte, um die Zellmembran nicht zu zerstören.

Erst mit Hilfe der Saugelektrodentechnik wurde es möglich, die Kapazität von Zellen mit größerer Genauigkeit zu bestimmen [52].

### 2.1.2 Die Saugelektrodentechnik

Die Saugelektrodentechnik wurde 1976 von E. Neher und B. Sakmann vorgestellt. Sie ermöglicht es, Ströme durch einzelne Ionenkanäle und über die gesamte Zellmembran mit hoher Auflösung zu messen [77].

Dazu wird eine fein ausgezogene Glaskapillare (Pipette), deren Spitzendurchmesser etwa 1-3 µm beträgt, mit Elektrolytlösung gefüllt und vorsichtig auf die intakte Zellmembran aufgesetzt. Mit Hilfe einer Glasspritze wird von Hand ein leichter Unterdruck im Pipetteninnern erzeugt. Die Zellmembran legt sich dabei dicht an die Pipettenöffnung an und wölbt sich in Form eines Omegas (Ω) in das Lumen der Pipette vor [77]. Durch molekulare Wechselwirkungen zwischen der Glasoberfläche und der Zellmembran bildet sich dann spontan ein äußerst hochohmiger Abdichtwiderstand (1-100 Gigaohm) zur Zelloberfläche aus.

Um eine elektrische Verbindung zum Zellinneren zu schaffen, wird mit einem kurzzeitigen, kräftigeren Unterdruck die Membran im Pipettenlumen eingerissen, ohne jedoch die seitliche Abdichtung zwischen Glas und Membran zu verringern. Hierdurch entsteht eine direkte Verbindung zwischen dem Pipettenlumen und dem Zellinneren. Diese Konfiguration wird als sogenannte Ganzzell-Ableitung bezeichnet.

Zwischen der Elektrolytlösung im Pipetteninneren und dem Zytoplasma bildet sich ein Ohmscher Widerstand aus, der an der engen Übergangsstelle von der Zelle in die Pipette lokalisiert ist [77]. Er wird als Zugangswiderstand $R_a$, sein Kehrwert als Zugangsleitfähigkeit $G_a$ bezeichnet. Der Zugangswiderstand liegt meistens zwischen 10 und 100 Megaohm und ist damit um etwa 3 Zehnerpotenzen kleiner als der Abdichtwiderstand. Ein hoher Abdichtwiderstand in Verbindung mit einem niedrigen Zugangswiderstand ermöglicht es, mit Hilfe empfindlicher Meßverstärker den Strom über die Zellmembran mit hohem Auflösungsvermögen zu messen.

Eine weitere Variation stellt die Nystatinmethode dar. Dabei wird die Zellmembran im Innern der Saugelektrode nicht mechanisch eingerissen, sondern durch Einlagerung von Nystatinmolekülen elektrisch permeabel gemacht [39,46]. Sie konnte jedoch an der Kolonkrypte nicht eingesetzt werden, da die erreichten Zugangsleitfähigkeiten zu gering (<5 nS in 10 Experimenten) zur Kapazitätsmessung waren.

### 2.1.3 Elektrische Eigenschaften der Zelle mit Saugelektrode

Zur Beschreibung einer Zelle mit angelegter Saugelektrode wurde von M. Lindau und E. Neher das in Abb.6 dargestellte elektrische Ersatzschaltbild vorgeschlagen [65,76]. Neben der Leitfähigkeit $G_m$ und der Kapazität $C_m$ der Zellmembran sind auch die Zugangsleitfähigkeit $G_a$ und die Pipettenkapazität $C_p$ berücksichtigt.

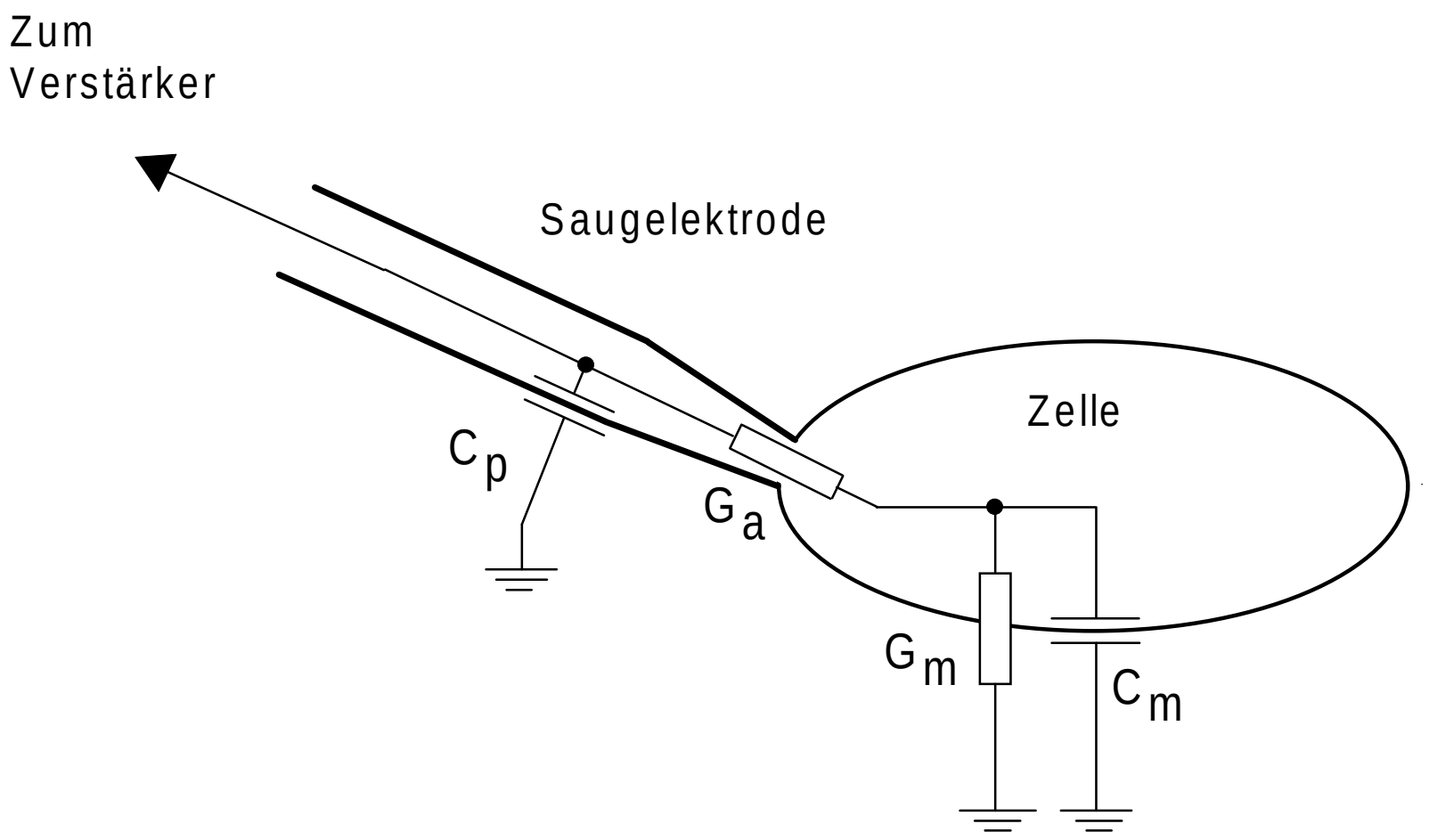

Abb.6: *Ersatzschaltbild einer Zelle mit angelegter Saugelektrode.*
*$C_p$: Pipettenkapazität, $G_a$: Zugangsleitfähigkeit, $G_m$: Membranleitfähigkeit,*
*$C_m$: Membrankapazität.*

Während die Ohmsche Membranleitfähigkeit $G_m$ durch das Anlegen einer Gleichspannung an die Pipette und die Messung des Pipettenstromes bestimmt werden kann, muß zur Kapazitätsmessung stets ein zeitabhängiges Pipettenpotential U(t) vorgegeben werden. Daher stellt man die an die Pipette angelegte Wechselspannung U(t) und den gemessenen Pipettenstrom I(t) als komplexwertige, zeitabhängige Funktionen dar:

$$U(\omega,t) = \hat{U} \cdot e^{i\omega t}$$
$$I(\omega,t) = \hat{I} \cdot e^{i\omega t} \qquad (5)$$

$\hat{U}$: *Amplitude der Wechselspannung*
$\hat{I}$: *Amplitude des Wechselstromes*
$\omega = 2\pi f$: *Kreisfrequenz des Wechselstromes*

Der Zusammenhang zwischen Strom und Spannung wird durch die komplexe Funktion Z(ω), die sogenannte Impedanz, beschrieben. Ihr Kehrwert wird als komplexer Leitwert oder Admittanz Y(ω) bezeichnet:

$$U(\omega,t) = I(\omega,t) \cdot Z(\omega)$$
$$I(\omega,t) = U(\omega,t) \cdot Y(\omega) \qquad (6)$$

$Z(\omega)$: *Impedanz des Schaltkreises*
$Y(\omega) = \frac{1}{Z(\omega)}$: *komplexer Leitwert (Admittanz) des Schaltkreises*

Aus dem Ersatzschaltbild einer Zelle (Abb.6) läßt sich der komplexe Leitwert Y(ω) des Systems aus Zelle und Saugelektrode berechnen, indem man die Kirchhoff-schen Regeln für die Parallel- und Serienschaltung von Widerständen anwendet. Der Blindwiderstand des Kondensators $C_m$ ist durch

$$Z_C = \frac{1}{i\omega\, C_m} \tag{7}$$

gegeben. Die Parallelschaltung aus $C_m$ und $R_m$ hat dann die Impedanz

$$\begin{aligned} Z_{C,R} &= \frac{1}{\frac{1}{R_m} + \frac{1}{Z_C}} \\ &= \frac{1}{G_m + i\omega\, C_m} \end{aligned} \tag{8}$$

*$G_m$ : Membranleitfähigkeit*
*$C_m$ : Membrankapazität*
*$\omega$ : Kreisfrequenz der Wechselspannung*

In Serie zu dieser Impedanz liegt der Zugangswiderstand $R_a$. Vernachlässigt man zunächst die Pipettenkapazität (siehe Kapitel 2.5), so ergibt sich für die Impedanz der Gesamtschaltung:

$$\begin{aligned} Z_{ges} &= R_a + \frac{1}{G_m + i\omega\, C_m} \\ &= \frac{1}{G_a} + \frac{1}{G_m + i\omega C_m} \\ &= \frac{G_a + G_m + i\omega C_m}{G_a\left(G_m + i\omega C_m\right)} \end{aligned} \tag{9}$$

*$G_a$ : Zugangsleitfähigkeit*
*$G_m$ : Membranleitfähigkeit*
*$C_m$ : Membrankapazität*
*$Z_{ges}$ : Impedanz der Gesamtschaltung*

Für den komplexen Leitwert $Y_i$ der Ersatzschaltung bei der Kreisfrequenz des Wechselstromes $\omega_i$ erhält man daher als Ergebnis:

$$Y_i = G_a \frac{G_m + i\omega_i C_m}{G_a + G_m + i\omega_i C_m} \tag{10}$$

Separiert man den Real- und den Imaginärteil dieser komplexen Gleichung, so erhält man nach kurzer Rechnung:

$$\mathrm{Re}(Y_i) = \frac{G_a G_m^{\ 2} + G_a^{\ 2} G_m + G_a \omega_i^{\ 2} C_m}{(G_m + G_a)^2 + \omega_i^{\ 2} C_m^{\ 2}}$$

$$\mathrm{Im}(Y_i) = \frac{\omega_i C_m G_a^{\ 2}}{(G_m + G_a)^2 + \omega_i^{\ 2} C_m^{\ 2}} \qquad (11)$$

$\mathrm{Re}(Y_i)$ : *Realteil des komplexen Leitwertes der Zelle*
$\mathrm{Im}(Y_i)$ : *Imaginärteil des komplexen Leitwertes der Zelle*
$C_m$ : *Membrankapazität*
$G_m$ : *Membranleitfähigkeit*
$G_a$ : *Zugangsleitfähigkeit*
$\omega_i$ : *Kreisfrequenz der verwendeten Wechselspannung*

### 2.1.4 Die Kapazitätsmessung mit Hilfe eines Lock-In-Verstärkers („phase tracking“)

Zur Kapazitätsmessung nach der 1982 von E. Neher und A. Marty vorgeschlagenen Methode [76] wird in einem Versuchsaufbau, wie er in Abb.6 dargestellt ist, an die Pipette eine sinusförmige Wechselspannung angelegt und die Phasenverschiebung zwischen dem Pipettenstrom und der angelegten Spannung mit Hilfe eines Lock-In-Verstärkers bestimmt. Der Anteil des Stromes in Phase zur angelegten Spannung ist ein Maß für die Ohmsche Leitfähigkeit der Zelle; der um $\pi/2$ phasenverschobene Strom gibt die Membrankapazität wieder. Die Zugangsleitfähigkeit $G_a$ wird dabei vor der Messung mit Hilfe einer elektronischen Schaltung im Saugelektrodenverstärker kompensiert [96]. Mit dieser Methode konnten an Mastzellen Kapazitätsänderungen in der Größenordnung von einigen Femtofarad aufgelöst werden [76].

Ein entscheidender Nachteil dieser Methode ist der systematische Fehler in den Meßergebnissen durch Änderungen der Zugangsleitfähigkeit $G_a$, die in Serie zu der zu messenden Kapazität $C_m$ liegt. Da sich der Zugangswiderstand jedoch häufig im Laufe eines Experimentes ändert, können dadurch auch die gemessenen Kapazitätswerte systematisch beeinflußt werden [17].

### 2.1.5 Die Pseudo-Random-Binärsequenz-Methode

Eine Methode, die den systematischen Fehler durch Änderungen der Zugangsleitfähigkeit $G_a$ ausschaltet, wurde 1986 von M. Lindau und J. Fernandez [64] entwickelt. Hierzu werden an die Pipette keine sinusförmigen Spannungen, sondern sogenannte Pseudo-Random-Binärsequenzen (PRBS) angelegt, ein durch einen Zufallsgenerator erzeugtes und mit einem Digital-Analog-Wandler in Spannungen umgesetztes Rauschen, das im Mittel alle Frequenzen in einem bestimmten Intervall mit gleicher Amplitude enthält. Untersucht man nun im gemessenen Stromsignal das Frequenzspektrum durch Fourieranalyse, so lassen sich daraus die Impedanz des komplexen Schaltkreises ableiten und alle drei Größen $C_m$ und $G_m$ und auch $G_a$ berechnen. Der Nachteil dieser Methode liegt jedoch in einem ungünstigen Signal-Rausch-Verhältnis, woraus ein niedriges Auflösungsvermögen resultiert [17].

### 2.1.6 Die Zweifrequenz-Synchrondetektionsmethode

Im Jahre 1993 haben V. Rohliçek und J. Rohliçek eine neue Methode, die Zweifrequenz-Synchrondetektionsmethode, zur Zellkapazitätsmessung eingeführt [87,89, 90]. Sie wurde im Physiologischen Institut der Universität Freiburg weiterentwickelt und auch anfangs für die Messungen in der vorliegenden Arbeit verwendet. Der neuen Meßmethode liegt die folgende Idee zu Grunde:

In einer experimentellen Anordnung, wie sie auch bei der konventionellen Saugelektrodentechnik verwendet wird (Abb.6), werden zwei Sinusspannungen mit unterschiedlichen Frequenzen gleichzeitig an die Pipette angelegt. Die Phasenverschiebung und die Amplitude des dann fließenden Wechselstromes werden gleichzeitig mit Hilfe zweier Lock-In-Verstärker bestimmt (Abb.7). Die Phasenverschiebung $\alpha$ und die Amplitude $\hat{I}$ des Wechselstromes stehen dann in folgendem Zusammenhang mit dem komplexen Leitwert Y der Zelle:

$$
\begin{aligned}
\tan\alpha &= \frac{\mathrm{Im}(Y)}{\mathrm{Re}(Y)} \\
\hat{I} &= \hat{U} \cdot |Y| \\
Y &= \mathrm{Re}(Y) + i\,\mathrm{Im}(Y)
\end{aligned}
\qquad (12)
$$

$\mathrm{Re}(Y)$ : *Realteil des komplexen Leitwertes*
$\mathrm{Im}(Y)$ : *Imaginärteil des komplexen Leitwertes*
$\hat{I}$ : *Amplitude des gemessenen Wechselstromes*
$\alpha$ : *Phasenverschiebung des gemessenen Wechselstromes*

Aus dem so gemessenen Real- und Imaginärteil des komplexen Leitwertes bei zwei unterschiedlichen Frequenzen $\omega_1$ und $\omega_2$ können dann aus Gleichung (11) die drei unbekannten Größen Membranleitfähigkeit $G_m$, Membrankapazität $C_m$ und Zugangsleitfähigkeit $G_a$ berechnet werden.

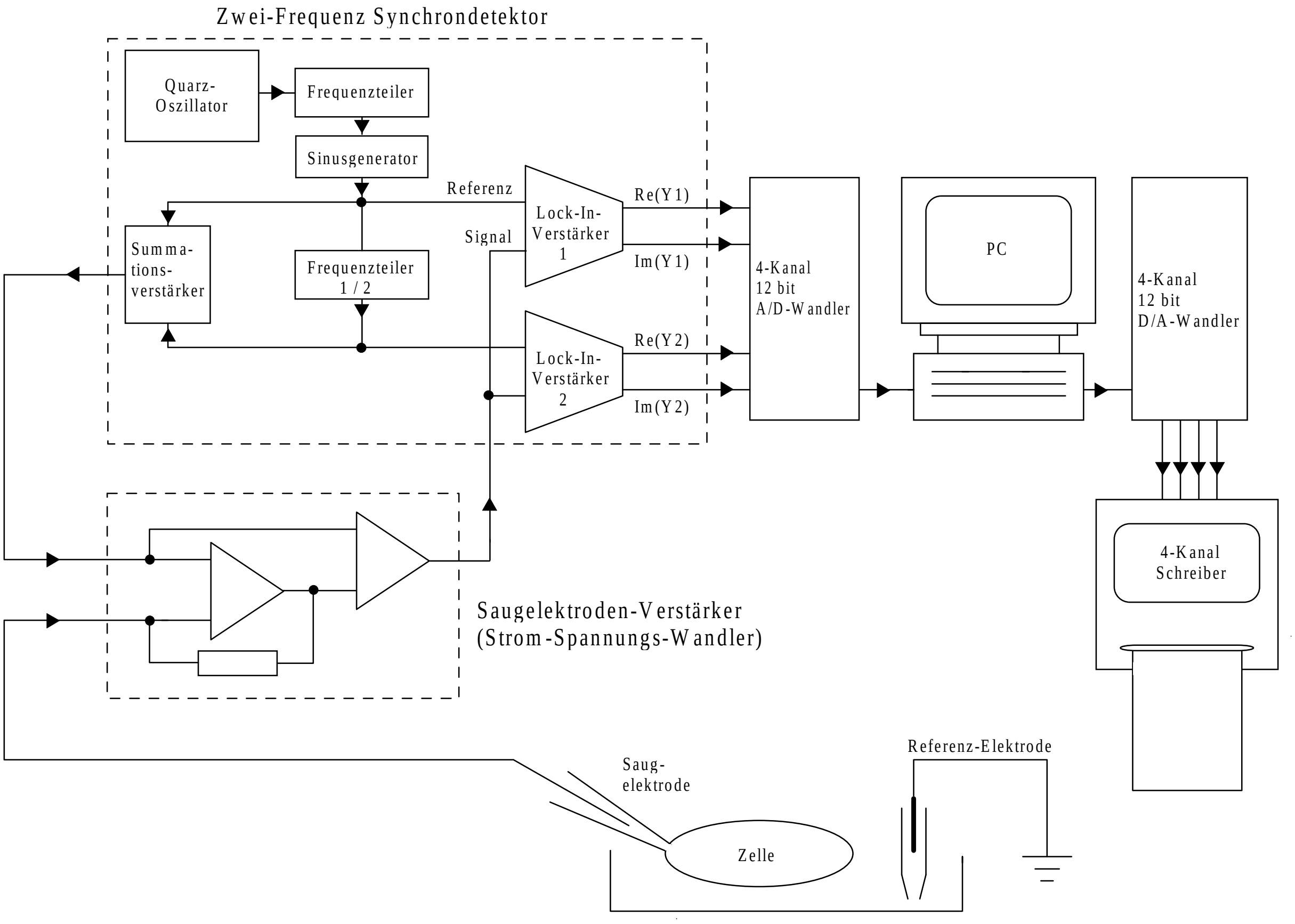

Abb.7: *Schaltplan des verwendeten Zweifrequenz-Synchrondetektors.*

### 2.1.7 Die Berechnung der Membrankapazität $C_m$, Membranleitfähigkeit $G_m$ und Zugangsleitfähigkeit $G_a$ aus den gemessenen komplexen Leitwerten $Y_i$

Das System von vier Gleichungen, das man aus Gleichung (11) nach Einsetzen der vier Meßwerte Re( $Y_1$), Im($Y_1$) und Re($Y_2$), Im($Y_2$) erhält, ist einfach überbestimmt, da es nur drei Unbekannte $G_a$, $G_m$ und $C_m$ enthält:

$$\mathrm{Re}(Y_1) = \frac{G_a G_m^{\ 2} + G_a^{\ 2} G_m + G_a \omega_1^{\ 2} C_m^{\ 2}}{\left(G_m + G_a\right)^2 + \omega_1^{\ 2} C_m^{\ 2}}$$

$$\mathrm{Im}(Y_1) = \frac{\omega_1 C_m G_a^{\ 2}}{\left(G_m + G_a\right)^2 + \omega_1^{\ 2} C_m^{\ 2}}$$

$$\mathrm{Re}(Y_2) = \frac{G_a G_m^{\ 2} + G_a^{\ 2} G_m + G_a \omega_2^{\ 2} C_m^{\ 2}}{\left(G_m + G_a\right)^2 + \omega_2^{\ 2} C_m^{\ 2}} \qquad (13)$$

$$\mathrm{Im}(Y_2) = \frac{\omega_2 C_m G_a^{\ 2}}{\left(G_m + G_a\right)^2 + \omega_2^{\ 2} C_m^{\ 2}}$$

*Real- und Imaginärteil des komplexen Leitwertes einer Zelle in Abhängigkeit von den Parametern $G_m$, $G_a$ und $C_m$. $\omega_1$: Kreisfrequenz der ersten Sinusspannung, $\omega_2$: Kreisfrequenz der zweiten Sinusspannung.*

Zur Bestimmung der gesuchten Größen $C_m$, $G_m$ und $G_a$ kann man daher auf eine der vier Gleichungen verzichten und das System nach den verbleibenden drei Unbekannten auflösen. Dieses Verfahren hat jedoch einen entscheidenden Nachteil: Setzt man nur drei der vier gemessenen Größen $Y_i$ in das Gleichungssystem (13) ein und löst es dann nach den drei Unbekannten $C_m$, $G_m$ und $G_a$ auf, so verliert man die Information, die im vierten Meßwert $Y_i$ enthalten ist. Durch den Verlust eines Viertels der Meßdaten nimmt die Genauigkeit der berechneten Meßergebnisse ab, und das Auflösungsvermögen der Messung sinkt.

Eine andere Methode zur Berechnung der gesuchten Größen ist die von V. Rohliçek [89] aufgrund einer Fehlerrechnung vorgeschlagene Lösung (14), die alle vier Meßwerte verwendet und sich durch Einsetzen in (13) leicht verifizieren läßt ( $\omega_2 = 2\omega_1$ ).

$$C_m = \frac{2\frac{\left(\mathrm{Re}^2\,Y_2 + \mathrm{Im}^2\,Y_2\right)}{\mathrm{Im}\,Y_2} - \frac{\left(\mathrm{Re}^2\,Y_1 + \mathrm{Im}^2\,Y_1\right)}{\mathrm{Im}\,Y_2}}{3\omega_1}$$

$$G_m = \frac{2\left(\mathrm{Re}^2\,Y_2 + \mathrm{Im}^2\,Y_2\right)\mathrm{Re}\,Y_1 - \left(\mathrm{Re}^2\,Y_1 + \mathrm{Im}^2\,Y_1\right)\mathrm{Re}\,Y_2}{3\,\mathrm{Im}\,Y_1\,\mathrm{Im}\,Y_2} \qquad (14)$$

$$G_a = \frac{2\left(\mathrm{Re}^2\,Y_2 + \mathrm{Im}^2\,Y_2\right)\mathrm{Im}\,Y_1 - \left(\mathrm{Re}^2\,Y_1 + \mathrm{Im}^2\,Y_1\right)\mathrm{Im}\,Y_2}{2\left(\mathrm{Im}\,Y_1\,\mathrm{Re}\,Y_2\right) - \left(\mathrm{Re}\,Y_1\,\mathrm{Im}\,Y_2\right)}$$

*Auflösung des Gleichungssystems (13) nach den Unbekannten $C_m$, $G_m$ und $G_a$ nach einem Vorschlag von V. und J. Rohliçek [89].*

Die von V. Rohliçek vorgeschlagenen Gleichungen (14) zur Berechnung der Zellparameter Membrankapazität $C_m$, Membranleitfähigkeit $G_m$ und Zugangswiderstand $G_a$ verwenden zwar alle vier gemessenen komplexen Leitwerte $Y_i$ [87], haben aber einen anderen Nachteil: Für bestimmte Zahlenwerte von $Y_i$ weisen sie Divergenzen auf, da eine Linearkombination der Meßwerte in den Nenner eines Bruches eingesetzt wird. In der Nähe dieser Polstellen kann eine kleine Abweichung eines der gemessenen komplexen $Y_i$-Werte einen großen Fehler in den berechneten Ergebnissen hervorrufen. So wurden mit Gleichung (14) in einigen Experimenten Zugangsleitfähigkeiten $G_a$ größer als 400 nS errechnet. Dieses Ergebnis steht jedoch im Widerspruch zu der vor der Messung bestimmten Eigenleitfähigkeit der Pipetten von 200-300 nS. Die Meßfehler in den komplexen Leitwerten gehen also nicht im gesamten Meßbereich mit gleicher Gewichtung in die berechneten Meßergebnisse $C_m$, $G_m$ und $G_a$ ein.

## 2.2 Eine neuentwickelte Methode zur Berechnung der Membrankapazität $C_m$, Membranleitfähigkeit $G_m$ und Zugangsleitfähigkeit $G_a$ mit Hilfe einer $\chi^2$-Minimierung

Da alle angewendeten Rechenverfahren die in den vorangegangenen Abschnitten erläuterten Nachteile aufweisen, wurde im Rahmen dieser Diplomarbeit eine neue Methode zur Berechnung der Membrankapazität $C_m$, der Membranleitfähigkeit $G_m$ und der Zugangsleitfähigkeit $G_a$ aus dem gemessenen komplexen Leitwerten $Y_i$ eingeführt. Das Verfahren der $\chi^2$-Minimierung erlaubt es, alle vier Meßwerte mit gleicher Gewichtung in die Berechnung der Werte $G_m$, $C_m$ und $G_a$ einzubringen.

Das neue Verfahren läßt sich auch dann anwenden, wenn mehr als zwei Frequenzen verwendet werden oder mehr als drei Zellparameter bestimmt werden sollen. Im Falle von mehr als zwei Frequenzen oder weiteren Parametern des Zellmodells läßt sich das Gleichungssystem (11) nicht mehr explizit nach den Variablen auflösen. Damit war die Verwendung der mathematischen Methode einer $\chi^2$-Minimierung Voraussetzung für die Entwicklung des in Kapitel 2.5 beschriebenen Vierfrequenz-Synchrondetektors.

Die $\chi^2$- oder Least-Squares-Minimierung gehört zu den sogenannten Maximum-Likelihood-Abschätzungen [6,85]. Ganz allgemein werden Maximum-Likelihood Schätzungen dann eingesetzt, wenn man eine Reihe verschiedener Meßdaten dazu verwenden möchte, unbekannte Parameter eines vorgegebenen Modells zu bestimmen [5].

Das Zellmodell, das zu Beginn des 2. Kapitels in Abb.6 dargestellt wurde, enthält die unbekannten Parameter Membrankapazität $C_m$, Membranleitfähigkeit $G_m$ und Zugangswiderstand $G_a$. Als Maß für die Übereinstimmung zwischen den Meßwerten und den Parametern des Modells dient eine sogenannte „figure-of-merit"-Funktion: Sie wird üblicherweise so gewählt, daß kleine Werte eine gute Übereinstimmung repräsentieren. In unseren Fall wählten wir die sogenannte $\chi^2$-Funktion, die in Gleichung (15) definiert ist.

$$\chi^2 = \sum_{i=1}^{N} \frac{\left| Y_{i,mess} - Y_i(C_m, G_m, G_a) \right|^2}{\sigma_i^2} \tag{15}$$

$Y_{i,mess}$: *Gemessener komplexer Leitwert bei den beiden Frequenzen $\omega_1$ und $\omega_2$,*
$Y_i(C_m,G_m,G_a)$: *berechneter komplexer Leitwert aus Gleichung (10),*
$\sigma_i$: *Angenommener Meßfehler des Wertes $Y_{i,mess}$,*
$N$: *Zahl der Frequenzen.*

Als Fehler der Einzelmessung $\sigma_i$ wählten wir die Gaußsche Fehlersumme aus dem Meßfehler des 12 Bit Analog-Digital-Wandlers $\sigma_{AD}=10^{-3} \cdot |Y_{max}|$ und dem geschätzten Meßfehler des Synchrondetektors $\sigma_{Syndet}= 0.01 \cdot |Y_i|$.

$$\sigma_i = \sqrt{\sigma_{AD}^2 + \sigma_{Syndet}^2} \tag{16}$$

$\sigma_i$ : *Fehler des Meßwertes* $Y_i$,
$\sigma_{AD}$ : *Meßfehler des Analog-Digital-Wandlers,*
$\sigma_{Syndet}$ : *Meßfehler des Synchrondetektors.*

Um die gesuchten Parameter $C_m$, $G_m$, und $G_a$ zu bestimmen, müssen die Werte von $C_m$, $G_m$ und $G_a$ so lange variiert werden, bis $\chi^2(C_m, G_m, G_a)$ minimal ist.

Hierzu verwendeten wir das FORTRAN-Programm „Valley“, das durch numerische Iteration das Minimum einer Funktion auffinden kann und von Prof. Königsmann zur Verfügung gestellt wurde. Da die vorhandenen Programme zur Datenaufnahme und -speicherung in der Programmiersprache $C^{++}$ geschrieben waren, wurde das Programm nach $C^{++}$ übersetzt und anschließend an die vorhandene Software angepaßt.

Zur Kontrolle des Experimentierablaufs ist es notwendig, die Zellparameter $C_m$, $G_m$, und $G_a$ unmittelbar zu verfolgen. Das Programm sollte also noch während der Datenaufnahme („online“) Meßwerte liefern. Hierzu mußte die Laufzeit des Programms an die Geschwindigkeit der Datenaufnahme angepaßt werden. Das kleinste sinnvolle Meßintervall ist bei der verwendeten Sinusfrequenz von $\nu$=406.9 Hertz etwa 2.5 Millisekunden.

$$\tau = \frac{1}{\nu} \approx 2.5\,\textit{Millisekunden}\,. \tag{17}$$

$\tau$: *Zeitintervall zwischen zwei Messungen.*
$\nu$: *Frequenz der verwendeten Sinusspannung.*

Das Minimierungs-Programm andererseits benötigte für die Berechnung der Parameter $G_m$, $C_m$ und $G_a$ mit einem relativen Fehler $<10^{-4}$ etwa 50 Iterationsschritte, die auf dem verwendeten Computer 3-4 Millisekunden beanspruchten. Daher wurde das Meßintervall auf fünf Millisekunden festgelegt. Ein zusätzlicher Zwischenspeicher für 500 Meßwerte im Analog-Digital-Wandler verhinderte den Datenüberlauf bei kurzen Verzögerungen.

## 2.3 Messung des Membranpotentials

Stimuliert man eine Zelle mit einem Agonisten, der eine spezifische Ionenleitfähigkeit der Zellmembran öffnet, kann man sowohl eine Erhöhung der Membranleitfähigkeit $G_m$ als auch eine Änderung des Membranpotentials $V_m$ in Richtung des Nernst-Potentials des jeweiligen Ions beobachten. Für unsere Untersuchungen an den Kolonzellen war es daher sinnvoll, neben der Membranleitfähigkeit auch das Membranpotential zu messen. Mit der vorhandenen Apparatur war dies nicht gleichzeitig möglich.

Um gleichzeitig Membranpotential, Membranleitfähigkeit und Membrankapazität zu messen, wurde im Rahmen dieser Diplomarbeit die folgende neue Idee entwickelt: Wenn der Zelle durch die Saugelektrode ein elektrisches Potential vorgegeben ist, das nicht mit ihrem Gleichgewichts-Membranpotential übereinstimmt, fließt über die Zellmembran ein Gleichstrom von Ionen, die ihrem elektrochemischen Gradienten über die Zellmembran folgen. Dieser Gleichstrom überlagert sich dem (doppel-)sinusförmigen Wechselstrom der Kapazitätsmessung. Integriert man den gemessenen Stromverlauf über eine Periode T, so heben sich die positive und die negative Halbwelle der Sinusfunktion gerade auf, als Wert des Integrals bleibt allein der Anteil des Gleichstromes.

Prinzipiell läßt sich damit aus der Größe des Gleichstromes bei bekannter Membranleitfähigkeit $G_m$ und Zugangsleitfähigkeit $G_a$ das Membranpotential nach dem Ohmschen Gesetz (18) berechnen.

$$V_m = I_{Gleich}\left(\frac{1}{G_a} + \frac{1}{G_m}\right) \qquad (18)$$

*Ohmsches Gesetz für das Membranpotential:* $V_m$: *Membranpotential,* $G_a$: *Zugangsleitfähigkeit,* $G_m$: *Membranleitfähigkeit.*

In die Berechnung gehen jedoch zwei Unbekannte $G_a$ und $G_m$ ein, die mit einer anderen Meßmethode bestimmt werden müssen. Zudem führt der Gleichstrom $I_{Gleich}$ über die Zellmembran im Laufe der Messung zu Elektrolytverschiebungen zwischen der Zelle und der extrazellulären Lösung, die die ionale Zusammensetzung des Zytoplasmas beeinflussen können.

Um diese beiden Fehlerquellen auszuschließen, wurde folgendes Konzept entwickelt und auch experimentell realisiert. Zur sinusförmigen Wechselspannung des Synchrondetektors wurde zusätzlich eine Gleichspannung variabler Größe addiert und die Gesamtspannung an die Saugelektrode angelegt. Die Größe der zugefügten

Gleichspannung ist abhängig vom gemessenen Gleichstrom durch die Zellmembran und wird durch den in Abb.8 dargestellten Rückkopplungs-Regelkreis gesteuert: Sobald ein positiver Nettostrom vom Zellinnern nach außen fließt, wird das Pipettenpotential erniedrigt, fließt er entgegengesetzt, so wird es erhöht.

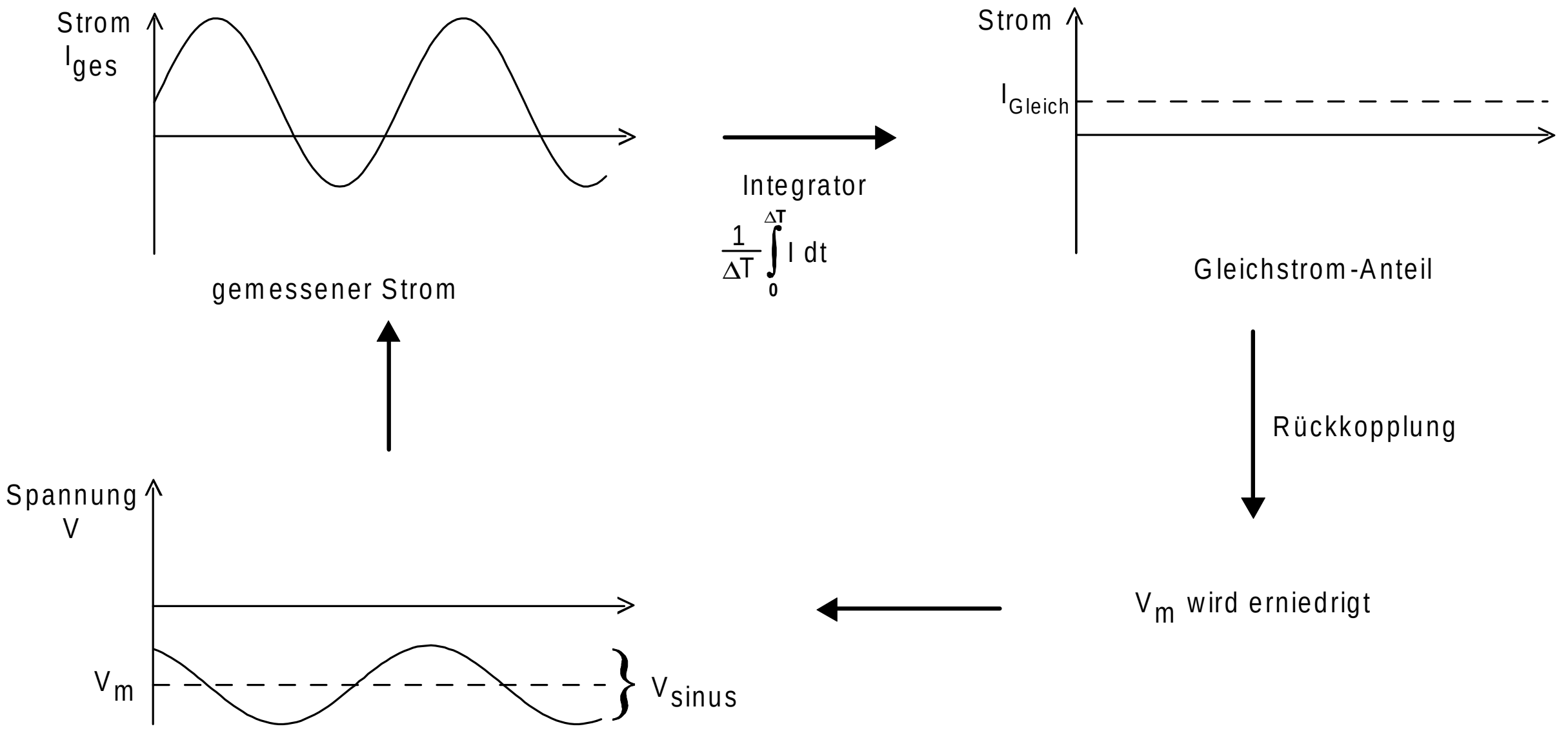

Abb.8: *Rückkopplung zur Potentialmessung (Erläuterungen im Text).*

Der Rückkopplungs-Regelkreis kompensiert damit ständig den Gleichstrom, der während der Kapazitätsmessung über die Zellmembran fließt, durch Veränderung des Pipettenpotentials. Wenn kein Gleichstrom mehr durch die Pipette fließt, entspricht die an die Pipette angelegte Spannung exakt dem Membranpotential der Zelle. Damit ist es möglich, ohne Kenntnis des Pipetten- oder Membranwiderstandes das Zellmembranpotential während der Kapazitätsmessung zu bestimmen. Zugleich fließt kein Netto-Ionenstrom mehr durch die Saugelektrode, der die Zusammensetzung des Zytoplasmas verändern könnte.

Als Integrator konnte ein bereits im Saugelektrodenverstärker vorhandener Analog-Integrationsschaltkreis („search-modus") verwendet werden [97]. Hierzu mußte nur noch die Länge T des Integrationsintervalls, die gleichzeitig auch die Zeitkonstante der Rückkopplung festlegt, an die Frequenz der Doppelsinusschwingung $\nu$ angepaßt werden. Um ein Mitschwingen des Rückkopplungs-Regelkreises mit den zur Kapazitätsmessung verwendeten Sinusspannungen mit Sicherheit zu verhindern, wurde die Zeitkonstante der Rückkopplung T sehr groß (100 Perioden der Sinusschwingung lang) gewählt. Die Zeitauflösung der Potentialmessung betrug damit 250 Millisekunden.

$$
\begin{aligned}
T &= 100 \cdot \tau \\
&= 100 \cdot \frac{1}{f_1} \\
&= 100 \cdot \frac{1}{406.9\ Hz} \\
&\approx 250\ msec
\end{aligned}
\qquad (19)
$$

*$T$: Zeitkonstante der Rückkopplungsschaltung,*
*$\tau$: Schwingungsdauer der Sinusschwingung,*
*$f_1$: Frequenz der (langsameren) Sinusspannung.*

Die gemessenen Membranpotentiale wurden über einen A/D-Wandler digitalisiert und zusammen mit $G_a$, $G_m$ und $C_m$ aufgezeichnet.

## 2.4 Testmessungen

Vor Beginn der Messungen an den Kolonzellen wurde die Apparatur zur Kapazitätsmessung auf die Meßgenauigkeit und auf systematische Meßfehler untersucht. Von besonderem Interesse war es, zu überprüfen, ob Änderungen eines der gemessenen Zellparameter einen systematischen Einfluß auf die anderen Meßwerte haben.

Um die Apparatur zu testen, wurden verschiedene Testschaltkreise entworfen und aufgebaut, die ein Abbild der elektrischen Eigenschaften der Zelle darstellen. Alle Schaltungen wurden zur elektrischen Abschirmung in Aluminiumgehäuse eingeschlossen und an Stelle des Pipettenhalters direkt mit dem Vorverstärker verbunden. Um die Unabhängigkeit der verschiedenen gemessenen Parameter voneinander zu überprüfen, ließ sich in jedem der Testschaltkreise während der Messung jeweils ein Parameter variieren, während die Konstanz der übrigen Meßgrößen geprüft wurde. Zunächst sollte der Einfluß von Änderungen der Membranleitfähigkeit $G_m$ auf die übrigen Parameter untersucht werden. Hierzu wurde die in Abb.9 dargestellte Schaltung entworfen.

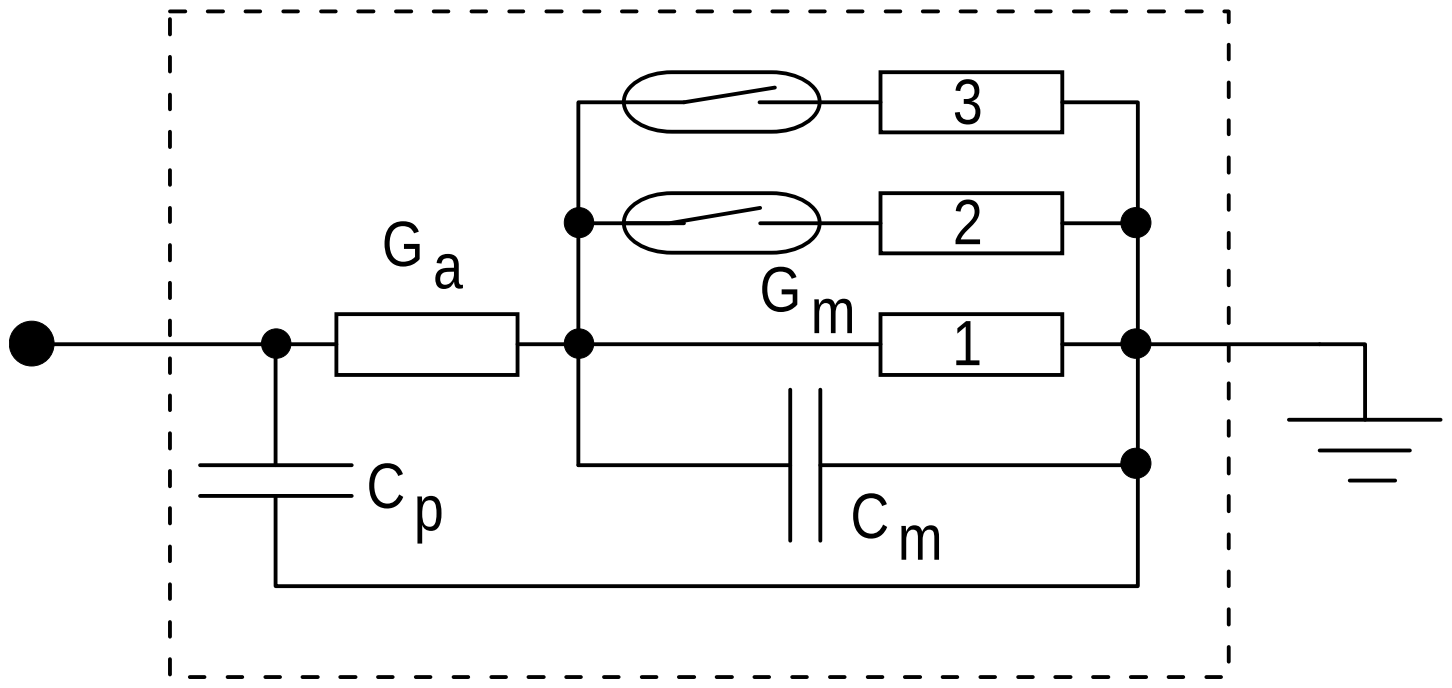

$C_m$ = 15.0 pF

$C_p$ = 4.7 pF

$G_m$ = 10.0 nS (1)
25.2 nS (1+2)
43.3 nS (1+3)
58.5 nS (1+2+3)

$G_a$ = 45.5 nS

Abb.9: *Schaltplan des $G_m$-Testschaltkreises.*

Die Schaltung besteht aus einem Eingangswiderstand, der die Ohmsche Zugangsleitfähigkeit $G_a$ simuliert. Parallel zu ihm liegt ein Kondensator $C_p$, der der Pipettenkapazität entspricht. Die Zellmembrankapazität wird durch eine Kondensator $C_p$, die Membranleitfähigkeit $G_m$ durch drei unterschiedliche Ohmsche Widerstände erzeugt. Durch Hinzuschalten der verschiedenen Widerstände kann eine Änderung der Membranleitfähigkeit simuliert werden.

Da die Kapazität eines mechanischen Schalters oder eines Drehwiderstandes (Potentiometers), etwa ein Pikofarad, auch die Kapazität der Schaltung beim Verändern von $G_m$ beeinflussen würde, wählten wir magnetgesteuerte Mikro-Reedkontakte als Umschalter. Sie werden von der Außenseite des Aluminium-Gehäuses aus mit Hilfe eines Permanentmagneten geöffnet und geschlossen. Mit zwei Reedkontakten kann damit zwischen $2^2 = 4$ verschiedenen Membranleitfähigkeiten gewählt werden. Die Eigenkapazität der verwendeten Mikro-Reedkontakte, 200 fF, stellt die untere Grenze für die Genauigkeit der Testschaltung dar. Die Genauigkeit des Absolutwertes der verwendeten Widerstände beträgt ± 5%. Die Messungen an der $G_m$-Testschaltung ergaben die in Tab.2 und Abb.10 dargestellten Ergebnisse.

| Sollwert der verwendeten Widerstände ± Toleranz (5%) | gemessene Membranleitfähigkeit $G_m$ |
|---|---|
| 10.0 ± 0.5 nS | 10.4 nS |
| 25.3 ± 1.3 nS | 25.1 nS |
| 43.3 ± 2.2 nS | 43.2 nS |
| 58.5 ± 2.9 nS | 56.9 nS |

Tab.2: *Größe der verwendeten Widerstände im Vergleich zu den gemessenen Werten für die Membranleitfähigkeit der $G_m$-Testschaltung. Alle gemessenen Leitfähigkeiten $G_m$ liegen im Toleranzbereich der verwendeten Widerstände.*

Die gemessenen Membranleitfähigkeiten $G_m$ liegen alle innerhalb des Toleranzintervalls der verwendeten Widerstände. Die relative Meßfehler $G_m$-Bestimmung ist daher sicherlich kleiner als das Toleranzintervall der verwendeten Widerstände, d.h. kleiner als 5%.

Zusätzlich wurde der Einfluß von Änderungen der Membranleitfähigkeit auf die Membrankapazitäts-Messung untersucht:

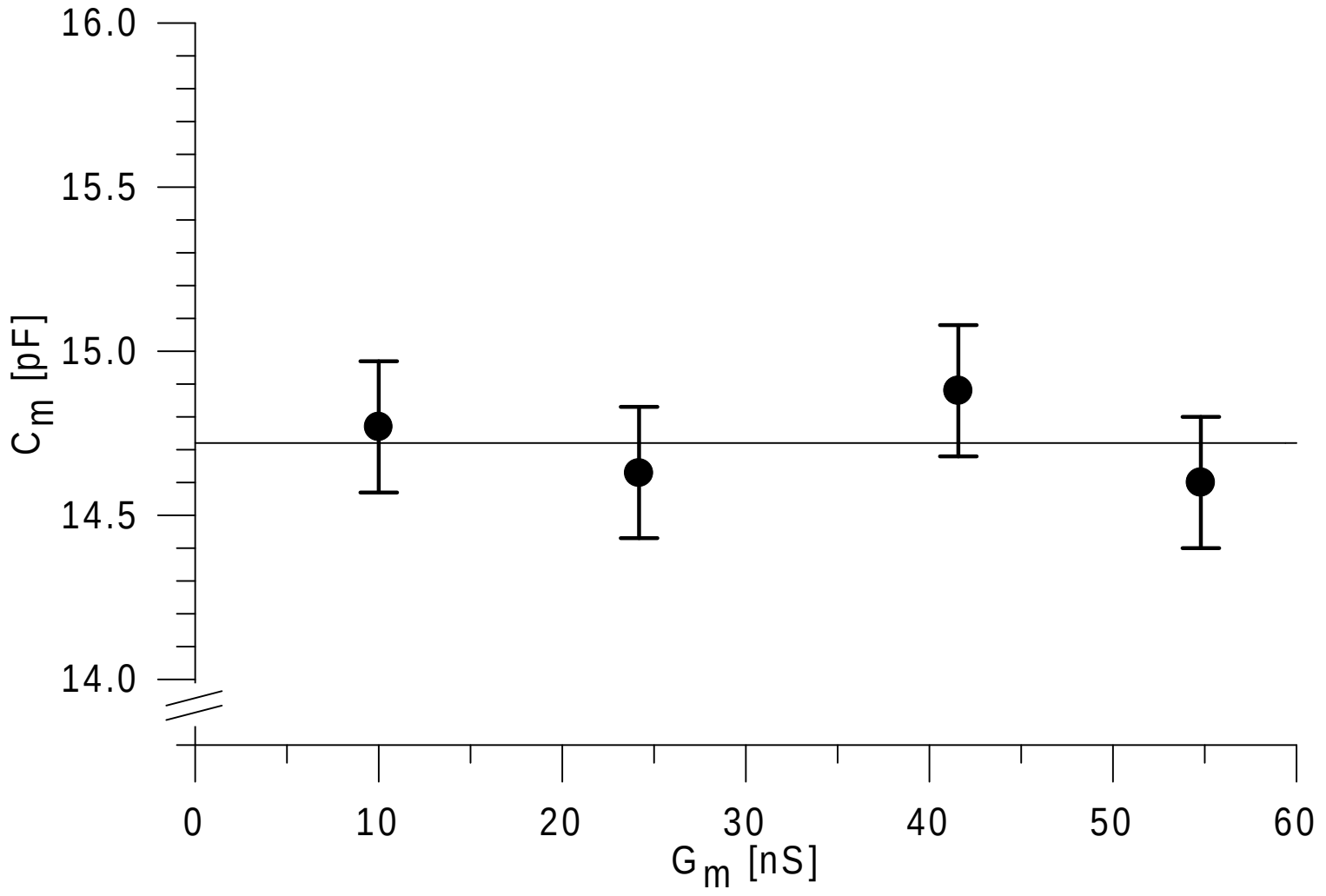

Abb.10: *Einfluß der Membranleitfähigkeit $G_m$ auf die Membrankapazitätsmessung $C_m$. Der Mittelwert der Membrankapazität beträgt 14.72 ± 0.13 pF (in n=4 Messungen). Die gemessenen Werte liegen im Rahmen der Toleranz des verwendeten Kondensators (15.0 ± 1.5 pF).*

Die im Experiment beobachtbaren Abweichungen vom Sollwert 15.0 Pikofarad werden durch die Kapazität der Reedkontakte von 200 fF erklärt. Der Einfluß einer Änderung der Membranleitfähigkeit auf die Kapazitätsmessung ist daher kleiner als $\frac{200\,fF}{46.9\,nS} = 4.26\,\frac{fF}{nS}$.

Mit einer zweiten Schaltung sollte der Einfluß von Kapazitätsänderungen auf die Messung der Membranleitfähigkeit $G_m$ geprüft werden. Die hierzu entworfene Schaltung ist in Abb.11 dargestellt. Eine Eingangsleitfähigkeit $G_a$ und ein Kondensator $C_p$ entsprechen der Ohmschen Leitfähigkeit und der Kapazität der Pipette. Die Membranleitfähigkeit $G_m$ wird durch einen Ohmschen Widerstand, die Membrankapazität durch einen Drehkondensator $C_p$ simuliert. Die Größe der Membrankapazität konnte durch Verstellen des Drehkondensators während der Messung verändert werden. Der Ohmsche Widerstand des Drehkondensators war vernachlässigbar groß (größer als 10 Gigaohm).

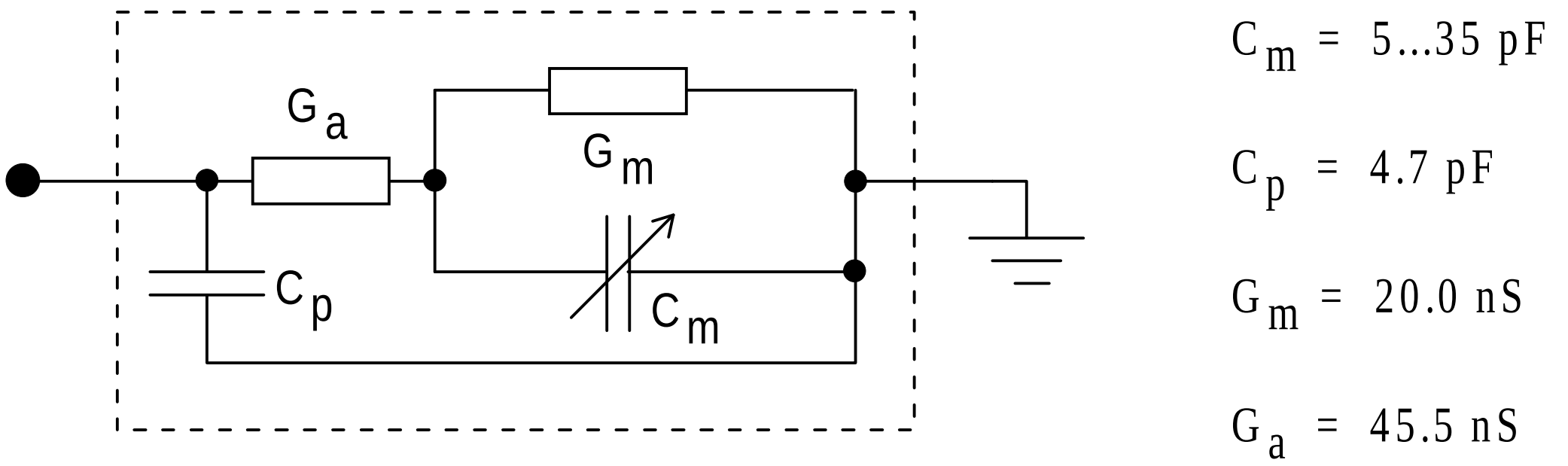

Abb.11: *Schaltplan des $C_m$-Testschaltkreises*

Die Messungen lieferten folgende Ergebnisse:

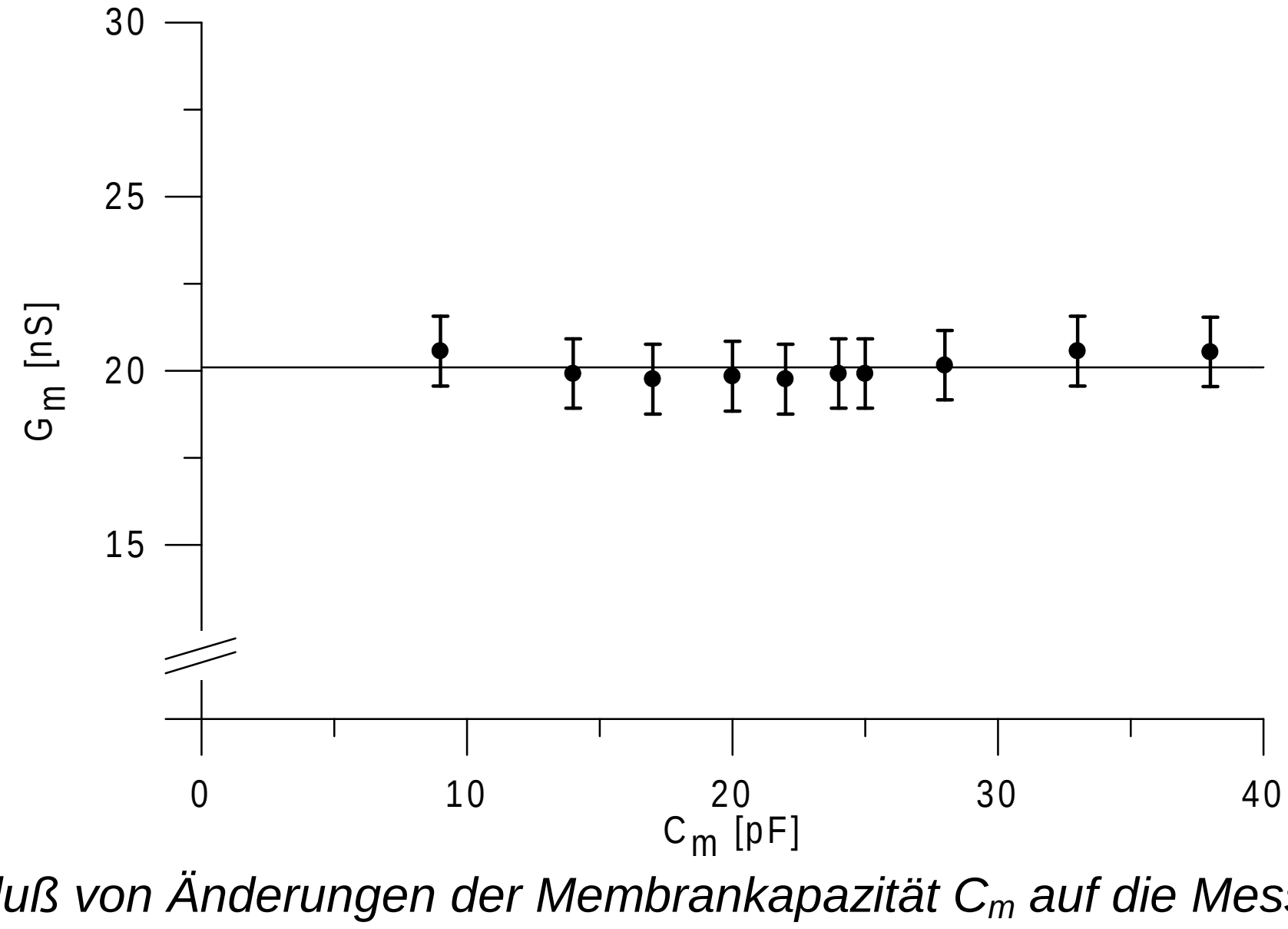

Abb.12: *Einfluß von Änderungen der Membrankapazität $C_m$ auf die Messung der Membranleitfähigkeit $G_m$. Die mittlere Membranleitfähigkeit beträgt 20.09 nS $\pm$ 0.33 nS (in n=10 Messungen).*

Die beobachteten Abweichungen der $G_m$- Messung sind kleiner als 1 nS bei Variationen von $C_m$ zwischen 5 und 35 Pikofarad. Der Einfluß einer Kapazitätsänderung auf die Leitfähigkeitsmessung ist daher kleiner als $\frac{1\,nS}{30\,pF} = 33.3\,\frac{pS}{pF}$.

Für alle oben genannten Testmessungen wurde der vierte Parameter des Testschaltkreises, die Pipettenkapazität, durch eine analoge Kompensationsschaltung im Saugelektrodenverstärker vor der eigentlichen Messung kompensiert. Diese Schaltung war wie in Abb.13 dargestellt aufgebaut und arbeitete nach folgendem, von J. Sigworth erstmals vorgeschlagenem Prinzip [96]:

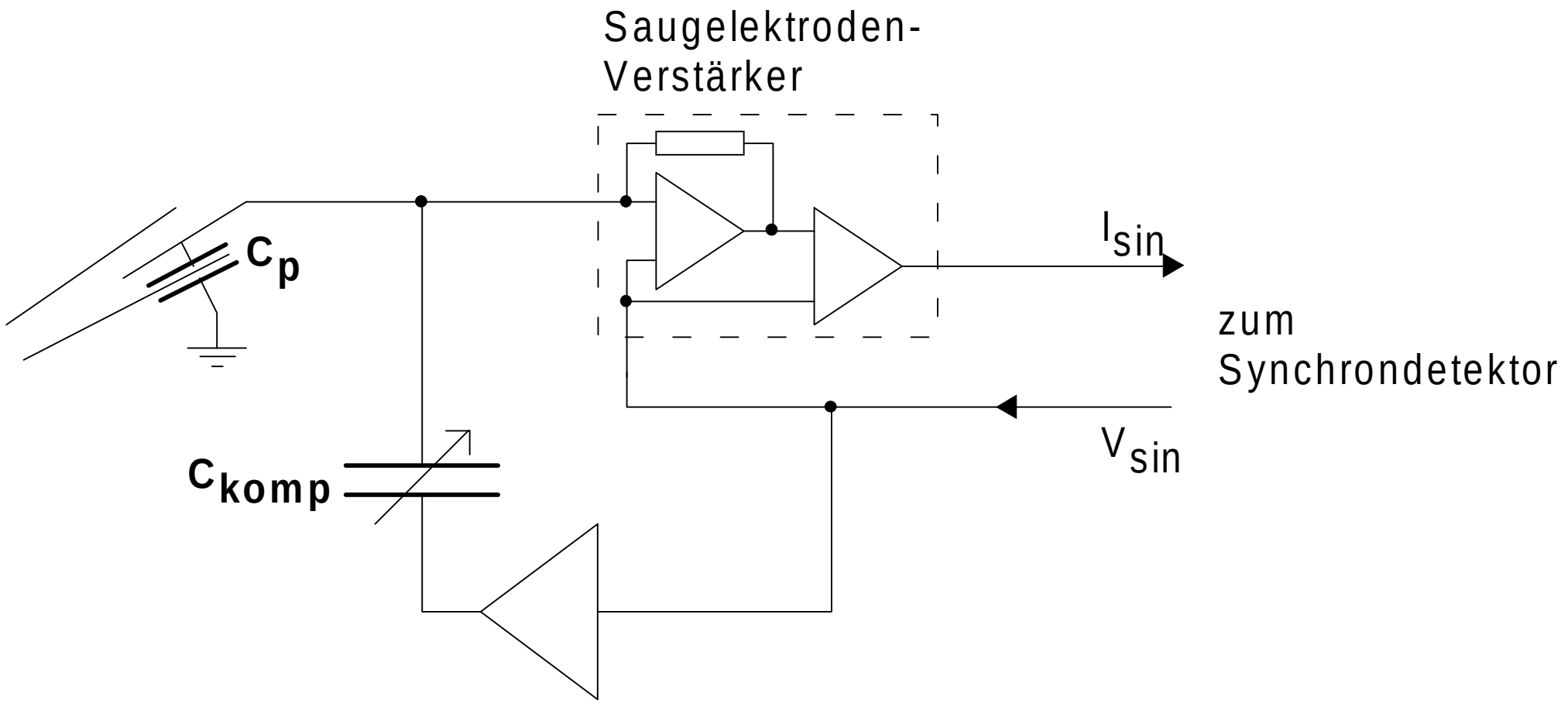

Abb.13: *Kompensationsschaltung für die Pipettenkapazität (nach [96]).*
*$C_p$: Pipettenkapazität, $C_{komp}$: Kompensationskondensator.*

Bei jeder Änderung des Pipettenpotentials $V_m$ wird durch Rückkopplung über einen zusätzlichen Verstärker und den Kompensations-Kondensator $C_{komp}$ ein Strom über die Pipette geleitet. Wenn die Kapazität des Kompensations-Kondensators gleich groß wie die der Pipette gewählt wird, so heben sich der Umlade-Strom des Pipettenkondensators und der Strom über den Ausgleichs-Kondensator gegenseitig auf.

Die Kapazität des Kompensations-Kondensators kann am Saugelektroden-Verstärker eingestellt und der Größe der Pipettenkapazität angepaßt werden. Hierzu wird eine sinusförmige Wechselspannung an die Pipette angelegt, bevor ein Zugang zum Zellinneren besteht und unter Beobachtung des Stromes auf einem Oszilloskop die Kapazität des Kompensationskondensators abgeglichen [35,65,97]. Da die Kapazität von Pipette zu Pipette variiert, wird der Kapazitätsabgleich für jedes Experiment neu durchgeführt.

Ist die Pipettenkapazität nicht richtig abgeglichen, zeigten sich bei unseren Testmessungen folgende systematische Fehler: Bei einer Unterkompensation der Pipettenkapazität ergab sich eine positive Korrelation zwischen Leitfähigkeit und Kapazitätsänderungen, d.h. eine Erhöhung von $G_m$ bewirkte auch einen parallelen Anstieg von $C_m$ und umgekehrt. Eine Überkompensation der Pipettenkapazität hingegen bewirkte eine negative Korrelation zwischen $G_m$ und $C_m$ .

Die Stärke der Korrelation war abhängig von der Höhe des Fehlabgleichs. Sie betrug 34.0 fF/nS bei Über- und 25.5 fF/nS bei Unterkompensation von $C_p$ um jeweils 200 Femtofarad.

In unseren Experimenten wurde die Pipettenkapazität nach der Ausbildung des Seal-Widerstandes zwischen Zelle und Pipette abgeglichen, indem eine Sinusspannung mit zehn Millivolt Amplitude an die Pipette gelegt wurde und die Kompensationskapazität so lange variiert wurde, bis der (dann fließende kapazitative) Strom minimal war. Nach dem Einreißen der Zellmembran und der Ausbildung der Ganzzell-Ableitung war eine Überprüfung des Pipetten-Abgleichs nicht mehr möglich, da die Pipettenkapazität nicht mehr von den anderen Zellparametern getrennt werden konnte.

Zusammenfassend konnte mit den Testmessungen gezeigt werden, daß:

- der Absolutwert der Leitfähigkeitsmessung einen Fehler kleiner als 5% aufweist.
- der relative Einfluß einer Leitfähigkeitsänderung auf die Kapazitätsmessung weniger als 4.26 fF/nS beträgt.
- der relative Einfluß einer Kapazitätsänderung auf die Leitfähigkeitsmessung kleiner als 33.3 pS/pF ist.
- Die Größe des systematischen Fehlers durch Pipettenfehlabgleich abgeschätzt werden kann.

## 2.5 Der Vierfrequenz-Synchrondetektor

Die Testmessungen am Zweifrequenz-Synchrondetektor im vorangegangenen Kapitel haben gezeigt, daß neben den drei durch die Messung bestimmten Zellparametern $C_m$, $G_m$ und $G_a$ auch die Pipettenkapazität $C_p$ einen systematischen Einfluß auf die Meßergebnisse hat. Um den Einfluß von $C_p$ ganz auszuschalten, wollten wir daher die Pipettenkapazität als vierten Parameter mit unserer Messung bestimmen. Dazu wurde das elektrische Modell, das den bisherigen Messungen zur Zellkapazität zugrunde liegt, um die Pipettenkapazität erweitert. In dem Ersatzschaltbild der Zelle, das in Abb.14 nochmals dargestellt ist, wurde die Pipettenkapazität als Meßgröße berücksichtigt.

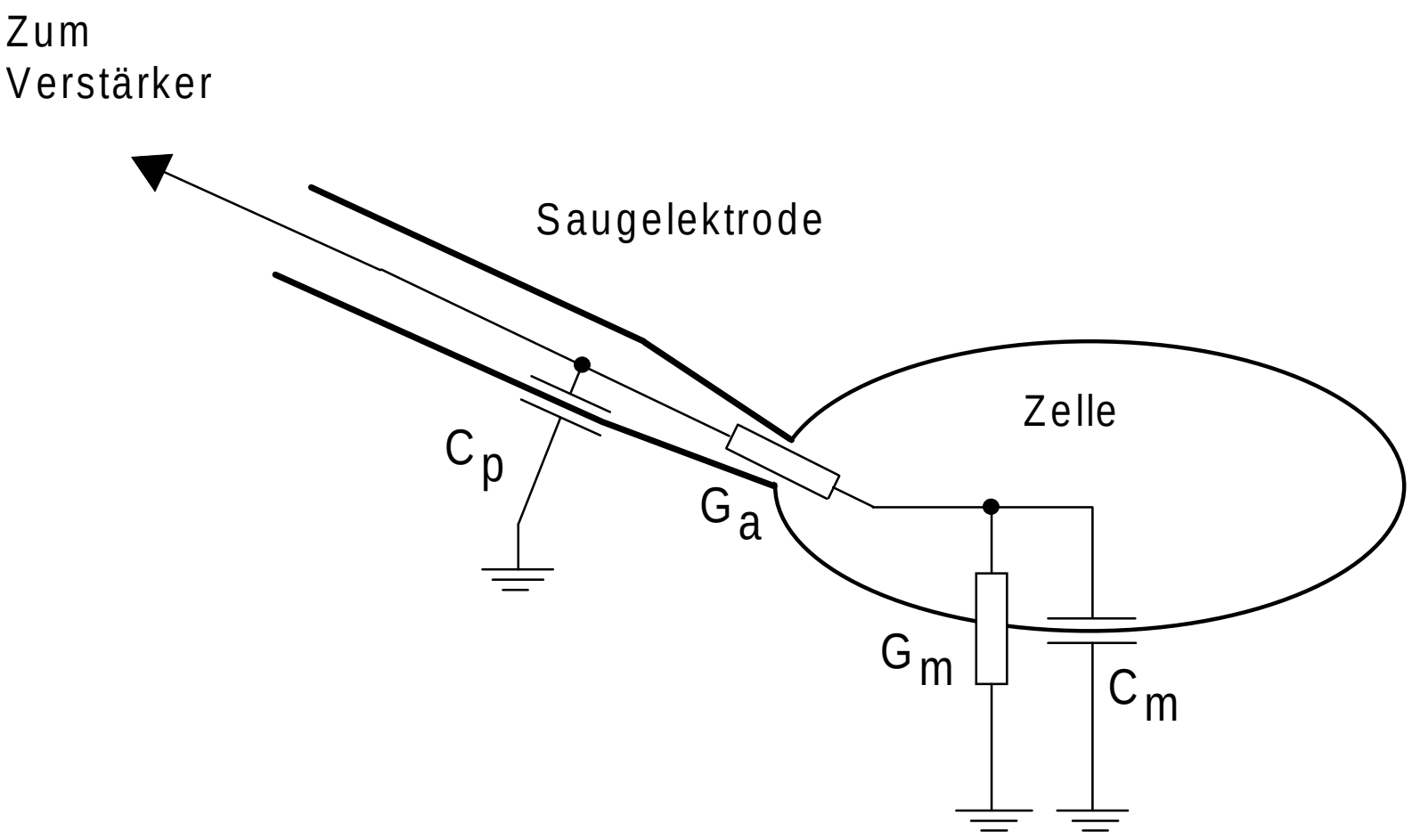

Abb.14: *Ersatzschaltbild einer Zelle mit angelegter Saugelektrode.*
*$C_p$: Pipettenkapazität, $G_a$: Zugangsleitfähigkeit, $G_m$: Membranleitfähigkeit, $C_m$: Membrankapazität.*

Wendet man die Kirchhoffschen Regeln für die Parallel- und Serienschaltung von Widerständen und Kapazitäten in Abb.14 an, erhält man Gleichung (20) für den komplexen Leitwert der Zelle.

$$\tilde{Y}_i = G_a \frac{G_m + i\omega_i C_m}{G_a + G_m + i\omega_i C_m} + i\omega_i C_p \qquad (20)$$

In die $\chi^2$-Minimierung, die in Kapitel 2.2 beschrieben ist, wurde nun als vierte Unbekannte die Pipettenkapazität $C_p$ hinzugefügt, wie in Gleichung (21) dargestellt ist.

$$\chi^2 = \sum_{i=1}^{N} \frac{\left| Y_{i,mess} - \tilde{Y}_i(C_m, G_m, G_a, C_p) \right|^2}{\sigma_i^2} \qquad (21)$$

*$Y_{i,mess}$: Gemessener komplexer Leitwert bei den beiden Frequenzen $\omega_1$ und $\omega_2$,*
*$\tilde{Y}_i(C_m, G_m, G_a, C_p)$: berechneter komplexer Leitwert aus Gleichung (20),*
*$\sigma_i$: Angenommener Meßfehler des Wertes $Y_{i,mess}$, $N$: Zahl der Frequenzen.*

Es zeigte sich jedoch, daß das Gleichungssystem, das nun mit vier Meßgrößen und vier Unbekannten nicht mehr überbestimmt ist, keine Berechnung aller vier gesuchten Parameter mit der geforderten Meßgenauigkeit zuläßt: Bei einer mittleren Kapazität der Zellen in der Größenordnung von 10 pF traten bereits Schwankungen von mehr als einem Pikofarad auf, so daß keine sinnvolle Kapazitätsmessung mehr möglich war.

Daher haben wir die Messungen auf insgesamt vier Sinusspannungen unterschiedlicher Frequenz erweitert: Vier Lock-In-Verstärker ermöglichen es, bei vier verschiedenen Frequenzen jeweils den Real- und Imaginärteil des komplexen Leitwerts von

Pipette und Zelle zu bestimmen. Als Frequenzen wählten wir 203.45 Hz, 406.9 Hz, 813.8 Hz und 1627.6 Hz. Alle Frequenzen stehen in ganzzahligen Verhältnissen zueinander und werden durch Frequenzteilung aus demselben Schwingquarz erzeugt. Damit kann eine stabile Phasenlage der vier Sinusschwingungen zueinander sichergestellt werden. Über acht 12 Bit A/D-Wandler werden die gemessenen Real- und Imaginärteile von $Y_i$ in den Computer aufgenommen. Die vier Unbekannten $C_p$, $C_m$, $G_a$ und $G_m$ werden durch eine $\chi^2$-Minimierung des nun vierfach überbestimmten Gleichungssystems aus acht Gleichungen für die Real- und Imaginärteile von $Y_1$ bis $Y_4$ ermittelt. Hierzu diente wiederum das Minimierungsprogramm „Valley“, das entsprechend modifiziert wurde. Um die erhöhte Rechenzeit auf dem PC auszugleichen, wurde die Datenaufnahme-Intervall auf 10 Millisekunden verdoppelt.

## 2.6 Testmessungen am Vierfrequenz-Synchrondetektor

Die folgenden Testmessungen, die bereits am Zweifrequenz-Synchrondetektor durchgeführt wurden, wurden nun mit den im Kaptitel 2.4 beschriebenen Test-Schaltkreisen durchgeführt. Zunächst wurde der Einfluß einer Membrankapazitätsänderung auf die gemessenen Leitfähigkeitswerte untersucht. Die Meßergebnisse sind in Abb.15 dargestellt.

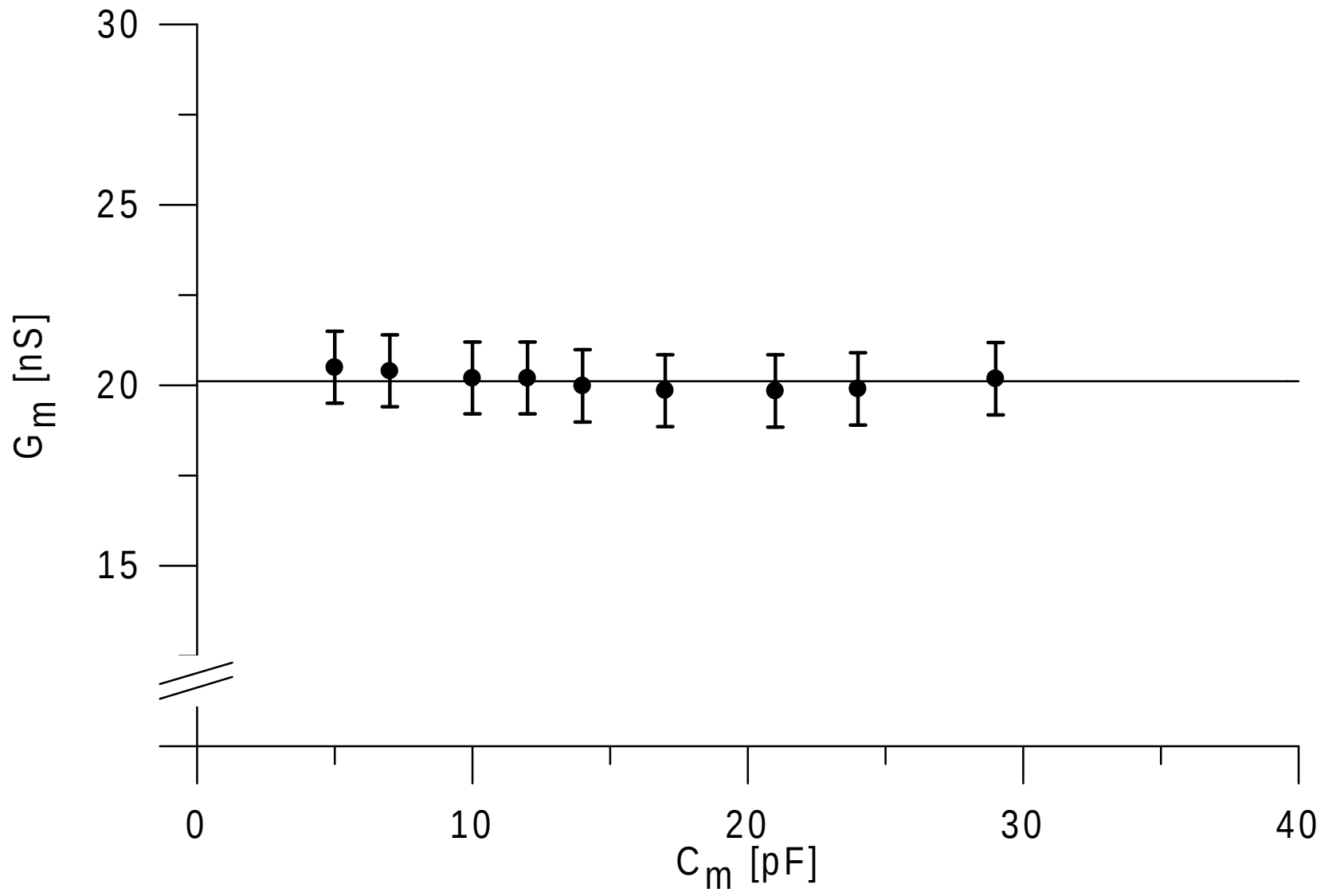

Abb.15: *Einfluß der Membrankapazität $C_m$ auf die gemessenen Membranleitfähigkeit $G_m$ . Die mittlere Membranleitfähigkeit beträgt 20.16 ± 0.23 nS (in n=9 Messungen).*

Mit der zweiten Testschaltung, die in Abb.11 dargestellt ist, wurde der Einfluß einer Änderung der Membranleitfähigkeit auf die gemessene Membrankapazität bestimmt.

Dazu wurde $G_m$ durch Hinzuschalten weiterer Widerstände mit Hilfe zweier Mikro-Reedkontakte variiert. Die Meßergebnisse sind in der Abb.16 wiedergegeben.

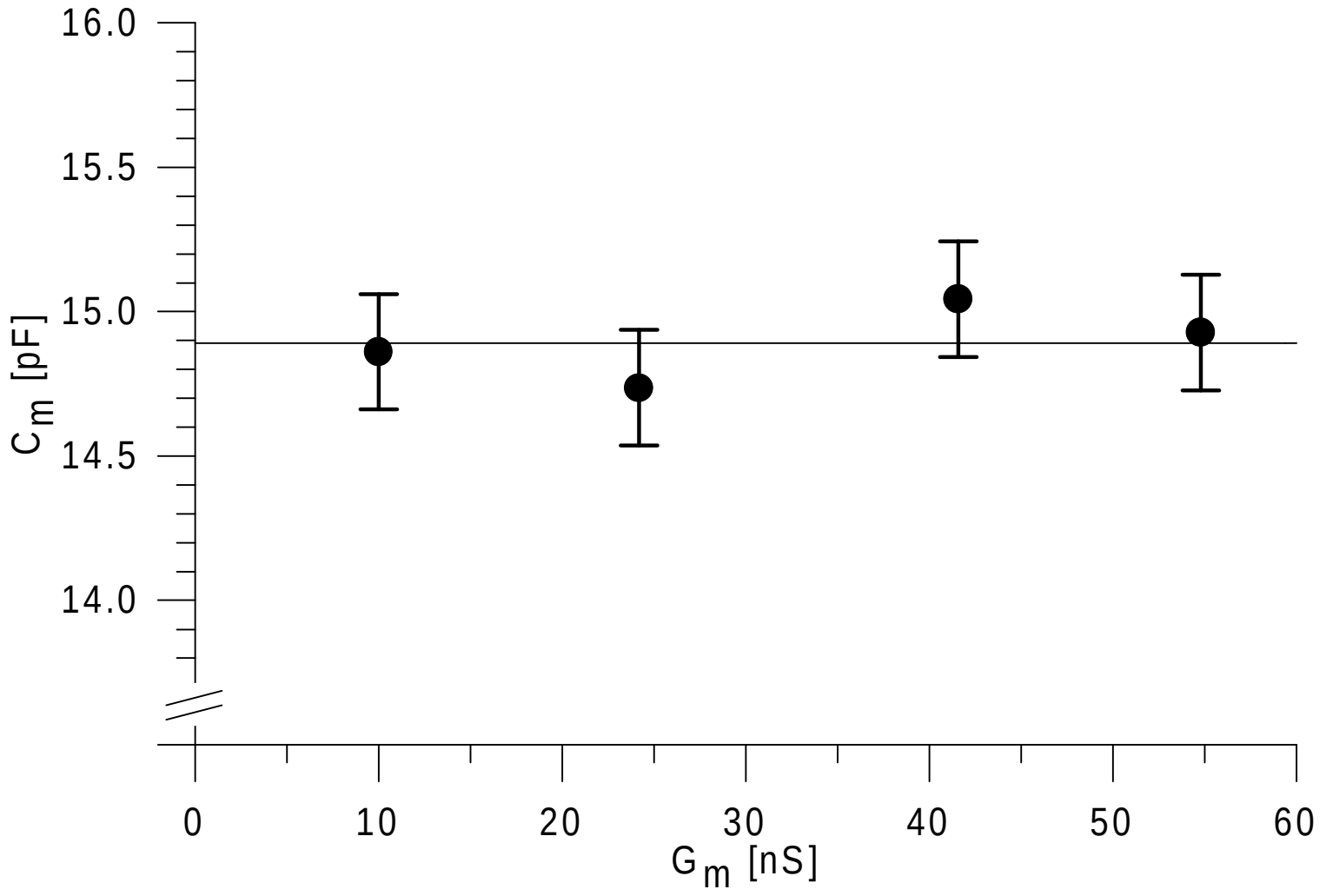

Abb.16: *Einfluß der Membranleitfähigkeit $G_m$ auf die gemessene Membrankapazität $C_m$ . Die mittlere Membrankapazität beträgt 14.89 ± 0.13 pF (in n=4 Messungen).*

Zusätzlich wurde eine Variation der Pipettenkapazität $C_p$ durch unterschiedliche Einstellung des im Kapitel 2.5 beschriebenen Kompensationsschaltkreises des Saugelektroden-Verstärkers für die Pipettenkapazität simuliert. Es ergab sich die in Abb.17 und Abb.18 dargestellte Abhängigkeit der gemessenen Membranleitfähigkeit $G_m$ und Membrankapazität $C_m$ von der Pipettenkapazität $C_p$.

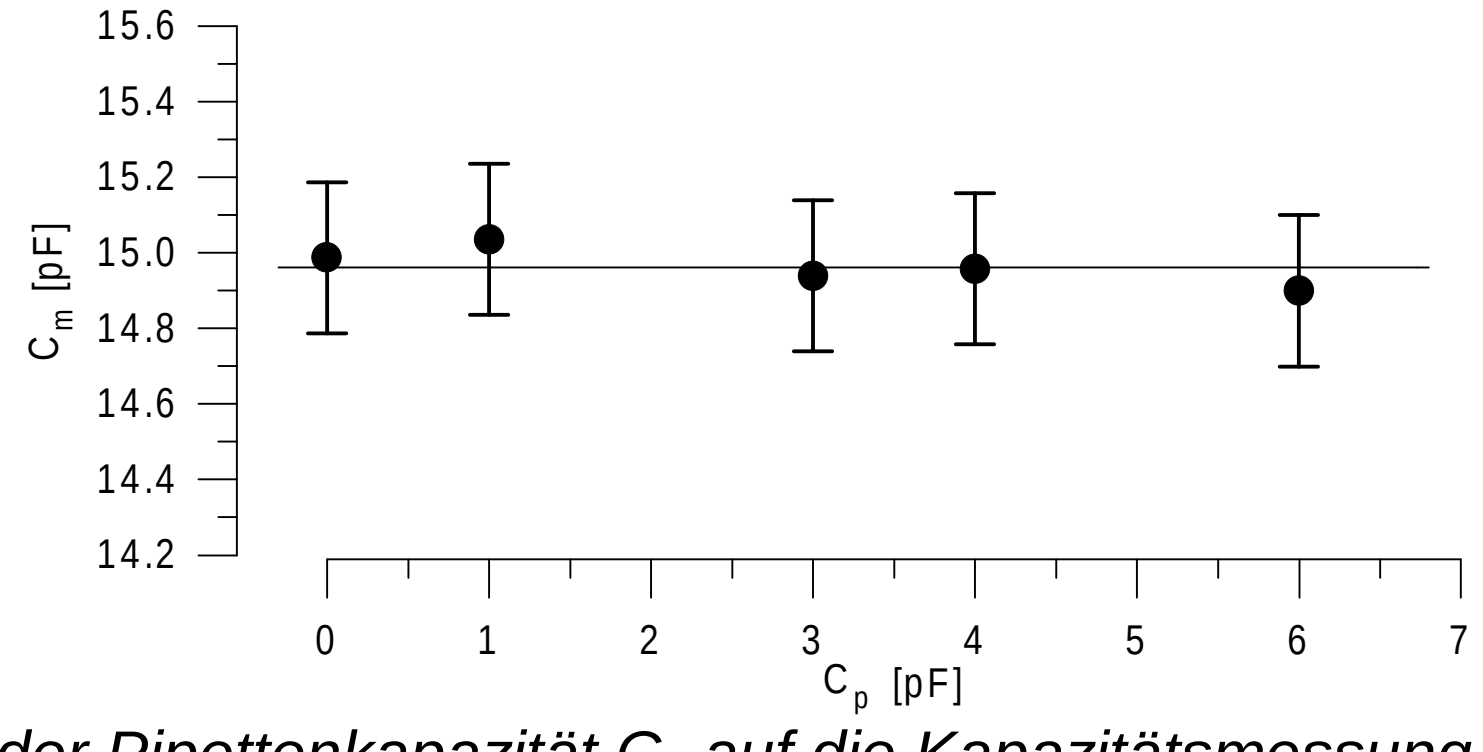

Abb.17: *Einfluß der Pipettenkapazität $C_p$ auf die Kapazitätsmessung (Experiment am $G_m$-Testschaltkreis aus Abb.9). Die mittlere Membrankapazität beträgt 14.96 ± 0.05 pF (in n=5 Messungen).*

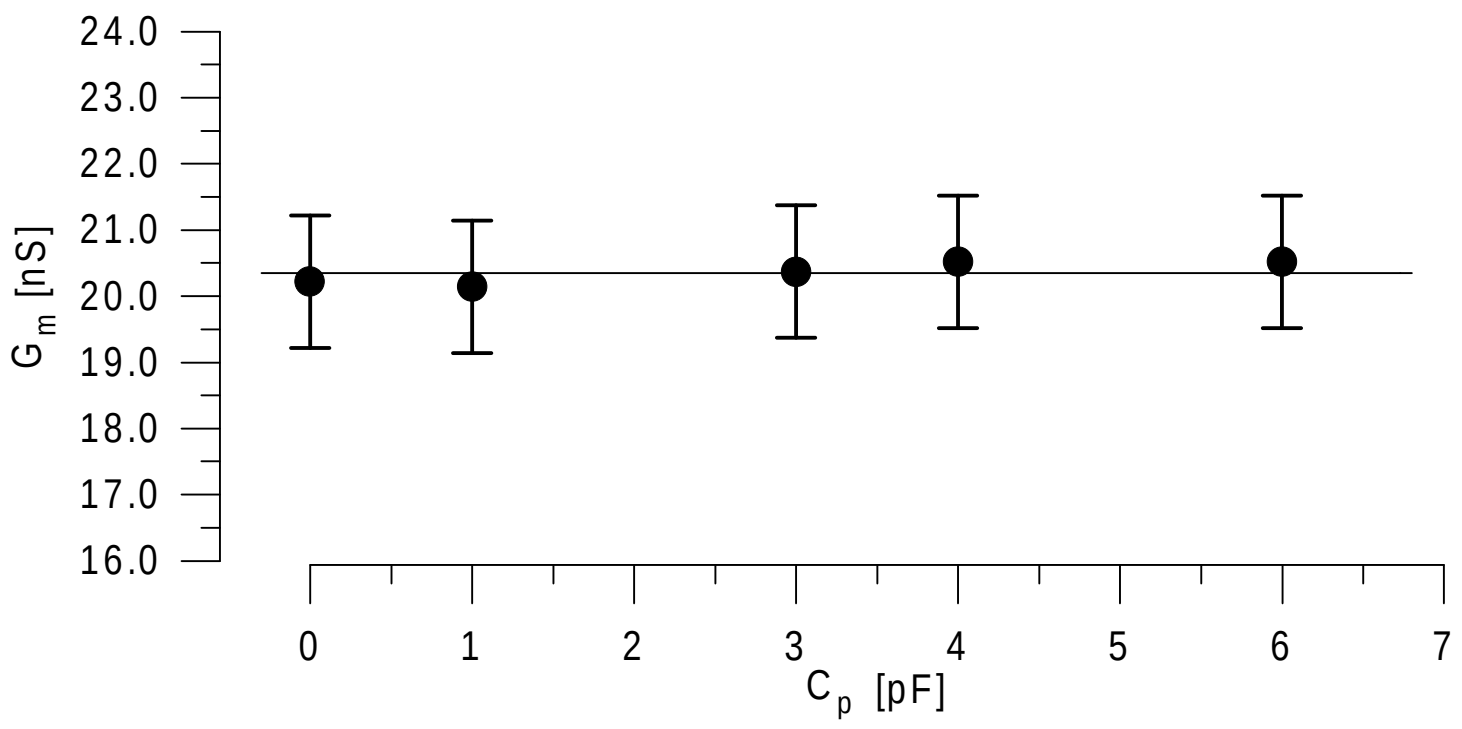

Abb.18: *Einfluß der Pipettenkapazität $C_p$ auf die Leitfähigkeitsmessung (Experiment am $C_m$-Testschaltkreis aus Abb.11). Die mittlere Membranleitfähigkeit beträgt 20.35 ± 0.17 nS (in n=5 Messungen).*

In Abb.17 und Abb.18 erkennt man, daß eine Variation der Pipettenkapazität $C_p$ um 6 pF keinen systematischen Einfluß mehr auf die Kapazitäts- und Leitfähigkeitsmessung hat.

Die Ergebnisse der Testmessungen am Vierfrequenz-Synchrondetektor kann man wie folgt zusammenfassen:

- Die Messung der Membrankapazität $C_m$ und der Membranleitfähigkeit $G_m$ ist unabhängig von einer Kompensation der Pipettenkapazität.
- Der Einfluß einer Leitfähigkeitsänderung auf die Kapazitätsmessung ist kleiner als 4.26 fF/nS.
- Der Einfluß einer Kapazitätsänderung auf die Leitfähigkeitsmessung ist kleiner als 33.3 pS/pF.

Da der Vierfrequenz-Synchrondetektor erst zum Ende der vorliegenden Arbeit von der Elektronik-Werkstatt fertiggestellt wurde, ist der Großteil der vorliegenden Daten noch mit dem Zweifrequenz-Verfahren aufgenommen worden. Die mit dem Vierfrequenz-Verfahren durchgeführten Experimente sind am Ende des Ergebnisteils aufgeführt. Eine qualitativer Unterschied zwischen den mit den beiden Verfahren aufgenommenen Daten war nicht zu erkennen.

## 2.7 Statistik

In allen Diagrammen wird jeweils der Mittelwert und die Standardabweichung des Mittelwertes (SEM) aus einer Zahl von *n* Experimenten angegeben. Der SEM wurde gemäß Gleichung (22) berechnet.

$$SEM = \frac{\sqrt{\sum_{i=1}^{N}\left(x_i - \bar{x}\right)^2}}{N-1} \quad (22)$$

Als statistischen Test für die Signifikanz einer Beobachtung wurde der einseitige *Student-t-Test* verwendet. Für „gepaarte“ Daten (Änderungen der Leitfähigkeit oder Kapazität *einer* Zelle nach der Gabe eines Agonisten) wurde die Hypothese getestet, daß der Mittelwert der Verteilung der Kapazitäts- oder Leitfähigkeitsänderungen $\Delta C_m$ bzw. $\Delta G_m$ verschieden von Null ist. Hierzu wurde die Testgröße t gemäß Gleichung (23) gebildet und anhand der tabellierten *Student-t-Verteilung* die Irrtumswahrscheinlichkeit der Testhypothese bestimmt [10].

$$t = \frac{\left|\bar{x}\right|}{\sqrt{\frac{\sigma^2}{n}}}$$

$\bar{x}$: Mittelwert der Verteilung der Meßwerte (23)
$\sigma$: Standardabweichung der Meßwerte
n: Anzahl der Messungen

Ein Ereignis mit einer Irrtumswahrscheinlichkeit kleiner als fünf Prozent bzw. einem Konfidenzlevel größer als 95% wurde als signifikant betrachtet und in den Ergebnisdiagrammen mit einem Stern (*) gekennzeichnet.

Zur statistischen Untersuchung „ungepaarter“ Meßdaten (*zwei* Verteilungen von Kapazitäten oder Leitfähigkeiten verschiedener Zellen, z.B. Zellen mit und Zellen ohne Toxininkubation) wurde der einseitige ungepaarte *Student-t-Test* verwendet. Als Testhypothese wurde angenommen, daß die Mittelwerte der beiden Verteilungen der Meßwerte verschieden voneinander sind. Zur Prüfung der Testhypothese wurde die Testgröße $\tilde{t}$ nach Gleichung (24) gebildet [10].

$$\tilde{t} = \frac{\left|\bar{x}_1 - \bar{x}_2\right|}{\sqrt{\left(n_1 - 1\right)\sigma_1^2 + \left(n_2 - 1\right)\sigma_2^2}}\sqrt{\frac{n_1 n_2 \left(n_1 + n_2 - 2\right)}{n_1 + n_2}} \quad (24)$$

*$\sigma_1$ : Standardabweichung der Meßwerte in der ersten Meßreihe*
*$\sigma_2$ : Standardabweichung der Meßwerte in der zweiten Meßreihe*
*$n_1$ : Zahl der Messungen in der ersten Meßreihe*
*$n_2$ : Zahl der Messungen in der zweiten Meßreihe*
*$\bar{x}_i$ : Mittelwert der Meßwerte der i - ten Meßreihe*

Anhand der tabellierten *Student-t-test*-Verteilung wurde das Konfidenzlevel und die Irrtumswahrscheinlichkeit für die Testhypothese bestimmt. Ein Ereignis mit einer Irrtumswahrscheinlichkeit kleiner als fünf Prozent (bzw. einem Konfidenzlevel > 95%) wurde wiederum als signifikant angesehen.

Um einen möglichen Zusammenhang zwischen den in Kapitel 4.6 beobachteten Kapazitätsänderungen $\Delta C_m$ und Leitfähigkeitsänderungen $\Delta G_m$ zu untersuchen, wurde der empirische Korrelationskoeffizient $r$ zwischen den beiden Größen nach Gleichung (25) aus den Meßdaten berechnet [11]. Er ist ein Maß für die Korrelation zwischen den beiden Größen $\Delta C_m$ und $\Delta G_m$ .

$$r = \frac{\sum_{i=1}^{n}(x_i - \bar{x})(y_i - \bar{y})}{\sqrt{\sum_{i=1}^{n}(x_i - \bar{x})^2 \sum_{i=1}^{n}(y_i - \bar{y})^2}} \qquad (25)$$

*r = empirischer Korrelationskoeffizient*
*$x_i$ = Leitfähigkeitsänderung $\Delta G_m$ im i - ten Experiments*
*$\bar{x}$ = Mittelwert der gemessenen Leitfähigkeitsänderungen*
*$y_i$ = Kapazitätsänderung $\Delta C_m$ im i - ten Experiments*
*$\bar{y}$ = Mittelwert der gemessenen Kapazitätsänderungen*
*n = Zahl der Experimente*

Um die Korrelation auf statistische Signifikanz zu testen, wurde das folgende Testverfahren angewandt [11]. Es wurde die Testhypothese überprüft, daß der Korrelationskoeffizient zwischen Kapazitätsänderungen $\Delta G_m$ und Leitfähigkeitsänderungen $\Delta C_m$ verschieden von 0 ist. Zur Überprüfung der Testhypothese wurde die Testgröße $t$ nach Gleichung (26) gebildet.

$$t = \frac{r}{\sqrt{1 - r^2}} \sqrt{n - 2} \qquad (26)$$

Die Testgröße *t* genügt, wie in der Literatur beschrieben ist [11], einer *Student-t-Verteilung*. Anhand der tabellierten *Student-t-Verteilung* wurde die Irrtumswahrscheinlichkeit bzw. das Konfidenzlevel für die Annahme $r \neq 0$ bestimmt. Eine Korrelation, die mit einer Irrtumswahrscheinlichkeit von 5% (95% Konfidenzlevel) verschieden von r = 0 ist, wurde als statistisch signifikant betrachtet.

## 2.8 Auswahlkriterien

Um Zellen, die bereits bei der Präparation oder durch mechanische Einwirkung beim Experimentieren geschädigt worden waren, von den Experimenten auszuschließen, wurden folgende Auswahlkriterien für die Kolonzellen zu Beginn jedes Experimentes aufgestellt:

- Das Membranpotential ist negativer als -50 mV.
- Die Ruhemembranleitfähigkeit liegt unter 20 nS.

Sowohl das Membranpotential als auch die Membranleitfähigkeit der unstimulierten Kolonkryptzelle sind Marker für eine Zellschädigung: Eine Zerstörung der Zellmembran führt zu einem Kurzschlußstrom über das Membran-Leck und erhöht $G_m$, gleichzeitig sinkt das durch die ATPasen aufgebaute Membranpotential der Zellen nach Erschöpfung ihres Energievorrates.

Als weitere Voraussetzung für jedes Experiment wurde eine Zugangsleitfähigkeit größer als 20 nS gefordert. Eine Zugangsleitfähigkeit kleiner als 20 nS führt zu großen Meßfehlern für $C_m$ und $G_m$ und einem ungünstigen Signal-Rausch-Verhältnis, da dann der Großteil der an die Pipette angelegten Sinusspannung am Zugangswiderstand und nicht mehr an der zu untersuchenden Zellmembran abfällt [17].

# 3. Durchführung der Versuche

## 3.1 Versuchstiere und Präparation

Für unsere Untersuchungen wurden Wistar-Ratten mit einem Körpergewicht zwischen 80 und 120 Gramm verwendet. Die Tiere wurden durch Dekapitation getötet. Das Kolon wurde freipräpariert, evertiert und anschließend mit $Ca^{2+}$-freier Kolonlösung I prall gefüllt. Bis zur Isolation der Krypten wurde es in $Ca^{2+}$-haltiger Kolonlösung II aufbewahrt. Zur Isolation der Einzelkrypten wurde das Kolon 10 Minuten lang in $Ca^{2+}$-freier Lösung im 37° C warmen Wasserbad mit einer Frequenz von ca. 3-5 Hz geschüttelt, um die Krypten vom submukösen Bindegewebe zu lockern. Anschließend wurde es wieder in $Ca^{2+}$-haltige Lösung II transferiert und von Hand etwa eine Minute lang kräftig geschüttelt, wobei sich die Krypten einzeln oder in kleinen Gruppen ablösten. Die absedimentierten Einzelkrypten wurden bei 4° C in Lösung II bis zum Experiment für maximal 3 Stunden aufbewahrt.

## 3.2 Experimenteller Aufbau

Die Experimente wurden unter einem invertierten Zeiss-Binokular-Mikroskop mit Differential-Interferenz-Optik und einer Vergrößerung von 20 bis 400fach durchgeführt. Die Krypten wurden einzeln in ein etwa 5 x 8 Millimeter großes Experimentierbad aus Plexiglas (Eigenbau des Physiologischen Instituts) mit einem gläsernen Badboden transferiert. Das Bad wurde in eine auf 37° C beheizte, temperaturstabilsierte Badhalterung (Eigenbau S. Pfitzinger) aus Metall eingesetzt. Der Zulauf der Experimentierlösungen erfolgte durch hydrostatischen Druck aus etwa einem Meter Höhe über einen auf 37° C temperierten, doppelwandigen Heizschlauch, der nach dem Prinzip eines Durchlauferhitzers arbeitete. Das Totraumvolumen des Bades betrug etwa 0.5 Milliliter, die Flußrate der Lösungen etwa 30-40 Milliliter/Minute, so daß die Zeit für einen Lösungswechsel unter einer Sekunde lag. Der Ablauf erfolgte auf der Badunterseite über eine druckluftgetriebene Vakuumabsaugung nach dem Prinzip einer Wasserstrahlpumpe.

## 3.3 Saugelektroden

Die Saugelektroden wurden aus Borsilikatglas-Kapillaren mit einem Innendurchmesser von 0.86 Millimetern und einem Außendurchmesser von 1.5 Millimetern mit ei-nem automatisierten Pipetten-Ziehgerät (Zeiz Instruments, Augsburg) hergestellt. Das Gerät wurde so justiert, daß der Durchmesser der Pipettenspitzen etwa 1-2 Mikrometer betrug. Anschließend wurden die Pipetten einzeln unter einem binokularen Mikroskop überprüft, und durch Erhitzen mit einer glühenden Platinschlinge wurde die Pipettenspitze geglättet. Um die Pipettenkapazität durch Vergrößerung ihrer Wanddicke zu verringern, wurden die Pipetten an der Spitze mit Bienenwachs ummantelt. Dazu wurden die Pipetten von Druckluft (etwa ein Bar) durchströmt und kurz mit der Spitze in das erhitzte (120° C) Wachs getaucht. Die mittlere Kapazität der gewachsten Pipetten betrug dann 5.58 ± 0.12 pF (Mittelwert aus 18 Experimenten). Die Pipetten wurden anschließend mit der Spitze in die Pipettenlösung getaucht und auf etwa 1/3 ihrer Länge von 4 cm mit Hilfe eines dünn ausgezogenen Kunststoffschlauchs mit der Pipettenlösung gefüllt. Durch vorsichtiges seitliches Klopfen auf das Pipettenende wurden die Luftblasen aus der Spitze entfernt. Als Ableitelektrode wurde eine Silberchlorid-Elektrode bis auf wenige Millimeter an die Pipettenspitze herangeführt. Der Widerstand der verwendeten Pipetten bei offener Pipettenspitze in der Badlösung betrug etwa 3-5 Megaohm.

## 3.4 Haltepipetten

Eine einzelne Kolonkrypte wurde unter einem Binokular-Mikroskop aus der auf 4° C gekühlten Kolon II-Lösung mit Hilfe einer Transferpipette entnommen und in das Experimentierbad überführt. Zur Fixierung der Krypte am Badboden dienten zwei Haltepipetten, die im Winkel von etwa 30° zueinander angebracht waren. Sie wurden wie die Saugelektroden hergestellt und ihre Spitze anschließend durch Aufsetzen am Badboden scharf abgebrochen. Die eine von ihnen wurde über einen hydraulischen Narishige-Manipulator gesteuert, die andere ließ sich über Mikrometerschrauben justieren. Die beiden Haltepipetten wurden unter dem Experimentiermikroskop vorsichtig an die Krypte herangeführt. Die eine Pipette wurde an der Kryptoberfläche angesetzt, die andere seitlich an der Kryptmitte. Mit zwei Glasspritzen wurde ein leichter Unterdruck an die Haltepipetten angelegt, der zum Anhaften der Krypte an den Haltepipetten führte. Die Krypten wurden so eingespannt, daß die Badlösung freien Zugang zum Kryptlumen hatte.

Sobald die Krypte fixiert war, wurde der Lösungszulauf geöffnet und die Vakuumabsaugung eingeschaltet, wodurch ein kontinuierlicher Flüssigkeitsaustausch im Experimentierbad stattfand.

## 3.5 Ausbildung des Abdichtwiderstandes zwischen Zelle und Saugelektrode

Die Saugelektrode wurde mit einem Schrittmotor-gesteuerten Manipulator (Eppendorf, Hamburg) von der den Haltepipetten gegenüberliegenden Seite an die Krypte herangeführt. Die kleinste Schrittweite des Manipulators betrug 200 Nanometer. Dabei wurde eine Rechteckspannung mit 2.5 mV Amplitude und 2 Hz Frequenz an die Pipette angelegt. Durch Beobachtung des Pipettenstromes auf einem Oszilloskop konnte der Sealvorgang überwacht werden. Sobald der Pipettenstrom bei Kontakt mit der Zellmembran abnahm, wurde ein leichter Unterdruck durch vorsichtiges Ziehen an einer Glasspritze an die Pipette angelegt. Nach einigen Sekunden konnte die Ausbildung eines „Seals“ durch die plötzliche Abnahme des Stromes um zwei bis drei Zehnerpotenzen beobachtet werden. Ein Sealwiderstand größer als ein Gigaohm wurde als ausreichend akzeptiert. Durch einen kräftigen kurzen Sog an der Pipette wurde anschließend die Membran innerhalb der Pipette eingerissen und der elektrische Zugang zum Zellinnern geschaffen.

## 3.6 Elektrischer Aufbau und Verstärker

Da die Ströme, die bei den Experimenten durch die Zellmembran fließen, extrem niedrig sind (einige Pikoampère), wurde die gesamte Meßanordnung zur Abschirmung von äußeren elektrischen Störeinflüssen in einen Faraday-Käfig (auf der Vorderseite halb offen) eingeschlossen, der zusammen mit den elektronischen Geräten auf einen gemeinsamen Bezugspunkt geerdet wurde. Zusätzlich wurde der Pipettenhalter von einem weiteren Faraday-Käfig abgeschirmt. Der verwendete Verstärker EPC 7 (Nachbau Dr. U. Fröbe und R. Busche) besteht aus einem Vorverstärker, der direkt am Pipettenhalter angebracht und vom Manipulator mitbewegt wird und einem Hauptverstärker außerhalb der Experimentieranordnung. Er besitzt drei verschiedene Betriebsarten, zwischen denen gewählt werden kann:

- Im „voltage clamp“- und „search“-Modus arbeitet der Verstärker als Strommeßgerät (Strom-Spannungs-Wandler) bei festem, von außen vorgegebenem Pipetten-potential. Die Verstärkung läßt sich zwischen 1 und 1000mV/pA einstellen. Im „search“-Modus ist zusätzlich der im Kapitel 2.5 beschriebene Integrationsschaltkreis zur Potentialmessung in Betrieb. In diesem Modus wurden auch die Kapazitätsmessungen durchgeführt (siehe Kapitel 2).

- Im „constant zero current“-Modus arbeitet der Verstärker als hochohmiges Spannungsmeßgerät mit einem Eingangswiderstand größer als 10 Gigaohm. Er kann zur Bestimmung des Membranpotentials verwendet werden.

## 3.7 Lösungen, Hormone und Agonisten

Die verwendeten Lösungen hatten folgende Zusammensetzung in mmol/l:

| **Substanz** | Kontroll-Lösung | hypotone Lösung | hypertone Lösung | Kolon I ohne $Ca^{2+}$ | Kolon II mit $Ca^{2+}$ | Pipetten-Lösung |
|---|---|---|---|---|---|---|
| NaCl | 145 | 72.5 | 145 | 127 | 127 | |
| Na-Pyruvat | | | | 5 | 5 | |
| $NaH_2PO_4$ | | | | | | 1.2 |
| $Na_2HPO_4$ | | | | | | 4.8 |
| KCl | | | | 5 | 5 | 30 |
| $KH_2PO_4$ | 0.4 | 0.4 | 0.4 | | | |
| $K_2HPO_4$ | 1.6 | 1.6 | 1.6 | | | |
| K-Gluconat | | | | | | 95 |
| Glucose | 5 | 5 | 5 | 5 | 5 | 5 |
| Ca-Gluconat | 1.3 | 1.3 | 1.3 | | | |
| $CaCl_2$ | | | | | 1.25 | |
| $MgCl_2$ | 1 | 1 | 1 | 1 | 1 | 1.034 |
| EGTA | | | | | | 0.5 |
| EDTA | | | | 5 | | |
| HEPES | | | | 10 | 10 | |
| Mannitol | | | 100 | | | |

Tab.3: Zusammensetzung der verwendeten Lösungen (Angaben der Konzentrationen in mmol/l). Der pH-Wert der Pipettenlösung wurde auf 7.20 eingestellt, der pH-Wert aller anderen Lösungen auf 7.40. EGTA: 1,2-Bis-(2-amino-ethoxy)-ethan-N,N,N´,N´-tetraessigsäure; EDTA: Ethylen-diamin-tetraessigsäure; HEPES: N-[2-Hyproxyethyl]piperazin-N´-2-ethansulfonsäure; ATP: Adenosiontriphosphat (als Magnesiumsalz).

Die zur Stimulation der Kolonepithelzellen eingesetzten Agonisten wurden in folgenden Konzentrationen verwendet und in der Kontroll-Lösung direkt oder mit Hilfe des Lösungsvermittlers DMSO (Dimethylsulfoxid) aufgelöst.

| Agonist oder Hormon | Carbachol (CCH) | Adenosintriphosphat (ATP) | Forskolin/IBMX (Isobutylmethylxanthin) |
|---|---|---|---|
| Konzentration in mol/l | $10^{-4}$ | $10^{-4}$ | $2 \cdot 10^{-6}$ / $10^{-4}$ |

Tab.4: *Konzentration der für die Stimulation der Kolonzellen zur Sekretion verwendeten Agonisten.*

# 4. Ergebnisse

## 4.1 Bestimmung der unstimulierten Membranleitfähigkeit und Membrankapazität

In Kontrollexperimenten bestimmten wir die Membrankapazität, das Membranpotential und die Membranleitfähigkeit der unstimulierten Kryptbasiszelle. Die Ergebnisse sind in Tab.5 zusammengefaßt.

| | |
|---|---|
| Membrankapazität $C_m$ | $5.80 \pm 0.13$ pF |
| Membranleitfähigkeit $G_m$ | $10.46 \pm 0.45$ nS |
| Membranpotential $V_m$ | $-71.01 \pm 1.07$ mV |

Tab.5: *Membrankapazität $C_m$, Membranleitfähigkeit $G_m$ und Membranpotential $V_m$ der unstimulierten Kryptbasiszelle. Die Daten wurden in 168 Experimenten erhoben.*

## 4.2 Wirkung von Agonisten, die $[Ca^{2+}]_i$ erhöhen

Zunächst wurde die Effekt von Agonisten untersucht, die ihre Wirkung über den „second messenger" $Ca^{2+}$ entfalten:

### 4.2.1 Carbachol (CCH)

Carbachol (CCH) ist ein Derivat des physiologischen Transmitters Acetylcholin. Es führt an der Kolonkrypte ebenso wie jenes zu einem Anstieg der intrazellulären $Ca^{2+}$-Aktivität [7,55,79]. Unmittelbar nach der Zugabe von CCH in einer Konzentration von 100 µmol/l konnte eine Hyperpolarisation der Zellen beobachtet werden, wie in Abb.19 dargestellt. Zugleich stieg die Membranleitfähigkeit stark an und die Membrankapazität erhöhte sich. Die Wirkung von CCH hatte transienten Charakter, d.h. im Laufe der Zeit nahm die Hyperpolarisation, die Leitfähigkeitserhöhung und der Kapazitätsanstieg auch in Anwesenheit des Agonisten ab.

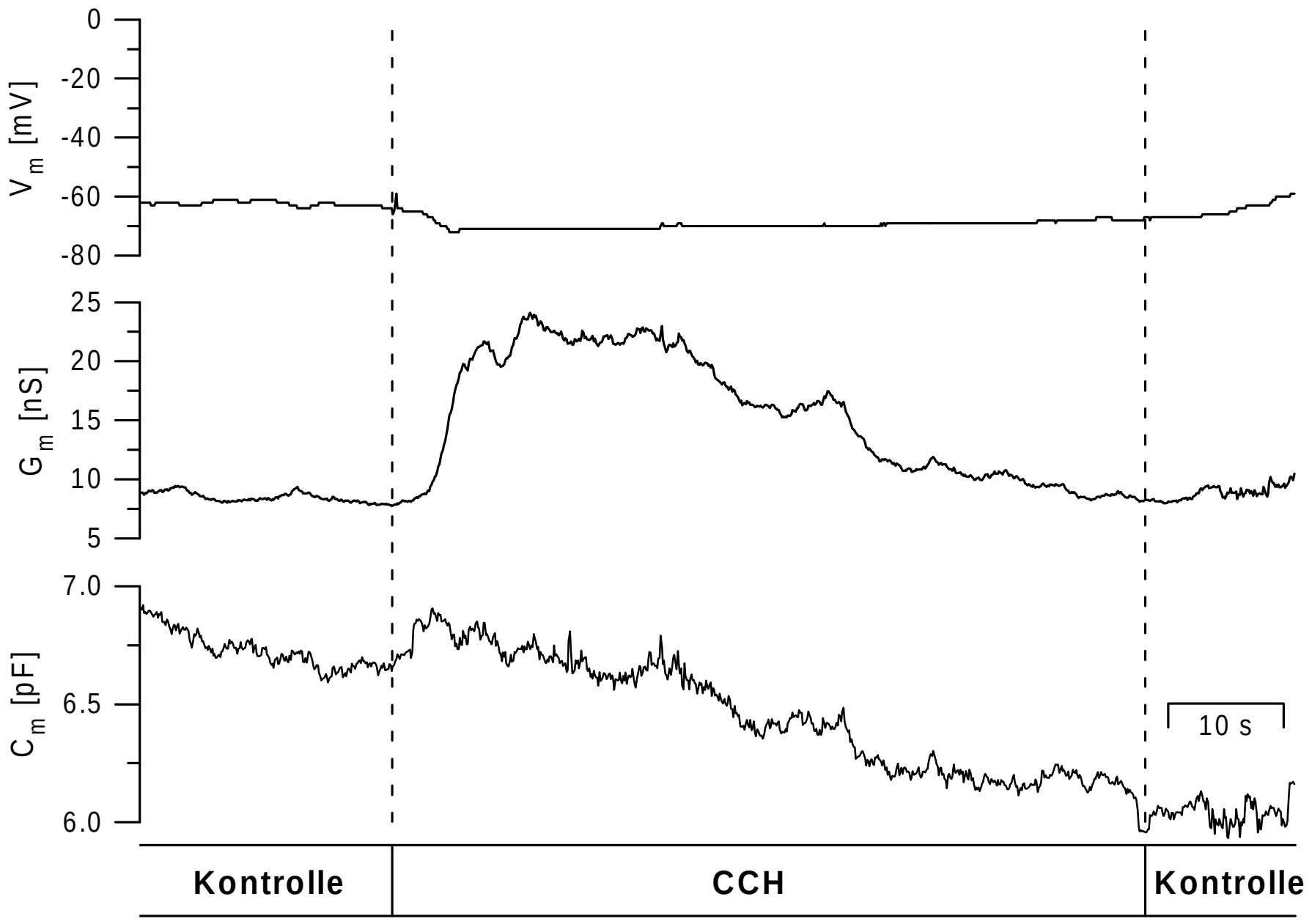

Abb.19: *Effekte von CCH (100 µmol/l) auf das Membranpotential $V_m$, die Membranleitfähigkeit $G_m$ und die Membrankapazität $C_m$ (Originalexperiment).*

In Abb.20 sind die Experimente, die an insgesamt 49 Kryptbasiszellen durchgeführt wurden, zusammengefaßt. Im Mittelwert zeigte sich eine statistisch signifikante Erhöhung der Membranleitfähigkeit um 7.1 ± 1.0 nS, eine signifikante Erhöhung der Membrankapazität um 0.32 ± 0.07 pF und eine ebenfalls signifikante Hyperpolarisation der Zellen um -7.1 ± 1.9 mV.

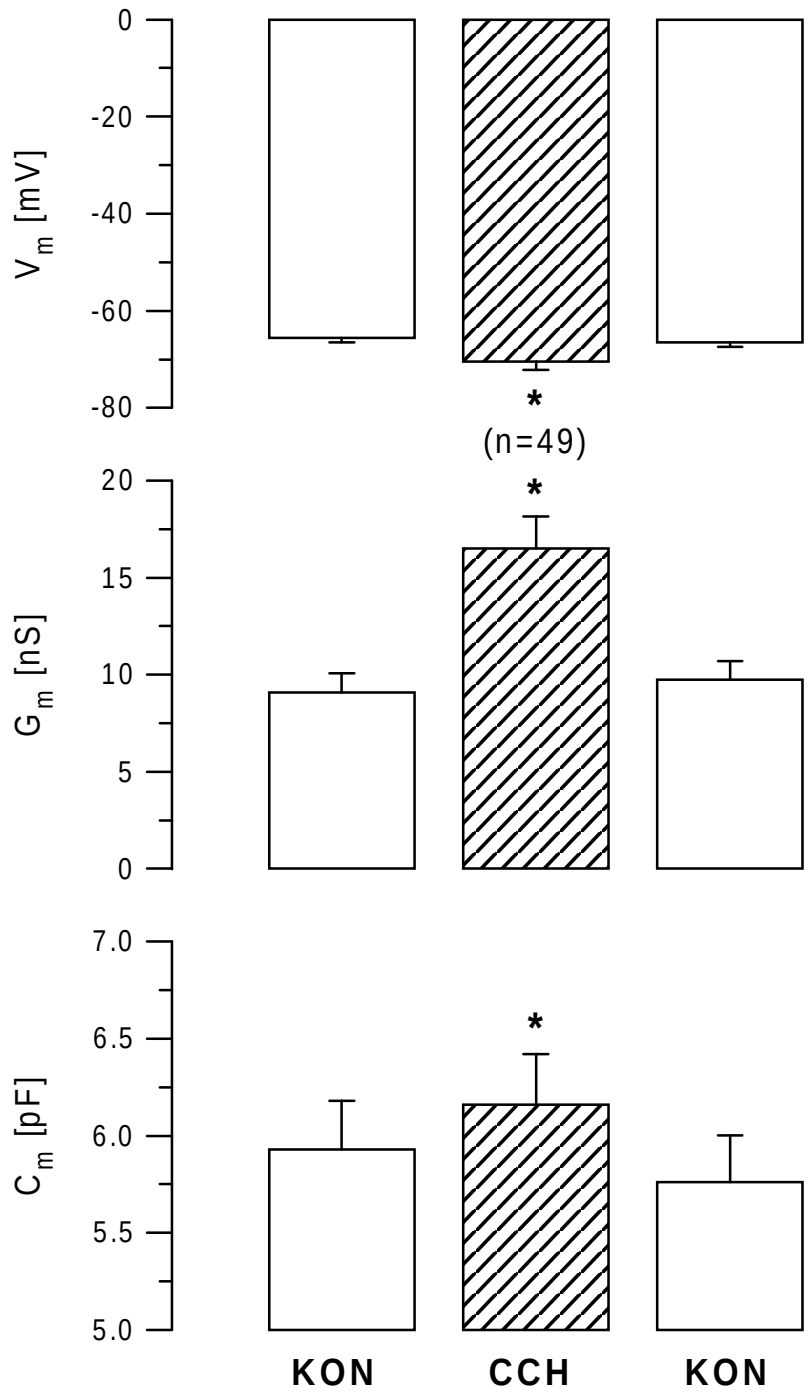

Abb.20: *Zusammenfassung des CCH-Effektes auf die Kolonkryptbasiszelle. KON: Kontrolle, CCH: Carbachol 100 µmol/l, n: Zahl der Experimente, *: im einseitigen Student-t-Test mit $p<0.05$ signifikante Änderung.*

## 4.2.2. Adenosintriphosphat (ATP)

Adenosintriphosphat ist ein physiologisch vorkommender Transmitter, der an der Kolonkryptzelle einen $[Ca^{2+}]_i$-Anstieg auslöst [62]. ATP wurde in einer Konzentration von 100 µmol/l eingesetzt. Unmittelbar nach Zugabe von ATP konnte eine Hyperpolarisation der Zellen, eine Leitfähigkeitserhöhung und ein Kapazitätsanstieg beobachtet werden. Die Wirkung von ATP hatte wie die von CCH transienten Charakter.

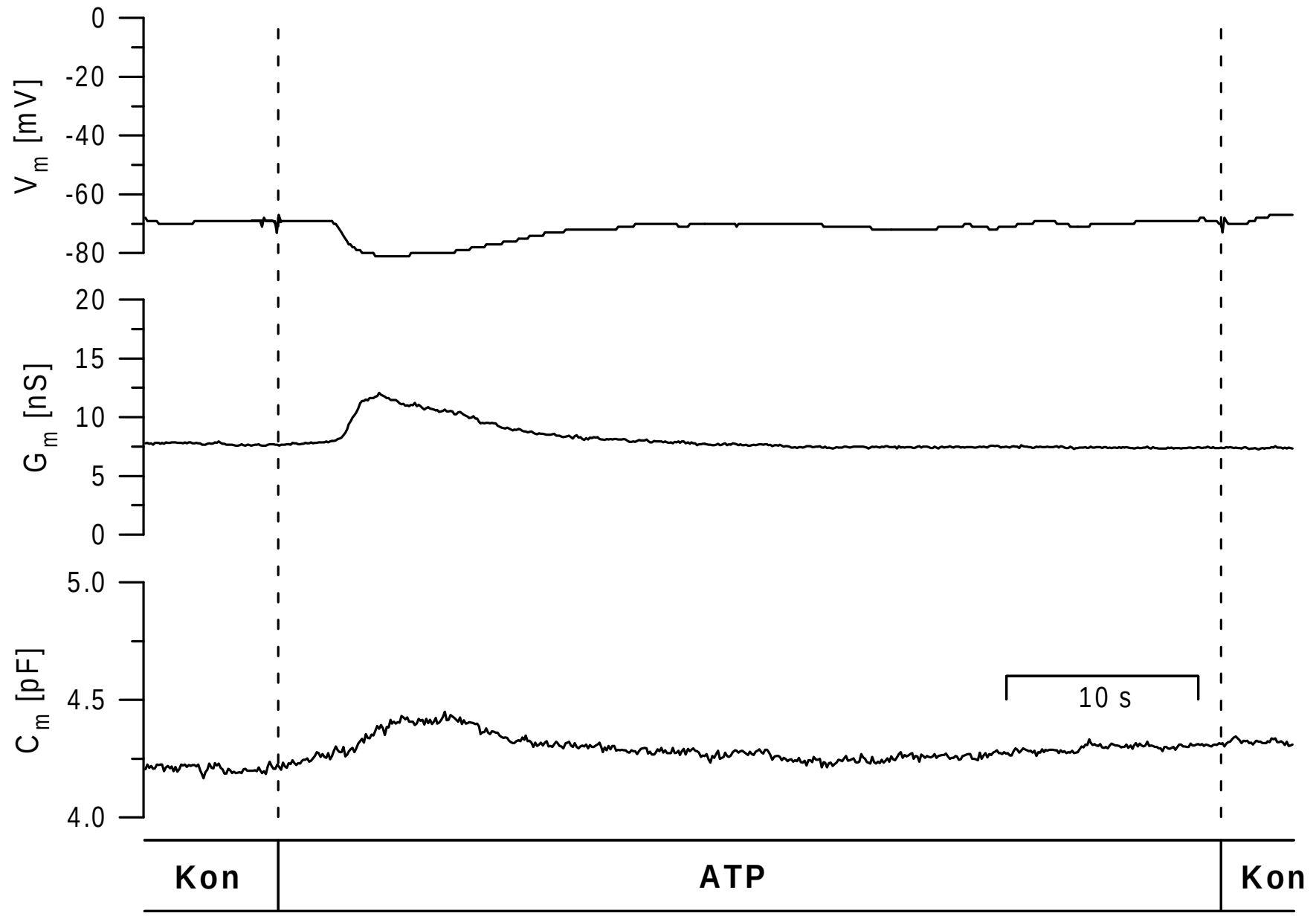

Abb.21: *Effekte von ATP (Adenosintriphosphat 100 µmol/l) auf das Membranpotential $V_m$, die Membranleitfähigkeit $G_m$ und die Membrankapazität $C_m$ (Originalexperiment).*

Die Zusammenfassung der 34 durchgeführten Experimente ergab im Mittel eine Hyperpolarisation der Zellen um -8.3 ± 1.4 mV, einen Leitfähigkeitsanstieg um 6.7 ± 1.2 nS und eine Kapazitätserhöhung um 0.39 ± 0.20 pF. Alle Änderungen waren statistisch signifikant.

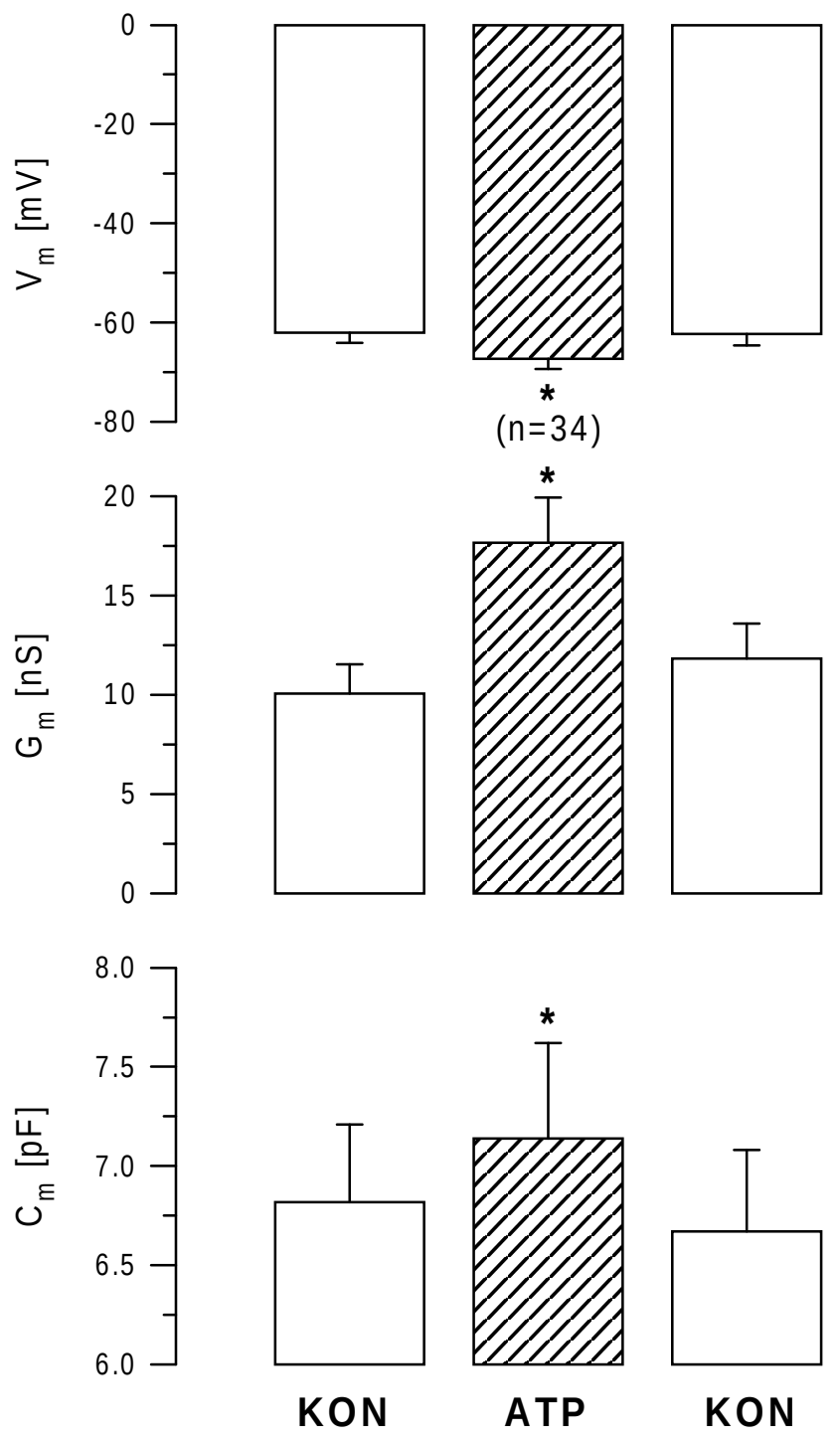

Abb.22: *Zusammenfassung des ATP-Effektes. ATP: Adenosintriphosphat 100 µmol/l, weitere Erläuterungen in* Abb.20.

## 4.3 Wirkung von Agonisten, die den cAMP-Weg aktivieren: Forskolin (FSK) und Isobutylmethylxanthin (IBMX)

Als nächstes untersuchten wir die Einflüsse von Agonisten, die über den *second messenger* cAMP ihre Wirkung entfalten. Hierzu setzten wir eine Mischung aus Forskolin und IBMX in den Konzentrationen 2 µmol/l bzw. 100 µmol/l ein.

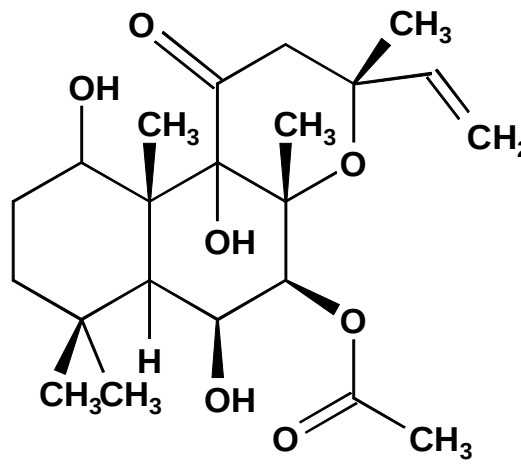

Abb.23: *Strukturformel von Forskolin (nach [29]).*

Forskolin, ein Alkaloid aus der Pflanze *Coleus forskolii*, stimuliert die katalytische Untereinheit der Adenylatcyclase. Dadurch kommt es zu einer gesteigerten cAMP-Synthese [29]. IBMX hingegen hemmt, ebenso wie das Genußmittel Koffein, die Phosphodiesterase, die den Abbau von cAMP katalysiert. Beide Substanzen wirken also synergistisch in der Steigerung der cAMP-Konzentration [40]. Wie in Abb.24 dargestellt, depolarisierten die Zellen etwa 10 Sekunden nach der Gabe der Agoni-sten. Gleichzeitig begann sich die Membranleitfähigkeit und die Membrankapazität zu erhöhen. Das Maximum der Membrankapazität und Membranleitfähigkeit wurde etwa 2 Minuten nach Agonistengabe erreicht.

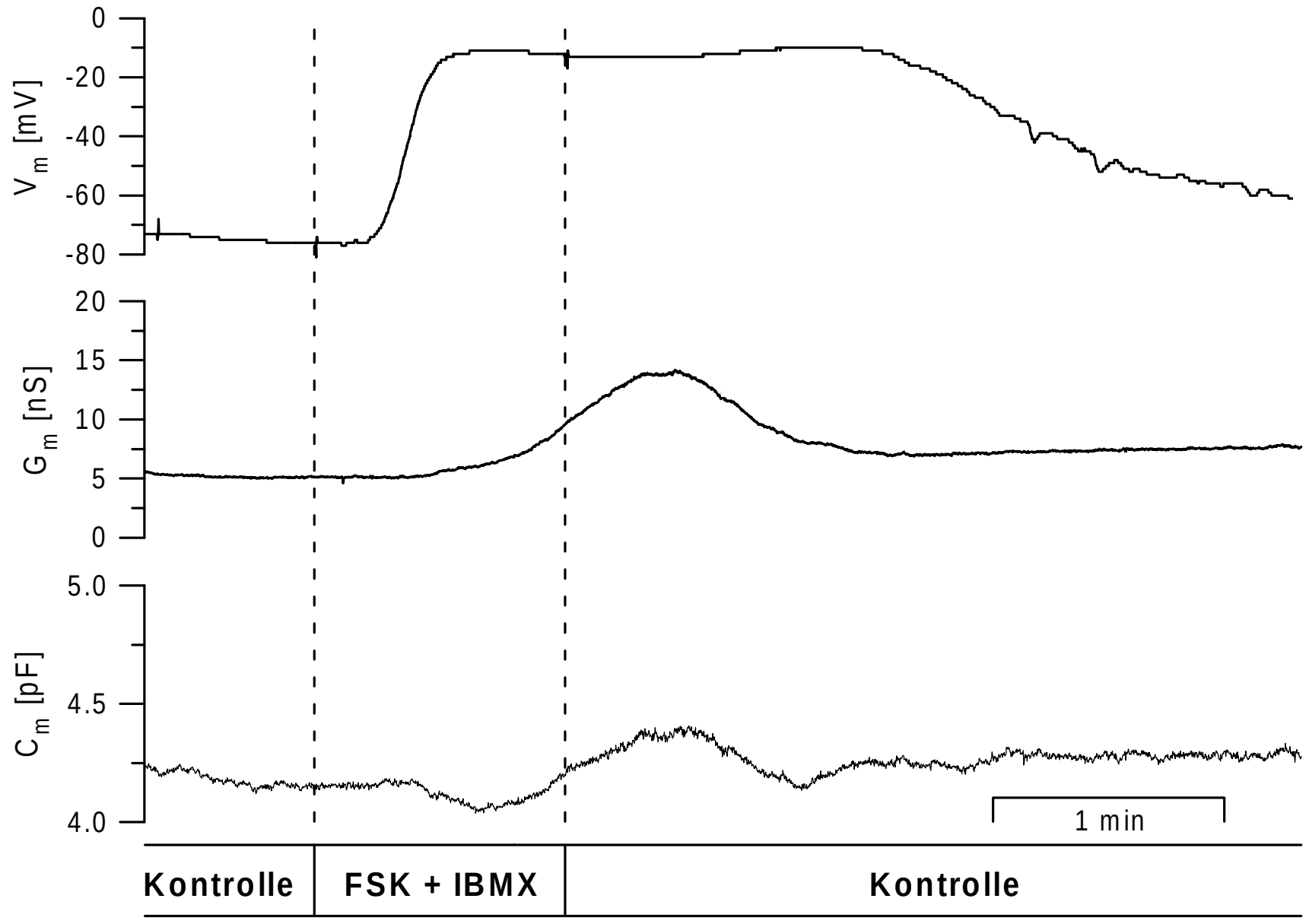

Abb.24: *Effekt von FSK/IBMX (Forskolin 2 µmol/l, Isobutylmethylxanthin 100 µmol/l) auf das Membranpotential $V_m$, die Membranleitfähigkeit $G_m$ und die Membrankapazität $C_m$ (Originalexperiment).*

In Abb.25 sind die Ergebnisse aus 34 Experimenten mit den Agonisten Forskolin/IBMX zusammengefaßt. Die Kryptzellen antworteten auf die Gabe von FSK/IBMX im Mittelwert mit einer signifikanten Depolarisation um 23.4 ± 2.9 mV, einer Zunahme der Membranleitfähigkeit um 7.3 ± 1.3 nS und einem Anstieg der Membrankapazität um 0.32 ± 0.09 pF.

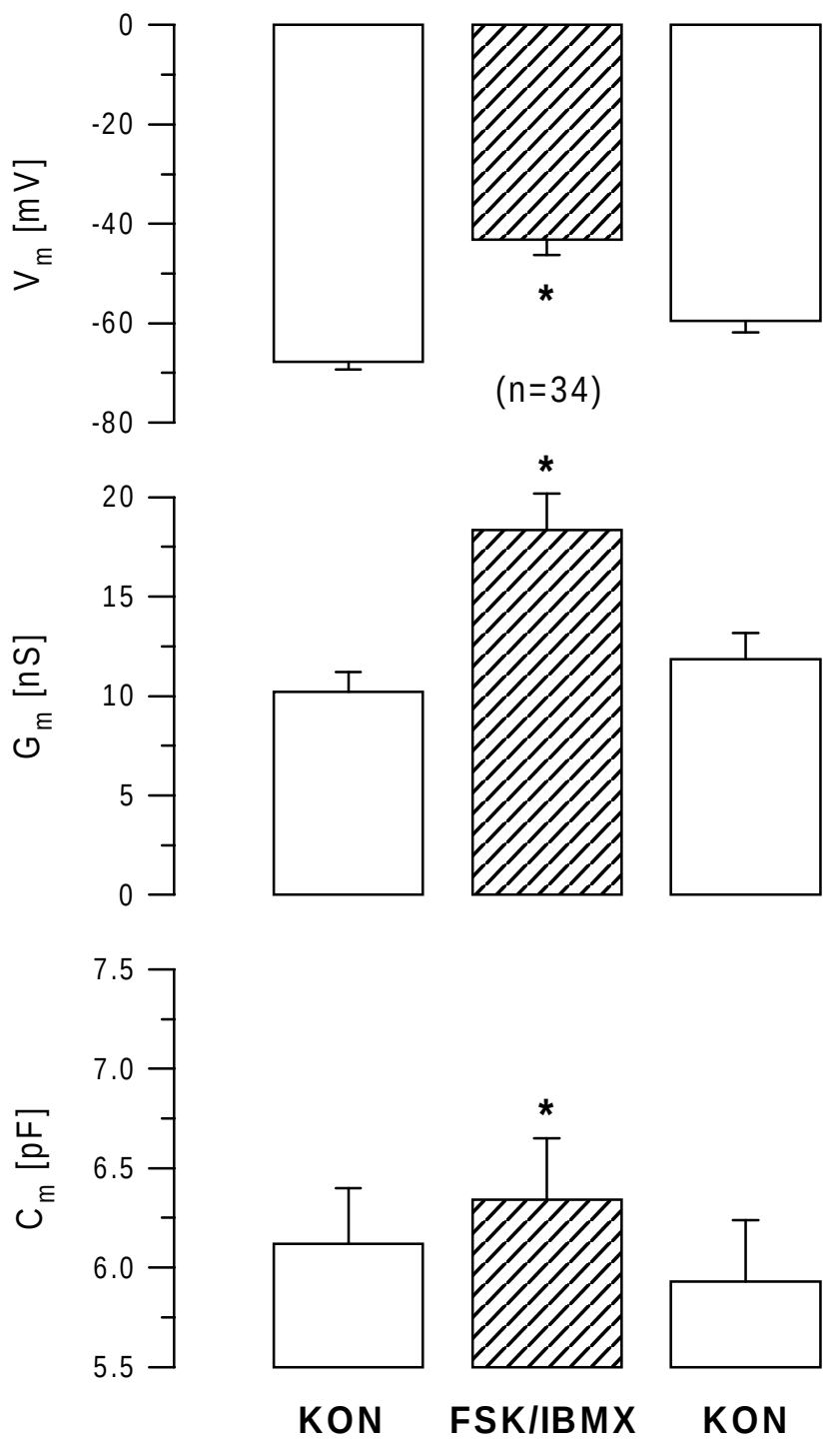

Abb.25: *Zusammenfassung der Effekte von FSK/IBMX (Forskolin 2 µmol/l und Isobutylmethylxanthin 100 µmol/l). Erläuterungen in Abb.20.*

## 4.4 Wirkung osmotischer Veränderungen

Viele Zellen, so auch die Kolonkrypte, reagieren auf osmotische Änderungen mit der Aktivierung oder Hemmung von Ionenleitfähigkeiten [60,81,102]. Wir untersuchten die Einflüsse einer Zellschwellung und von Zellschrumpfung auf das Membranpotential $V_m$, die Membranleitfähigkeit $G_m$ und die Membrankapazität $C_m$.

### 4.4.1 Hypotone Lösung

Zur Zellschwellung verwendeten wir eine hypotone (150 mosm/l) Ringerlösung, die einen osmotischen Wassereinstrom in die Kolonzellen verursacht [18,19]. Einige Sekunden nach Zugabe der hypotonen Lösung hyperpolarisierten die Zellen, die Membranleitfähigkeit und die Membrankapazität stiegen, wie in Abb.26 dargestellt, an.

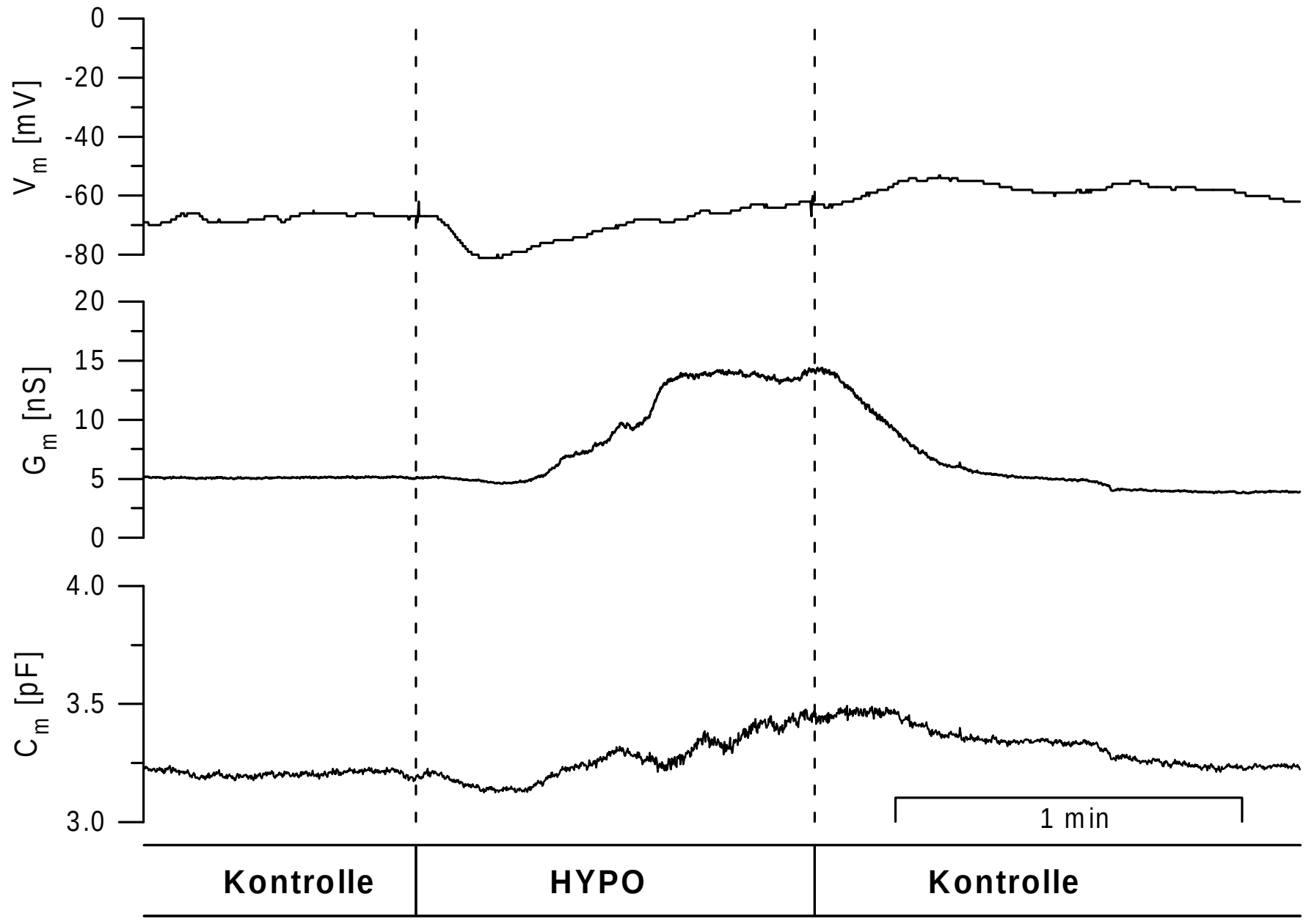

Abb.26: *Effekt von hypotoner (150 mosm/l) Lösung auf $V_m$, $G_m$ und $C_m$ (Originalexperiment).*

In Abb.27 sind die Ergebnisse aus 38 Experimenten zusammengefaßt. Die Zellen hyperpolarisierten statistisch nicht signifikant um -6.0 ± 3.3 mV, die Membranleitfähigkeit und die Membrankapazität stiegen signifikant um 6.9 ± 0.9 nS bzw. 0.85 ± 0.12 pF an.

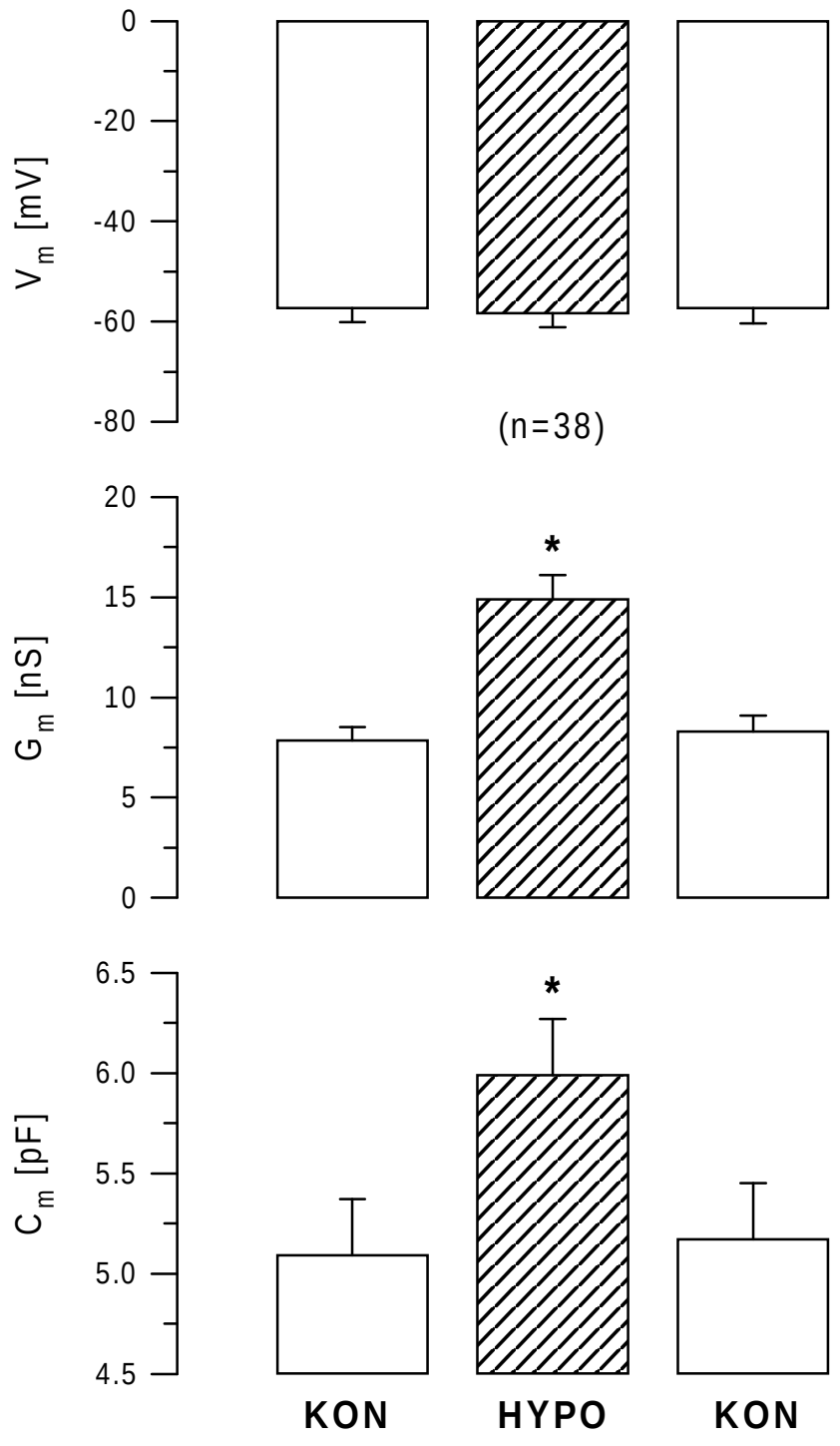

Abb.27: *Zusammenfassung des Effektes hypotoner Lösung auf $V_m$, $G_m$ und $C_m$ . HYPO: hypotone Ringer-Lösung (150 mosm/l). Erläuterungen in der Legende zu Abb.20.*

### 4.4.2 Hypertone Lösung

Eine hypertone Lösung, die durch Zugabe von 100 Millimol/Liter Mannitol erzeugt wurde, führte zu einer osmotischen Zellschrumpfung [107]. Wie in Abb.28 dargestellt, depolarisierten die Zellen unmittelbar nach der Zugabe der hypertonen Lösung, ihre Membranleitfähigkeit und ihr Membranpotential nahmen ab.

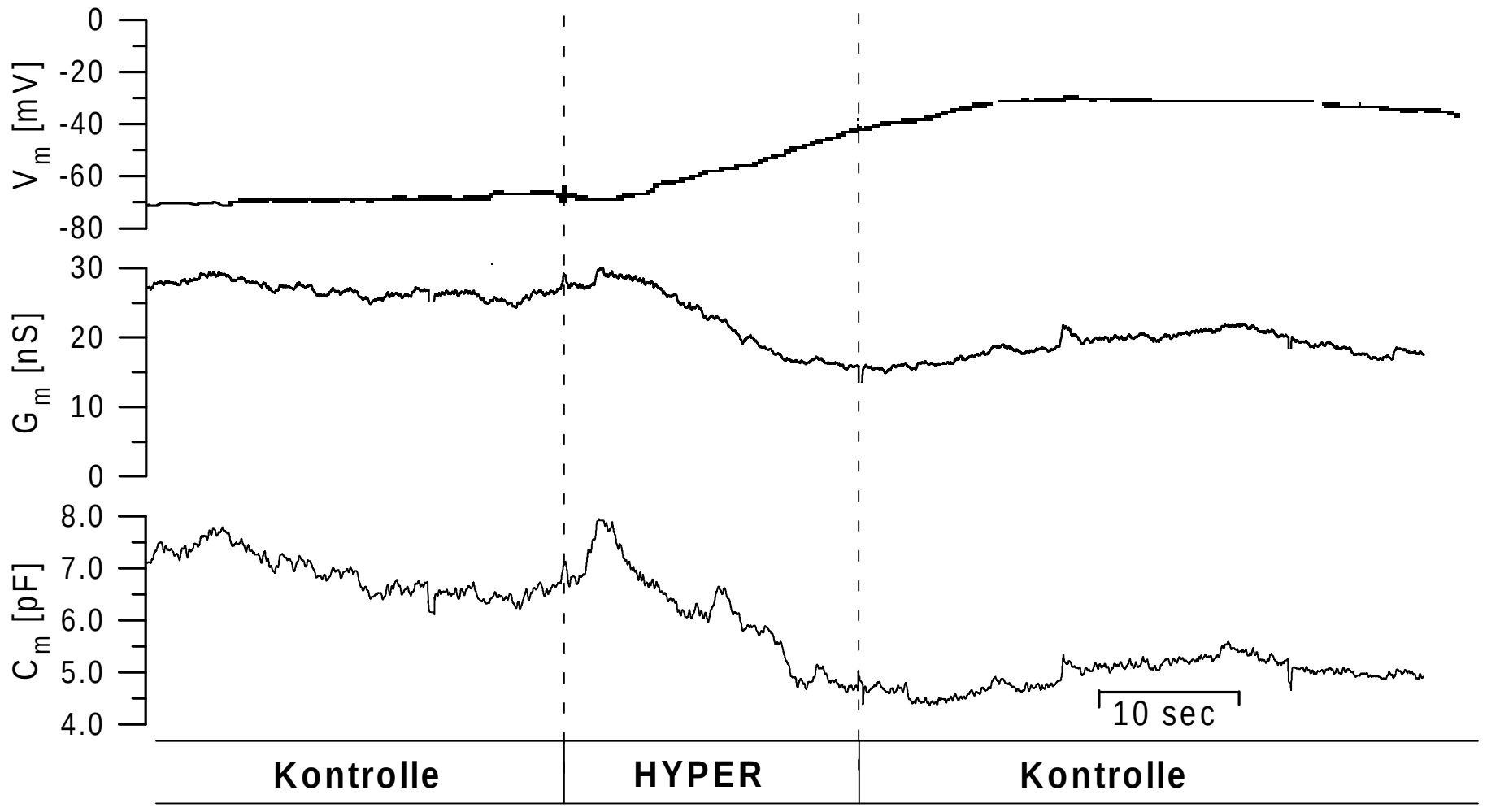

Abb.28: *Effekt von hypertoner Lösung (+150 mmol Mannitol) auf $V_m$, $G_m$ und $C_m$ (Originalexperiment).*

In Abb.29 sind die 12 durchgeführten Experimente zusammengefaßt. Die Zellen depolarisierten signifikant um 23.0 ± 4.7 mV, die Membranleitfähigkeit fiel signifikant um 5.6 ± 1.0 nS und die Membrankapazität nahm signifikant um 1.03 ± 0.23 pF ab. Die vorliegenden Daten wurden mit dem in Kapitel 2.5 beschriebenen, neuentwickelten Vierfrequenz-Synchrondetektor gemessen.

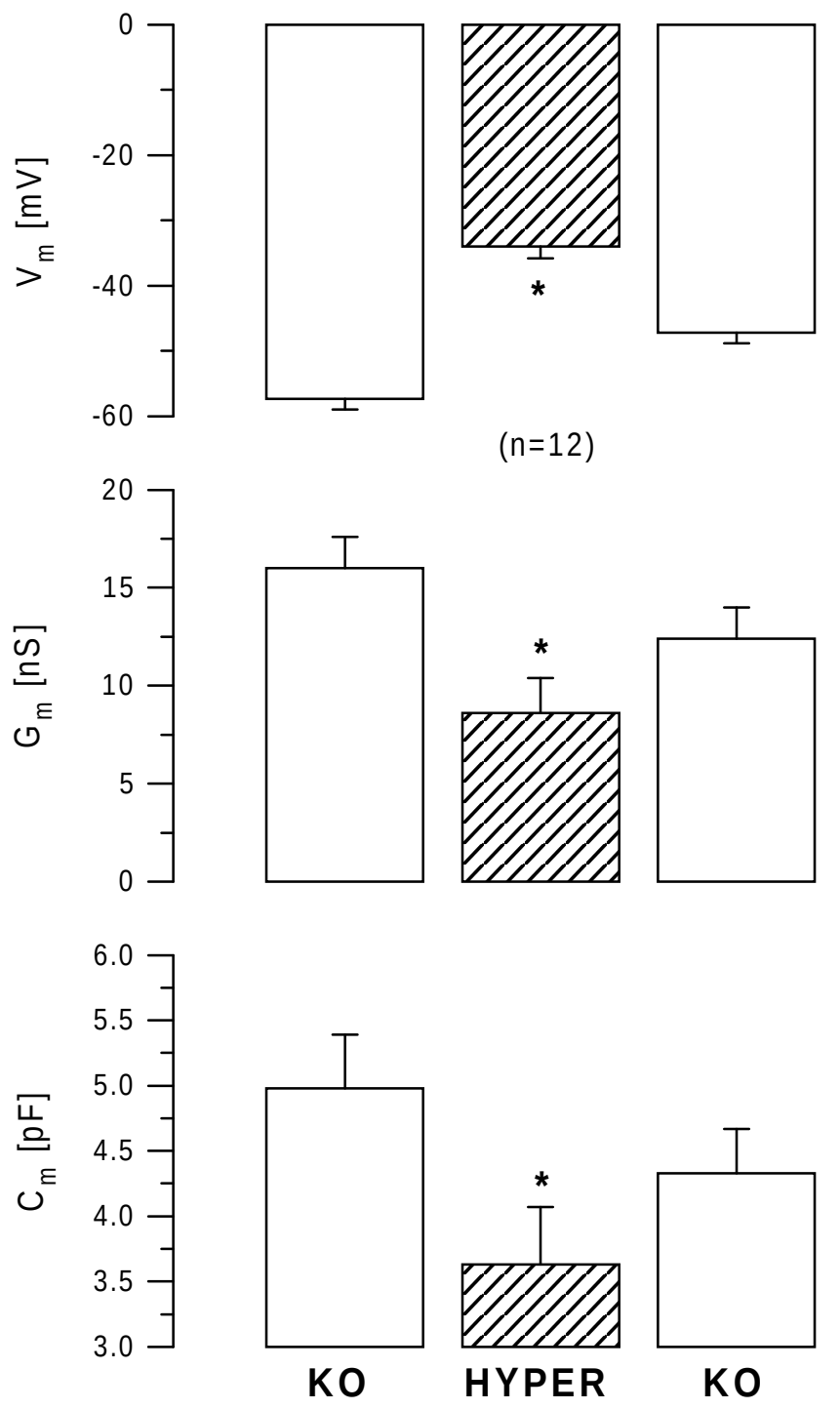

Abb.29: *Effekt von hypertoner Lösung (+150 mmol/l Mannitol) auf $C_m$, $G_m$ und $V_m$. Erläuterungen in der Legende zu Abb.20.*

## 4.5 Hemmung des $Na^+2Cl^-K^+$-Kotransporters durch Azosemid

Bei den Experimenten im vorangegangenen Abschnitt wurde eine osmotische Zellschrumpfung der Kolonepithelzellen durch hypertone Lösung hervorgerufen. Sie führte zu einer Erniedrigung der Membrankapazität. Um den beobachteten Zusammenhang zwischen einer Abnahme des Zellvolumens und der Membrankapazität weiter zu untersuchen, sollte nochmals eine Zellschrumpfung durch einen anderen, indirekten Mechanismus ausgelöst werden:

Der $Na^+2Cl^-K^+$-Kotransporter ist, wie in Kapitel 1.6 beschrieben, für die basolaterale Chlorid-Aufnahme des Kolonepithels bei der NaCl-Sekretion verantwortlich. Hemmt man den Kotransporter durch Azosemid (Abb.30), eine Substanz aus der Klasse der Schleifendiuretika mit einer hohen Affinität zum Kolon-Typ des Kotransporters [18], so verarmen die Zellen an Chlorid. Dies löst nachfolgend einen osmotischen Wasserausstrom aus der Epithelzellen aus, das Zellvolumen nimmt ab [18,105].

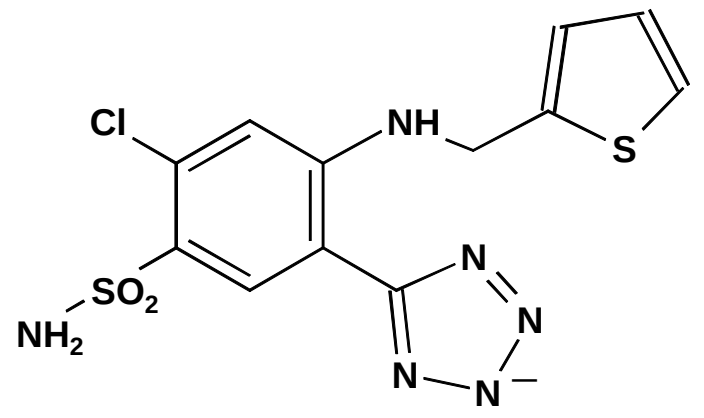

Abb.30: *Strukturformel von Azosemid.*

Zunächst wurden die Zellen mit Forskolin/IBMX in Konzentrationen von 2 µmol/l bzw. 100 µmol/l etwa eine Minute vorstimuliert, um ihre luminale Chlorid-Leitfähigkeit zu erhöhen. Die nachfolgende Gabe von Azosemid in einer Konzentration von 0.1 mmol/l führte zu den in Abb.31 dargestellten Reaktionen. Die Zellen hyperpolarisierten im Laufe von etwa 1-2 Minuten nach Azosemid-Gabe zunehmend. Gleichzeitig sanken die Membranleitfähigkeit und die Membrankapazität ab.

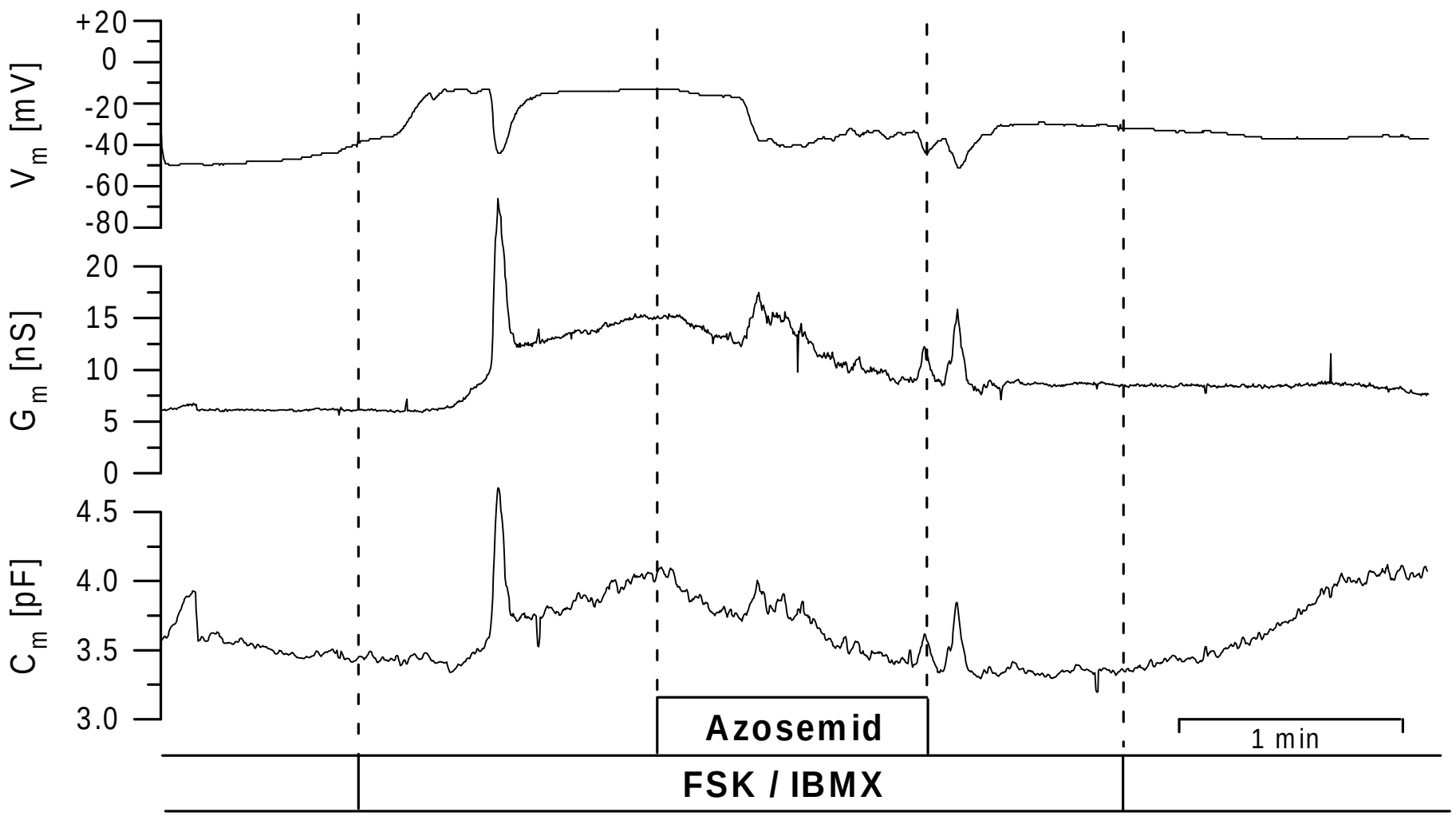

Abb.31: *Wirkung von Azosemid (0.1 mmol/l) auf $V_m$, $G_m$ und $C_m$ (Originalexperiment). FSK/IBMX: Forskolin 2 µmol/l, IBMX 100 µmol/l. (Der kurzzeitige Leitfähigkeitsanstieg während der FSK/IBMX-Gabe stellt wahrscheinlich eine spontane $Ca^{2+}$-Oszillation dar).*

Abb.32 faßt die Ergebnisse von insgesamt 25 Experimenten zusammen. Die Zellen hyperpolarisierten nach der Azosemid-Gabe signifikant um -15.7 ± 1.8 mV, die Membranleitfähigkeit $G_m$ sank signifikant um 3.9 ± 1.4 nS und die Membrankapazität $C_m$ fiel um 0.91 ± 0.20 pF.

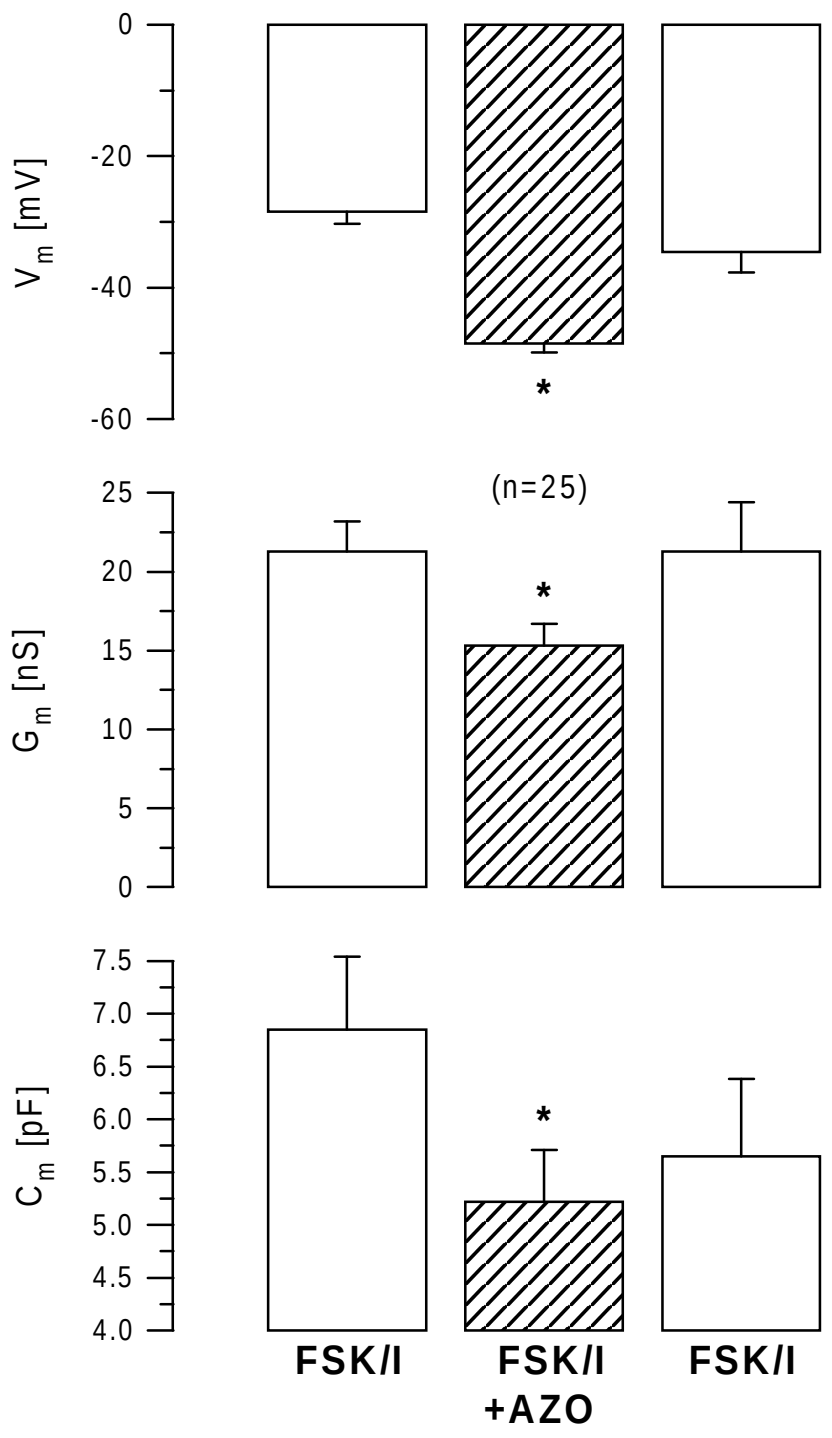

Abb.32: *Effekte von Azosemid (0.1 mmol/l) auf das Membranpotential $V_m$, die Membranleitfähigkeit $G_m$ und die Membrankapazität $C_m$. Erläuterungen in der Legende zu Abb.20.*

## 4.6 Korrelation zwischen Leitfähigkeits- und Kapazitätsänderungen

Um einen möglichen Zusammenhang zwischen den beobachteten Membrankapazitäts- und Membranleitfähigkeits-Änderungen in den bisher durchgeführten Experimenten aufzuzeigen, wurden für jeden zugegebenen Agonisten die an der Kolonkrypte gemessenen Leitfähigkeitsänderungen $\Delta G_m$ und Kapazitätsänderungen $\Delta C_m$ gegeneinander aufgetragen. Anschließend wurde eine Regressionsgerade durch die Datenpunkte gelegt und der empirische Korrelationskoeffizient *r* nach Gleichung (25) bestimmt [11].

### 4.6.1 Carbachol (CCH)

In Abb.33 sind die nach Stimulation der Zellen mit Carbachol beobachteten Kapazitäts- und Leitfähigkeitsänderungen gegeneinander aufgetragen. Der Korrelationskoeffizient beträgt r = 0.338, die Steigung der Regressionsgeraden 18.4 fF/nS. Die Korrelation ist statistisch signifikant nach dem in Kapitel 2.7 beschriebenen Testverfahren [11].

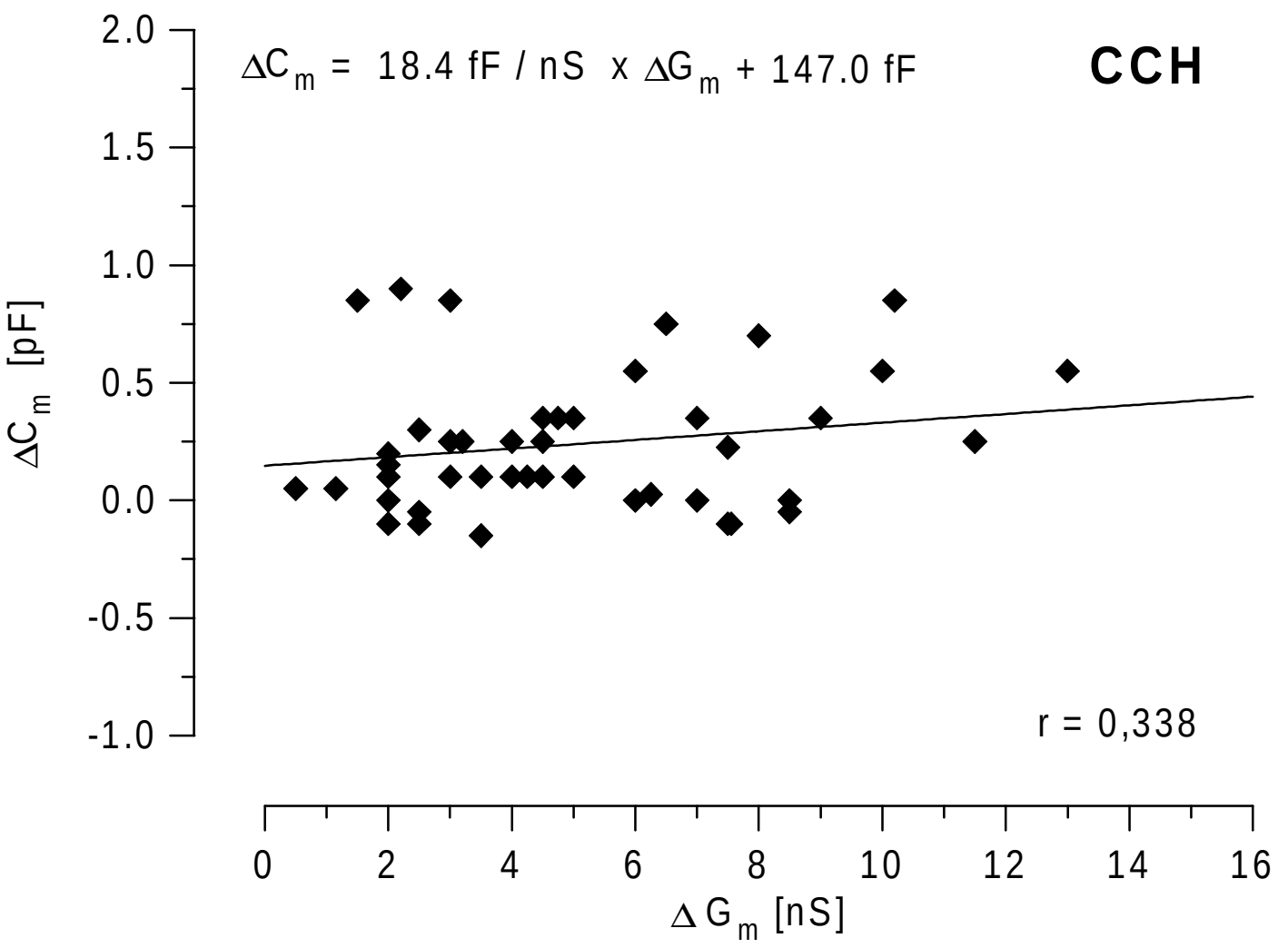

Abb.33: *Zusammenhang zwischen den beobachteten Leitfähigkeits- und Kapazitätsänderungen. Auf der Ordinate wurde die Änderung der Membranleitfähigkeit $\Delta G_m$, auf der Abszisse die Änderung der Membrankapazität $\Delta C_m$ nach der Stimulation der Zellen mit Carbachol aufgetragen. r: empirischer Korrelationskoeffizient nach (25), CCH: Carbachol 100 µmol/l.*

## 4.6.2 Adenosintriphosphat (ATP)

Nach Stimulation der Zellen mit ATP wurden die in Abb.34 dargestellten Kapazitäts- und Leitfähigkeitsänderungen gemessen. Die Korrelation ist statistisch nicht signifikant, der Korrelationskoeffizient beträgt r = 0.171, die Regressionsgerade hat eine Steigung von 10.9 fF/nS.

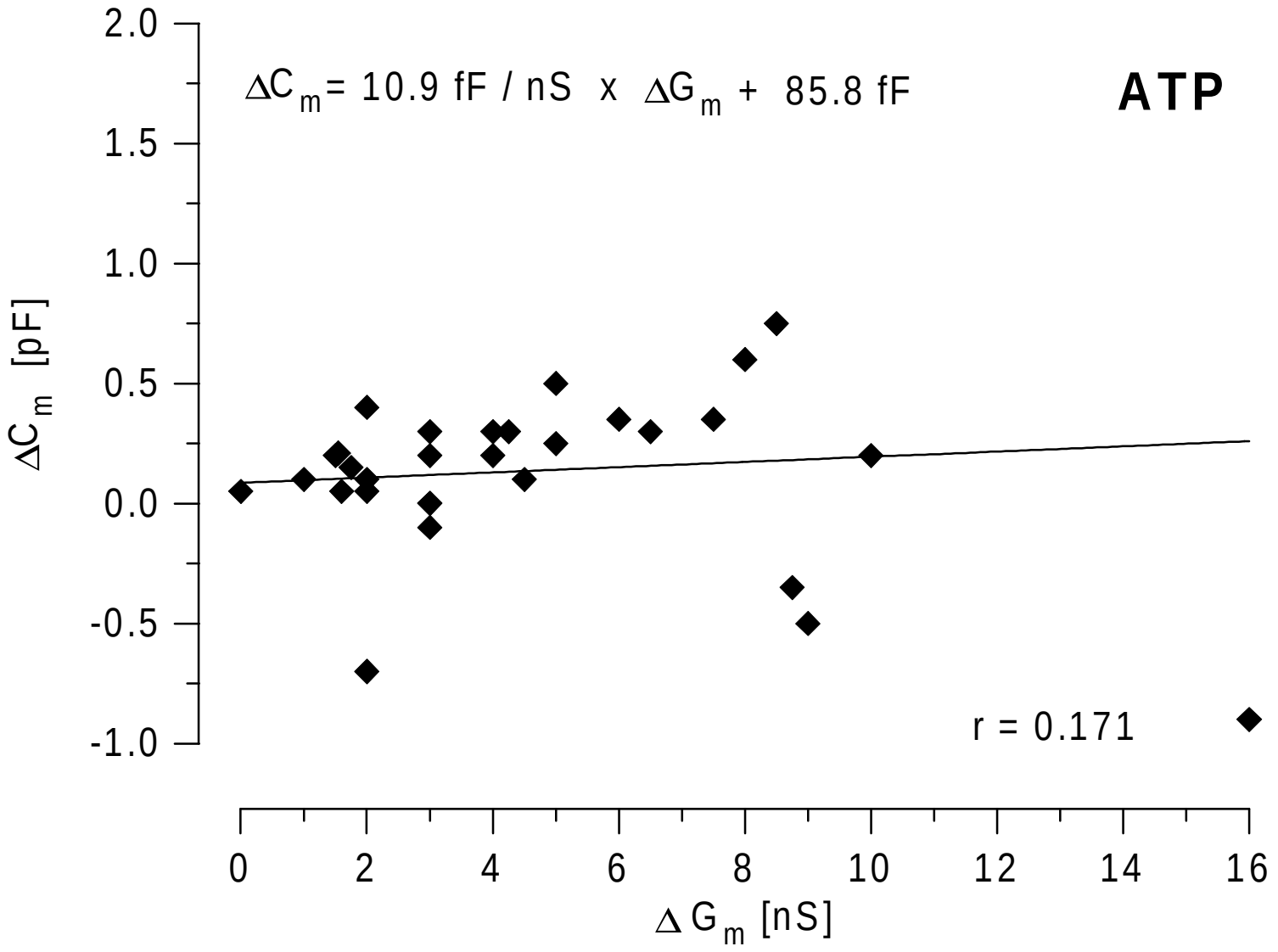

Abb.34: *Zusammenhang zwischen den beobachteten Leitfähigkeits- und Kapazitätsänderungen (Erläuterungen in Abb.33). ATP: Adenosintriphosphat 100 µmol/l.*

### 4.6.3 Forskolin/Isobutylmethylxanthin (FSK/IBMX)

Die nach Stimulation der Zellen mit Forskolin/Isobutylmethylxanthin beobachteten Kapazitäts- und Leitfähigkeitsänderungen sind in Abb.35 dargestellt. Es ergibt sich eine signifikante Korrelation zwischen $\Delta C_m$ und $\Delta G_m$ mit einem signifikanten Korrelationskoeffizienten r = 0.504 und einer Steigung der Regressionsgeraden von 32.6 fF/nS.

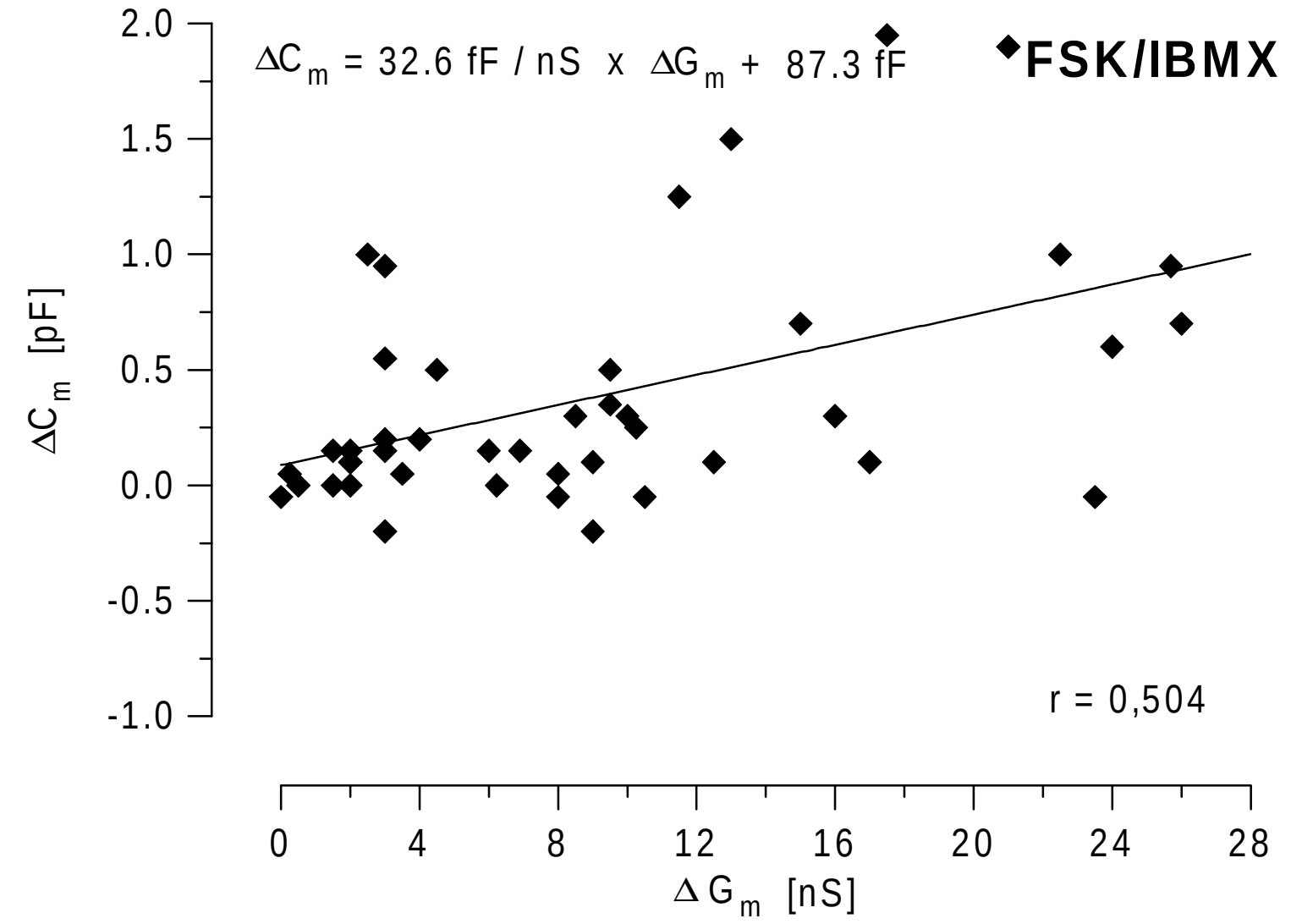

Abb.35: *Zusammenhang zwischen den beobachteten Leitfähigkeits- und Kapazitätsänderungen (Erläuterungen in Abb.33). FSK/IBMX: Forskolin 2 µmol/l / Isobutylmethylxanthin 100 µmol/l.*

### 4.6.4 Hypotone Badlösung (HYPO)

Viele Zellen, so auch die Kolonkrypte, regulieren ihr Zellvolumen bei osmotischer Zellschwellung durch das Öffnen von Ionenleitfähigkeiten [60,81,102]. Um die Verlauf der Gegenregulation zu untersuchen, wurden die Kapazitäts- und Leitfähigkeitsänderungen zu zwei unterschiedlichen Zeitpunkten nach Zugabe der hypotonen Badlösung untersucht. In Abb.36 sind die Kapazitäts- und Leitfähigkeitsänderungen 30 sec nach Zugabe der hypotonen Lösung, in Abb.37 nach 60 sec gegeneinander aufgetragen.

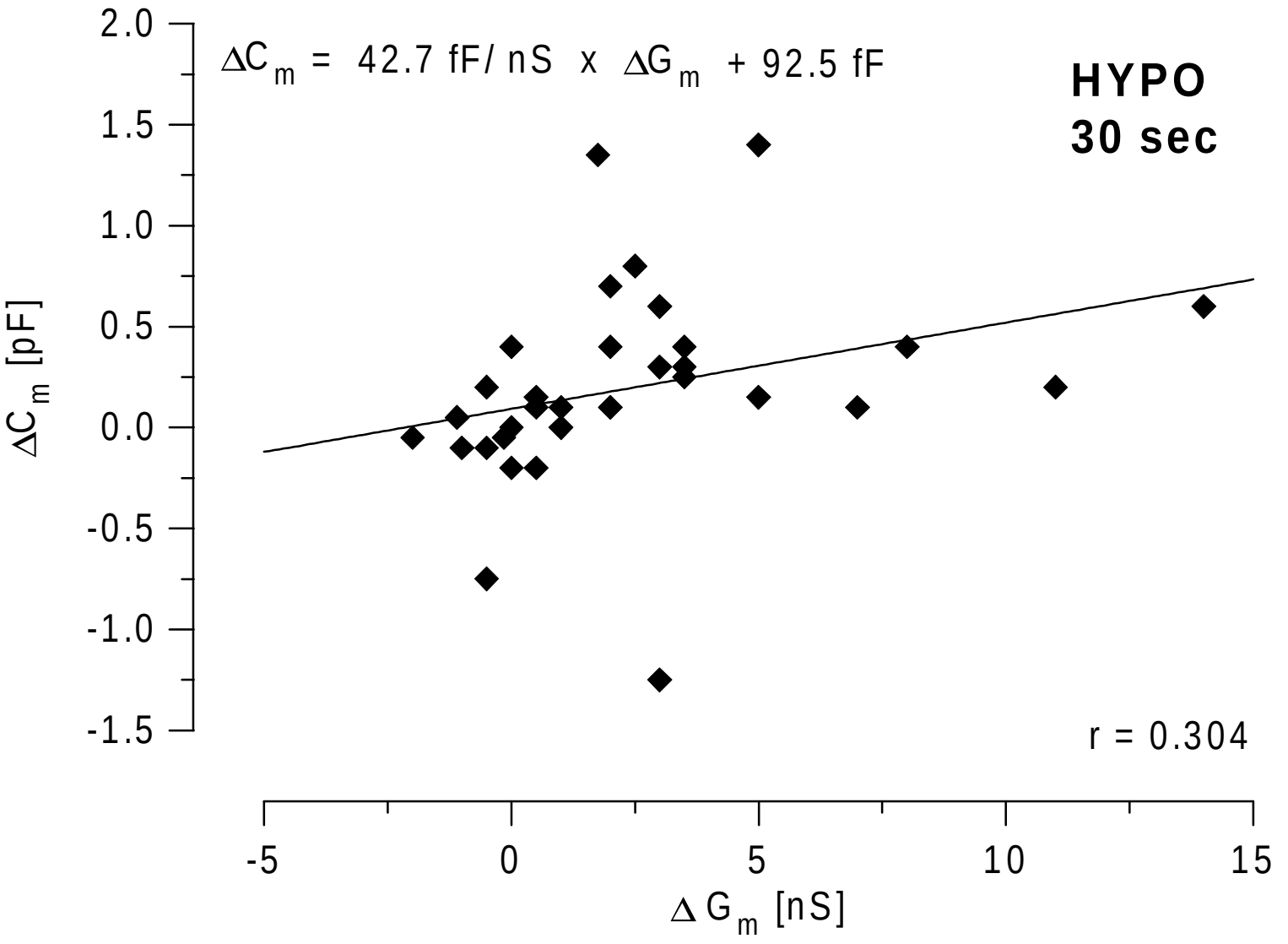

Abb.36: *Zusammenhang zwischen den beobachteten Leitfähigkeits- und Kapazitätsänderungen 30 sec nach Zugabe der hypotonen Badlösung (Erläuterungen in Abb.33). HYPO: hypotone Lösung 150 mmol/l.*

In Abb.36 erkennt man eine signifikante Korrelation zwischen den nach 30 sec beobachteten Kapazitäts- und Leitfähigkeitsänderungen. Der Korrelationskoeffizient beträgt r = 0.304, die Steigung der Regressionsgeraden ist 42.7 fF/nS.

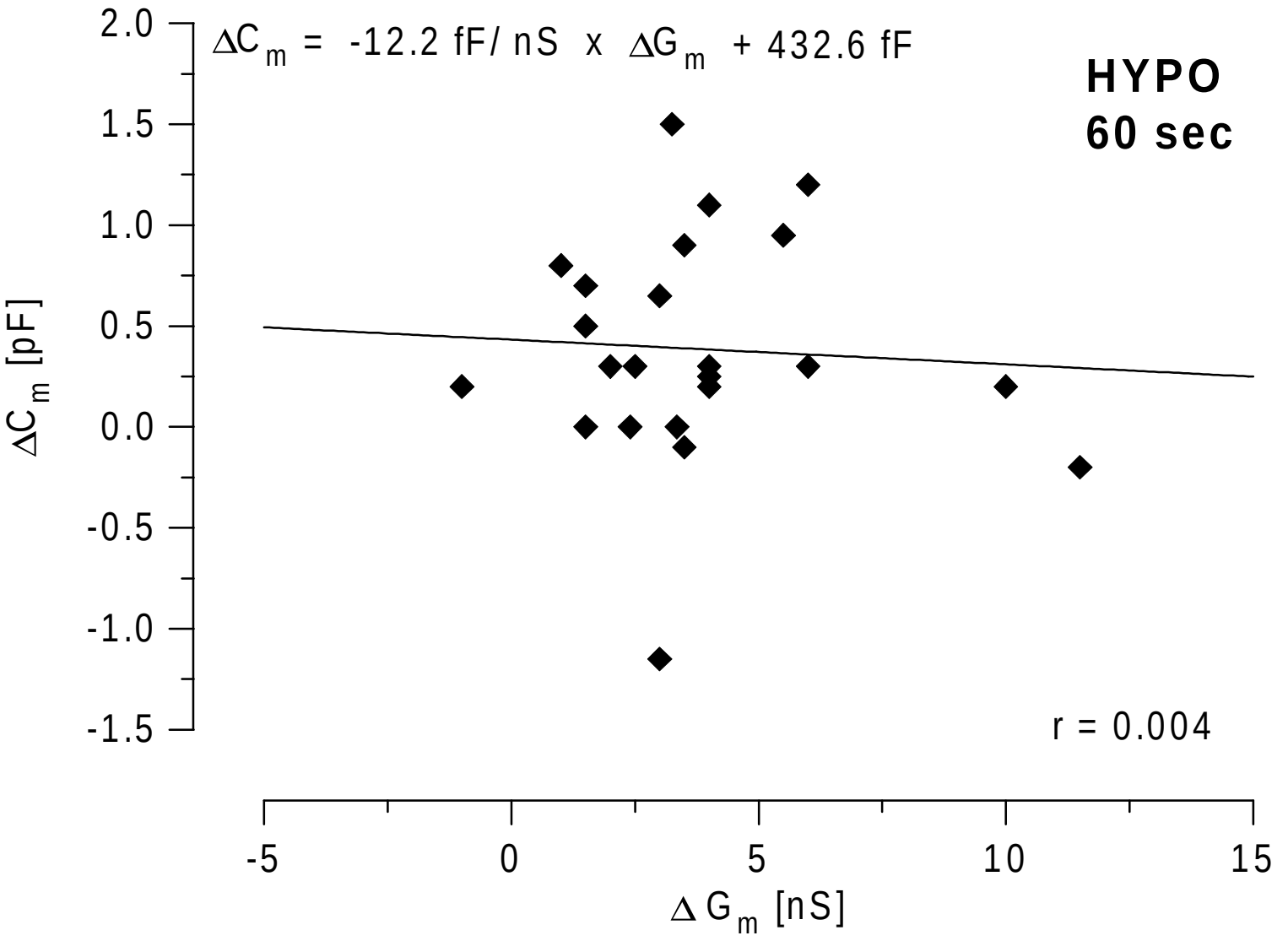

Abb.37: *Zusammenhang zwischen den beobachteten Leitfähigkeits- und Kapazitätsänderungen 60 sec nach Zugabe der hypotonen Badlösung (Erläuterungen in Abb.33). HYPO: hypotone Lösung 150 mmol/l.*

Nach 60 Sekunden hingegen ist keine signifikante Korrelation mehr zwischen den beobachteten Leitfähigkeits- und Kapazitätsänderungen zu erkennen. Der Korrelationskoeffizient beträgt 0.004, die Steigung der Regressionsgeraden ist -12.2 fF/nS.

### 4.6.5 Hypertone Badlösung (HYPER)

Die nach einer osmotischen Zellschrumpfung durch hypertone Badlösung beobachteten Veränderungen der Membranleitfähigkeit und Membrankapazität sind in Abb.38 gegeneinander aufgetragen.

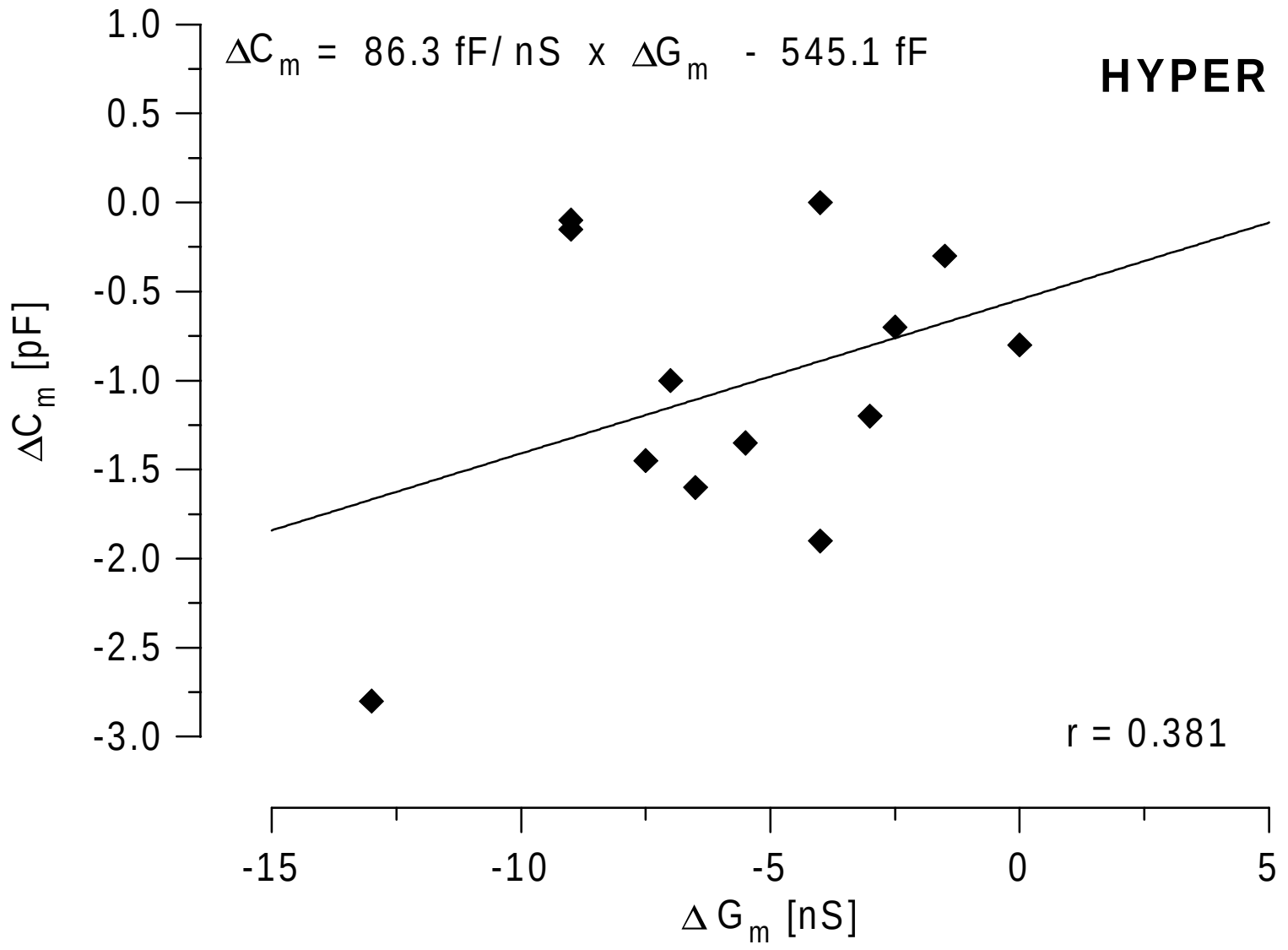

Abb.38: *Zusammenhang zwischen den beobachteten Leitfähigkeits- und Kapazitätsänderungen nach Zugabe der hypertonen Badlösung (Erläuterungen in Abb.33). HYPER: hypertone Lösung (+ Mannitol 150 mmol/l).*

Der Korrelationskoeffizient beträgt r = 0.381, die Korrelation ist statistisch nicht signifikant. Die Steigung der Regressionsgeraden ist 86.3 fF/nS.

## 4.7 Wirkung von Toxinen, die das Zytoskelett beeinflussen

In den in den Kapiteln 4.2 bis 4.4 beschriebenen Experimenten konnte nach einer Stimulation der Kolonepithelzellen zur Sekretion mit den Agonisten CCH, ATP und FSK/IBMX neben einer Erhöhung der Membranleitfähigkeit $G_m$ auch ein Anstieg der Membrankapazität $C_m$ beobachtet werden. Zwischen der Kapazitäts- und Leitfähigkeitserhöhung bestand in den durchgeführten Versuchsreihen jedoch nur eine schwache Korrelation. Es stellte sich nun die Frage, ob der beobachtete Kapazitätsanstieg bei der Stimulation der Zellen mit den Agonisten CCH, ATP oder FSK/IBMX durch einen Einbau von Exozytosevesikeln in die Zellmembran, wie er in Kapitel 1 beschrieben ist, hervorgerufen wurde.

Um diese Fragestellung zu beantworten, wurden die Zellen verschiedenen Zellgiften ausgesetzt, die unterschiedliche Proteine des Zytoskeletts zerstören und somit den Vesikeltransport in die Zellmembran unterbinden können. Innerhalb des Zytoskeletts gibt es zwei verschiedene Gruppen von Faserstrukturen, die Aktin-Filamente und die Mikrotubuli [103]. Während Mikrotubuli unter anderem bei der Zellteilung für die wohlgeordnete Trennung der Chromosomen in die beiden Tochterzellen verantwortlich sind [103], machen Experimente an Kulturzellen (16HBE-Bronchialepithelzellinien und $HT_{29}$- Kolonepithelzellinien) eine Beteiligung der Aktinfilamente an der Exozytose sehr wahrscheinlich [35,48].

Für die vorliegenden Untersuchungen wurden die Toxine A und B aus *Clostridium difficile*, *Cytochalasin B* und *Phalloidin* zur Zerstörung des Aktin-Zytoskeletts und das Zellgift *Colchicin* zum Abbau der Tubulin-Strukturen eingesetzt.

### 4.7.1 Toxin A aus *Clostridium difficile*

Toxin A, ein Gift des Bakteriums *Clostridium difficile*, ist an der Entstehung der pseudomembranösen Kolitis, einer schweren, oft tödlichen Durchfallerkrankung des Menschen beteiligt [27]. Es besteht aus zwei Untereinheiten, von denen die eine für die Aufnahme des Giftes in die Zellen, die andere für den Abbau des Aktin-Zytoskeletts verantwortlich ist [101]. Die in der Literatur beschriebene Wirkdosis des Toxins A beträgt etwa 15 µg/Liter [101].

Zunächst inkubierten wir die Kolonkrypten bei vier Grad Celsius in einer Kolonlösung II, die mit Toxin A in einer Konzentration von 1 bis 5 mg/l versetzt war. Es zeigten sich auch nach dreistündiger Inkubation keine morphologischen Veränderungen der Kolonkrypten. Alternativ dazu versetzten wir die Pipettenlösung mit Toxin A in einer Konzentration von 5 mg/l und warteten vor Gabe der Agonisten mindestens drei Minuten.

Da sich in den Saugelektrodenexperimenten kein Unterschied zwischen den beiden Applikationsarten des Toxins zeigte, wurden die Daten aus beiden Meßserien zusammengefaßt. Die Ergebnisse der Experimente sind in Abb.39 dargestellt.

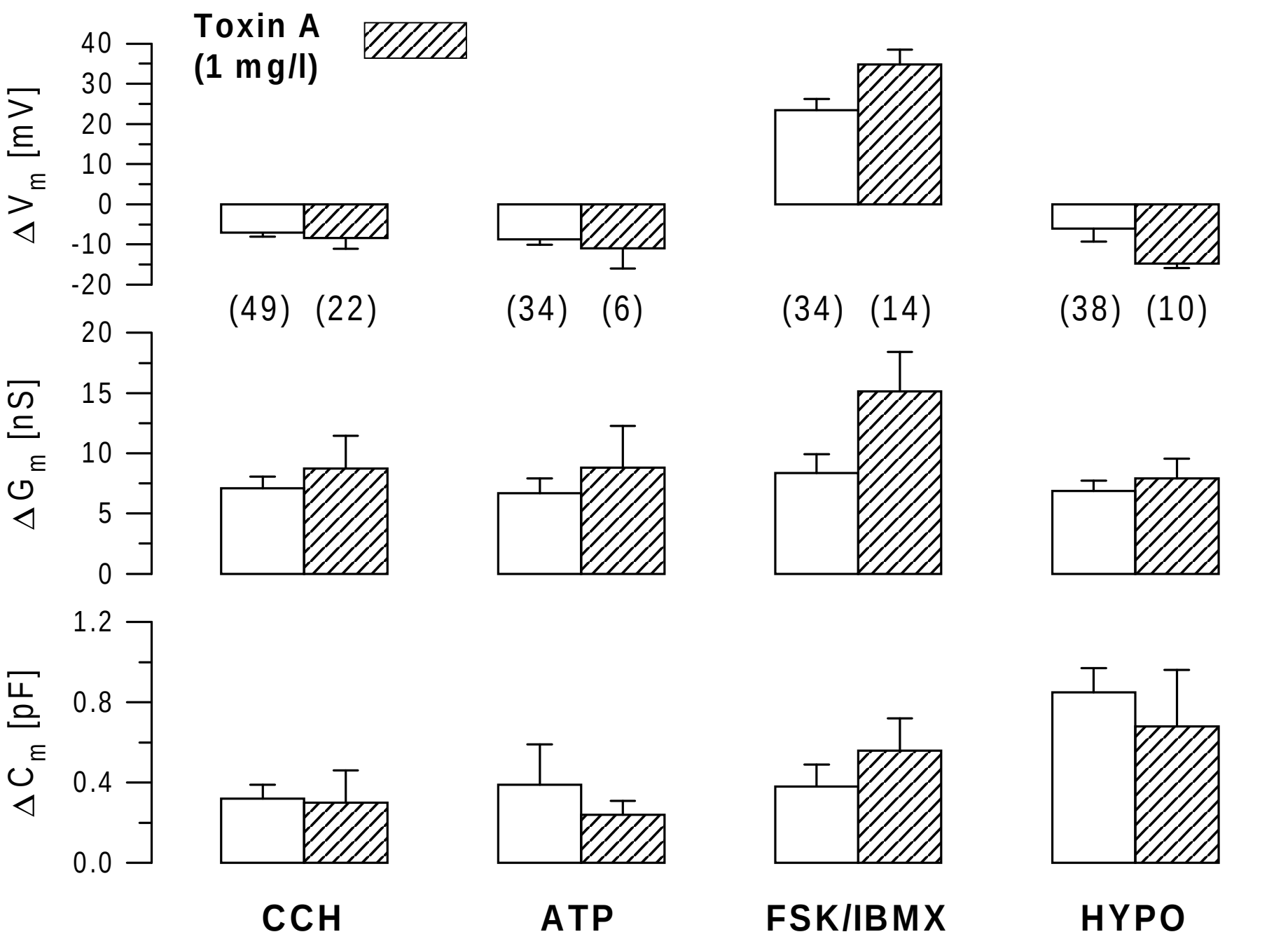

Abb.39: *Effekte von Toxin A (1 bis 5 mg/l) auf die Reaktion der Kolonkrypten nach Zugabe der Agonisten CCH, ATP, FSK/IBMX und der hypotonen Badlösung. Es wurde jeweils die Änderung des Membranpotentials, der Membrankapazität und der Membranleitfähigkeit gegenüber dem Mittelwert aus Vor- und Nachkontrolle bestimmt. Offene Balken: Kontrollexperiment, Schraffierte Balken: Toxinbehandelte Zellen, (n): Zahl der Experimente.*

Die Experimente zeigen, daß es keinen signifikanten Unterschied in der Reaktion der Kolonkryptzellen auf Agonisten zwischen den Toxin A-behandelten Zellen und unbehandelten Kontrollzellen gibt.

## 4.7.2 Toxin B aus *Clostridium difficile*

Auch das Toxin B aus Clostridium difficile entfaltet seine Wirkung durch eine Zerstörung des Aktin-Zytoskeletts. Es unterscheidet sich vom Toxin A durch eine etwa 10fach höhere Wirksamkeit. In der Literatur wird eine Wirkkonzentration von 1.5 µg/l beschrieben [101]. Wir setzten Toxin B sowohl in einer Konzentration von 100 µg/l in der Pipettenlösung als auch zur Vorinkubation der Zellen bei vier Grad Celsius in einer Konzentration von 100 µg/l für eine bis drei Stunden ein. Die Daten aus beiden Experimentierserien wurden zusammengefaßt, da sie keine Unterschiede zeigten. Die Ergebnisse der Experimente sind in Abb.40 dargestellt.

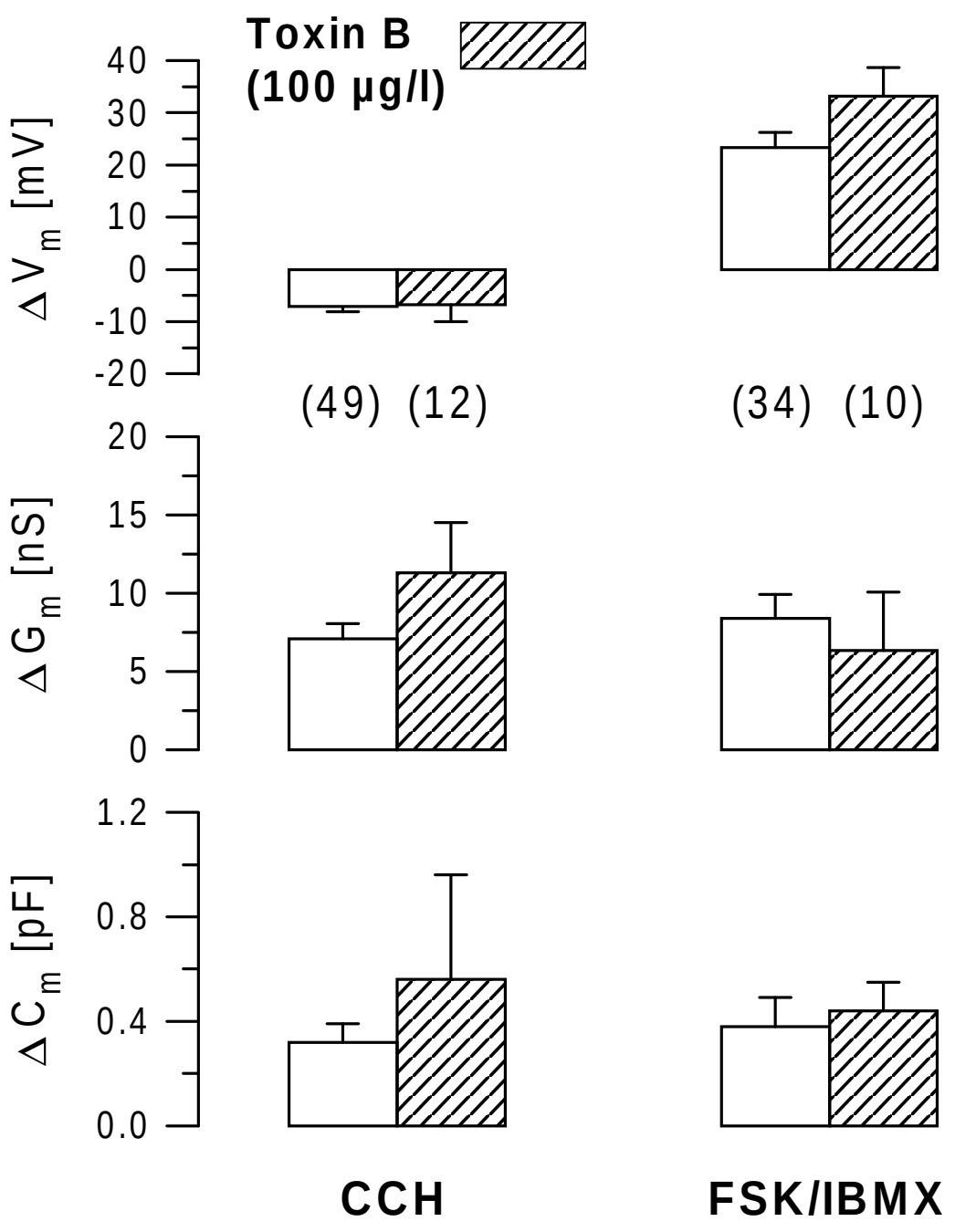

Abb.40: *Effekte von Toxin B (100 µg/l) auf die Reaktion der Kolonkrypten nach Zugabe der Agonisten CCH und FSK/IBMX. Erläuterungen in der Legende zur Abb.39.*

Es konnte kein signifikanter Unterschied in der Reaktion der Toxin B behandelten Kolonkryptzellen gegenüber der Kontrollgruppe beobachtet werden.

## 4.7.3 Cytochalasin B

Die Cytochalasine sind die Gifte der Schimmelpilze *Helminthosporum dematioideum* und *Phoma*. Sie hemmen das Anfügen von Aktin-Molekülen an das Ende bestehender Aktin-Filamente und führen so zu deren Depolymerisierung. Bekannte biologische Wirkungen der Cytochalasine sind die Hemmung der Bewegung von Fibroblasten, der Endozytose bei Makrophagen und der Sekretion von Schilddrüsen- und Wachstumshormonen [20].

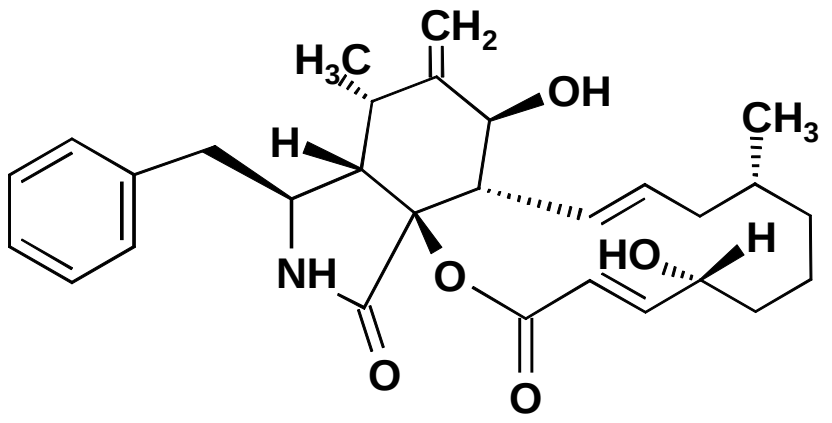

Abb.41: *Strukturformel von Cytochalasin B*

Cytochalasin B wurde in einer Konzentration von 10 µmol/l in der Pipettenlösung eingesetzt. Nach Stimulation der Zellen mit den Agonisten CCH, ATP und FSK/IBMX sowie hypotoner Badlösung zeigten sich die in Abb.42 zusammengefaßten Änderungen des Membranpotentials, der Membranleitfähigkeit und der Membrankapazität:

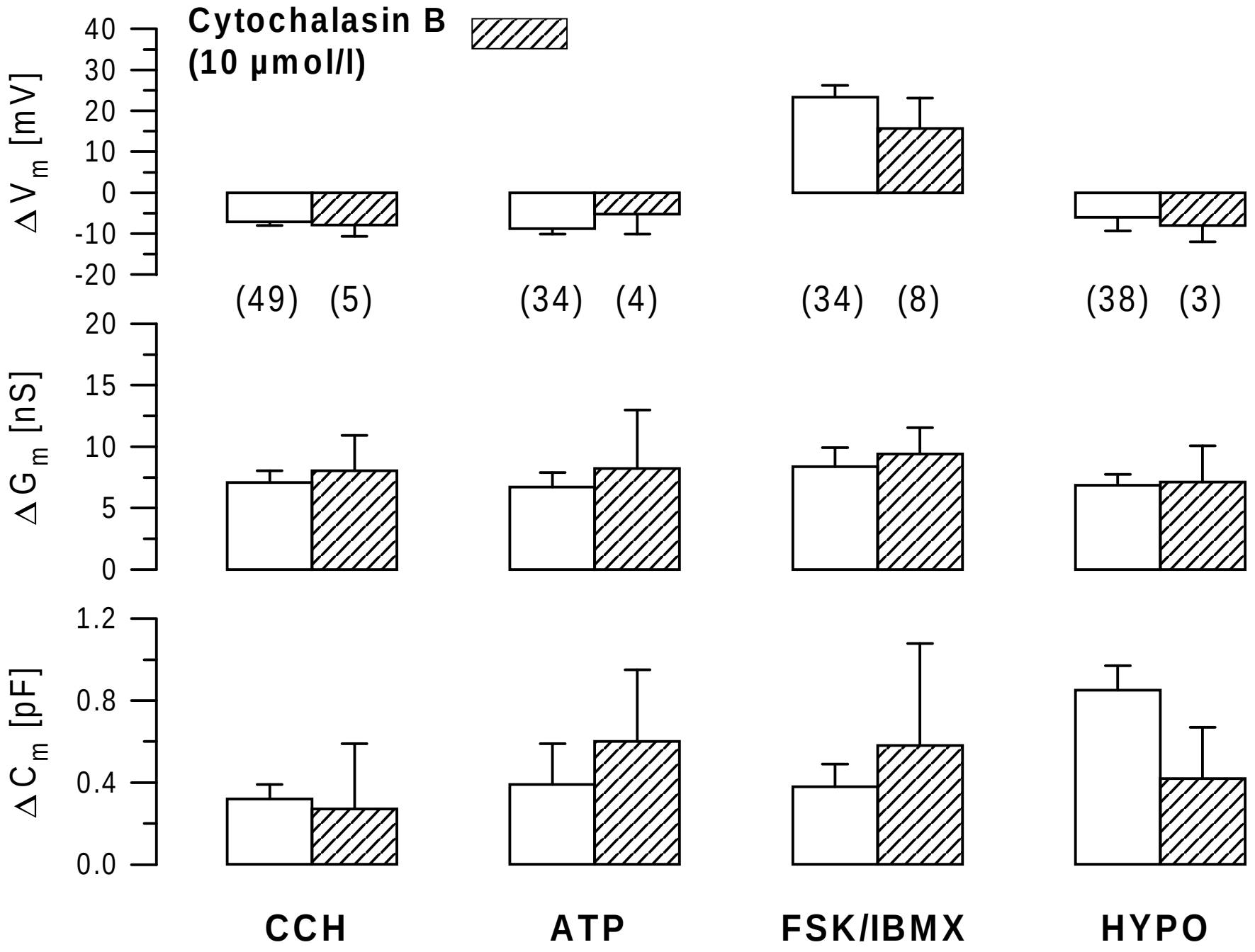

Abb.42: *Wirkung von Cytochalasin B (10 µmol/l) auf die Reaktion der Kolonkrypten nach Zugabe der Agonisten CCH, ATP, FSK/IBMX und von hypotoner Badlösung.*

Die mit Cytochalasin B behandelten Kolonkryptzellen zeigten keine signifikant verschiedene Reaktion auf die Stimulation mit Agonisten gegenüber der unbehandelten Kontrollgruppe.

## 4.7.4 Phalloidin

Phalloidin ist einer der Giftstoffe des grünen Knollenblätterpilzes *(Amanita muscaria)*. Es ist ein zyklisches Peptid mit einer Amanitin-ähnlichen Struktur. Bei Vergiftungen schädigt es die Zellen der Leber durch eine irreversible Bindung an F-Aktin [78]. Phalloidin wurde in der Pipettenlösung in einer Konzentration von 10 µmol/l eingesetzt. Zur Stimulation der Sekretion an den Kolonkryptzellen wurden wiederum die Agonisten CCH, ATP, FSK/IBMX sowie hypotone Badlösung eingesetzt. Die Ergebnisse der Experimente sind in Abb.43 zusammengefaßt.

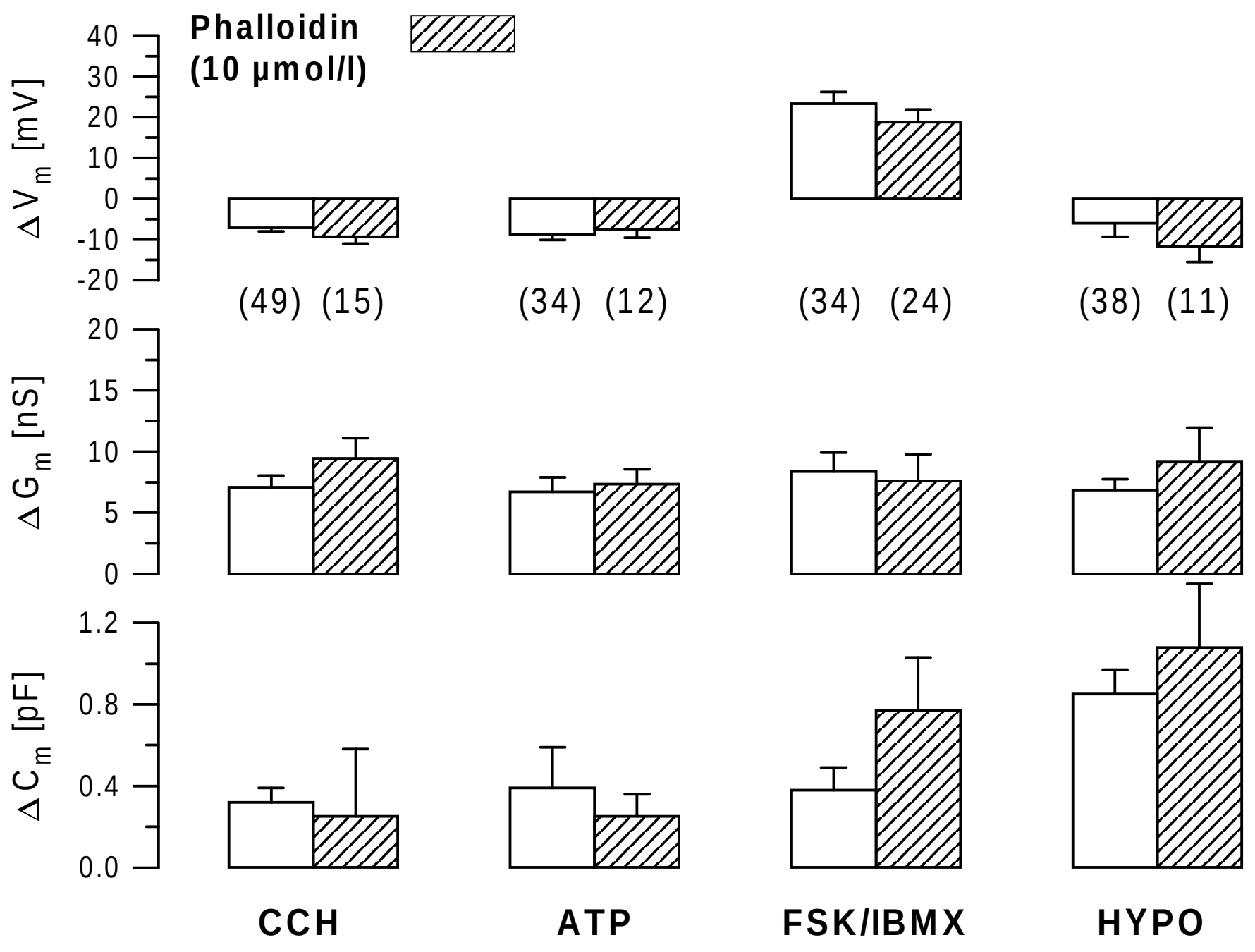

Abb.43: *Reaktion der Kolonkryptzellen auf die Agonisten CCH, ATP, FSK/IBMX und hypotone Badlösung nach Phalloidin-Applikation über die Saugelektrode. Erläuterungen in der Legende zu Abb.39.*

Die Phalloidin-behandelten Zellen zeigten keine signifikant verschiedene Reaktion auf die Agonisten CCH, ATP, FSK/IBMX und auf hypotone Badlösung gegenüber der unbehandelten Kontrollgruppe.

## 4.7.5 Colchicin

Das hochgiftige Colchicin kommt in der Natur zusammen mit anderen strukturverwandten Alkaloiden in der Herbstzeitlosen (*Colchicum autumnale)* vor. Es ist ein hochwirksames Zellteilungsgift und wirkt durch Hemmung der Aggregation der Mikrotubuli in der Metaphase. Die Vergiftung mit Colchicin ruft beim Menschen Lähmungen des ZNS und Atemstillstand hervor, die letale Dosis beträgt etwa 20 mg [20].

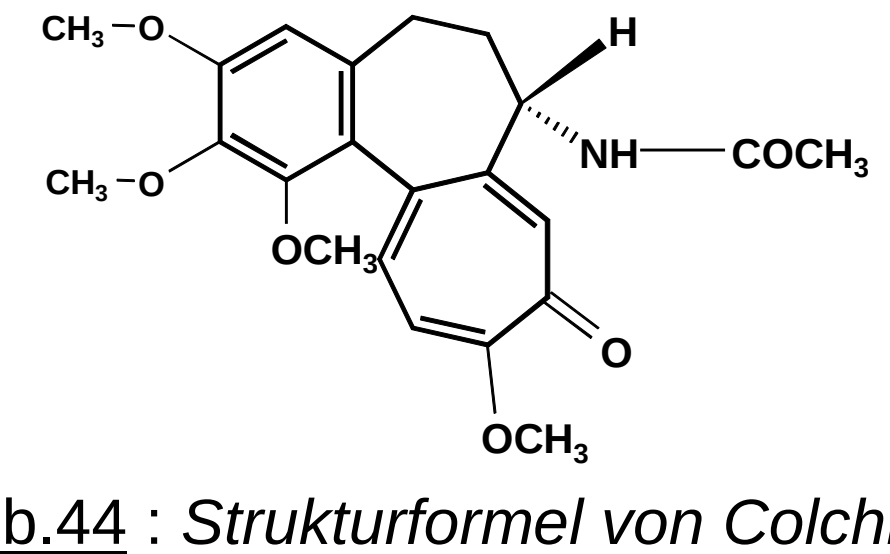

Abb.44 : *Strukturformel von Colchicin*

Zur vorliegenden Untersuchung an den Kolonepithelzellen wurde Colchicin in einer Konzentration von 10 µmol/l in die Pipettenlösung gegeben. Zur Stimulation der Kryptzellen wurden die Agonisten CCH, ATP, Forskolin/IBMX und hypotone Badlösung verwendet. Die Ergebnisse der Experimente sind in Abb.45 wiedergegeben.

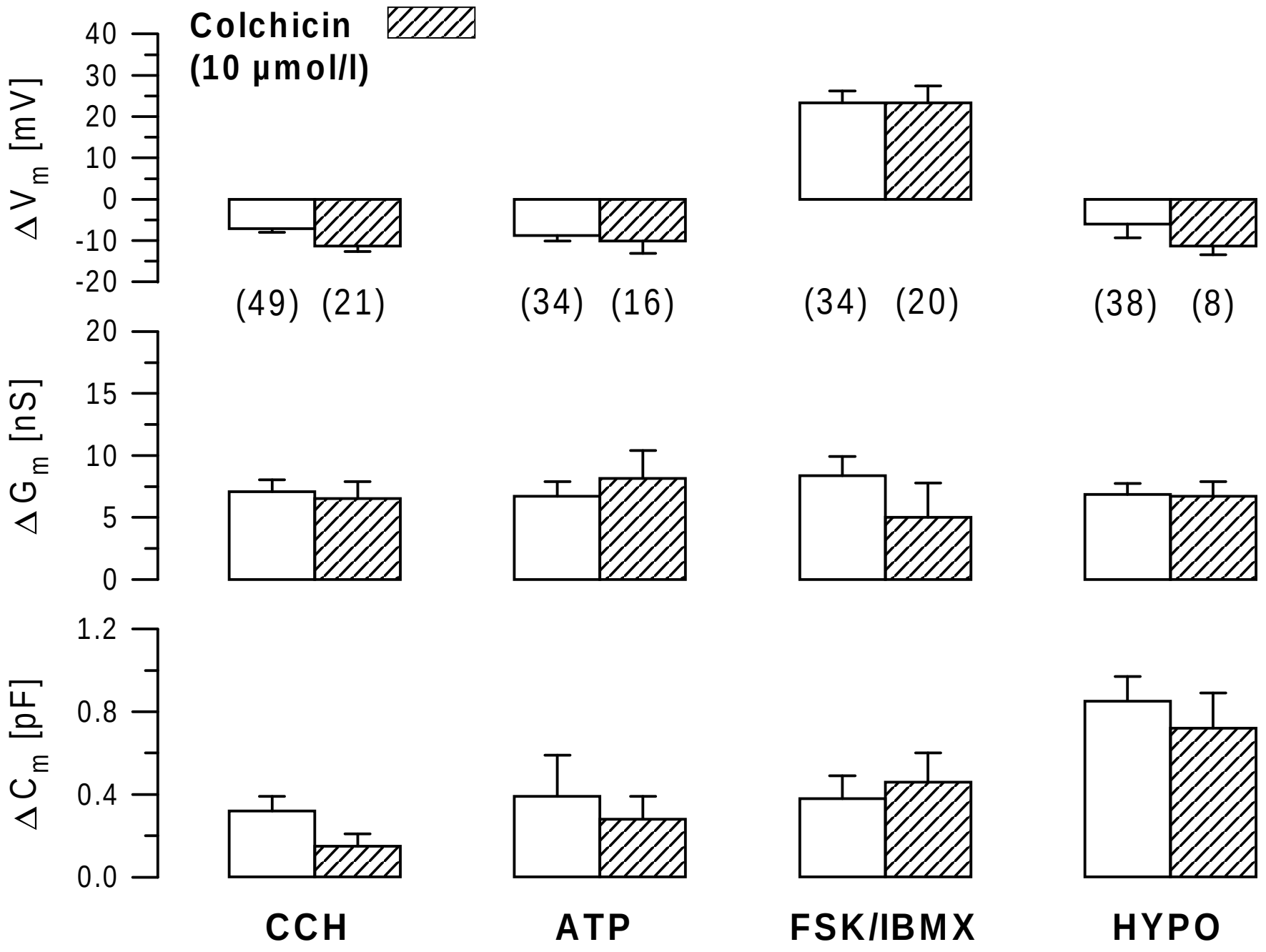

Abb.45: *Effekt von Colchicin (10 µmol/l) auf die Reaktion der Kryptbasiszelle nach Zugabe von CCH, ATP, FSK/IBMX und hypotoner Badlösung. Erläuterungen in der Legende zu Abb.39.*

Die mit Colchicin behandelten Kolonepithelzellen zeigten keine signifikant verschiedene Reaktion auf die Stimulation mit Agonisten gegenüber der Kontrollgruppe.

# 5. Diskussion

Die vorliegende Arbeit befaßte sich mit der Fragestellung, ob bei der Regulation der NaCl-Sekretion in Kryptbasiszellen des Kolons eine Exozytose von Ionenkanälen in die Zellmembran beteiligt ist. Neuere Untersuchungen haben gezeigt, daß an unterschiedlichen epithelialen Geweben, wie z.B. Bronchialepithelzell-Kulturen [48] oder $HT_{29}$-Kolonkarzinomzellen [24,35] die Aktivierung der NaCl-Sekretion mit Exozytoseprozessen verbunden ist.

Zunächst wurde mit Hilfe von Kapazitätsmessungen an der Kolonepithelzelle untersucht, ob eine Aktivierung ihrer Ionenleitfähigkeiten durch die *second messenger* Calcium ($Ca^{2+}$) und cyclisches Adenosinmonophosphat (cAMP) von Exozytoseereignissen begleitet ist und damit zu einem Membrankapazitäts-Anstieg führt.

Im zweiten Teil der Arbeit wurde dann die Exozytose im Kolonepithel mit verschiedenen Inhibitoren des Zytoskeletts blockiert und anschließend die Auswirkungen auf die Aktivierbarkeit der Ionenleitfähigkeit und auf die Zellmembrankapazität untersucht.

## 5.1 Methoden

Bei der Stimulation von Kolonzellen zur NaCl-Sekretion treten neben einer eventuellen Kapazitätsänderung durch Exozytose auch große Änderungen der Membranleitfähigkeit durch das Öffnen von Ionenkanälen auf [18,19]. Die zur Kapazitätsmessung an den Kolonkrypten eingesetzte Meßmethode muß daher erlauben, die Membrankapazität und die Membranleitfähigkeit unabhängig voneinander ohne gegenseitige Beeinflussung zu bestimmen.

### 5.1.1 Kapazitätsmessung mit Hilfe eines einzelnen Lock-In-Verstärkers („phase tracking“)

Das höchste Auflösungsvermögen bei Kapazitätsmessungen an lebenden Zellen konnte bisher mit der von E. Neher und A. Marty entwickelten Meßmethode mit Hilfe eines einzelnen Lock-In-Verstärkers erreicht werden [17,22,76]. Diese Methode erlaubt es, noch Kapazitätsänderungen im Bereich von einigen fF zu erfassen. Bisher kam diese Methode allerdings nur an Zellen zum Einsatz, bei denen die Membrankapazität $C_m$, aber nicht die Membranleitfähigkeit $G_m$ beeinflußt wurde.

Für Messungen an Zellen, deren Membranleitfähigkeit sich stark verändern kann oder bei denen $G_a$ schwankt, ist sie nicht geeignet, da eine Änderung eines der Parameter $G_a$ oder $G_m$ auch Einfluß auf die Kapazitätsmessung hat und zu systematischen Fehlern führt [17].

### 5.1.2 Kapazitätsmessung mit der Zweifrequenz-Synchrondetektionsmethode

Das Zweifrequenz-Synchrondetektionsverfahren, das erstmals von R. Rohliçek und V. Rohliçek vorgeschlagen wurde [89,90], erlaubt dagegen eine unabhängige Bestimmung von $C_m$ und $G_m$ auch bei wechselnden Zugangsleitfähigkeiten und großen Änderungen der Membranleitfähigkeit [87]. Obwohl das Auflösungsvermögen dieser Methode geringer ist als das der „phase tracking“-Technik [17], hat sie sich ihrer Vorteile wegen als eine häufig verwendete Methode zur Kapazitätsmessung in unterschiedlichen Geweben, z.B. CHO-Zellen [47,102], $HT_{29}$-Zellen [35] oder Zellen der Rektaldrüse des Dornhaies [43], durchgesetzt. Deshalb wurde auch in den vorliegenden Untersuchungen die Zweifrequenz-Synchrondetektionsmethode zur Kapazitätsmessung an Kolonkrypten verwendet.

### 5.1.3 Die Berechnung der Membrankapazität mit Hilfe einer $\chi^2$-Minimierung

Zur Berechnung der Membrankapazität $C_m$ und der Membranleitfähigkeit $G_m$ aus der gemessenen Admittanz der Zelle $Y_i$ bei zwei Frequenzen stand zunächst nur die von R. Rohliçek vorgeschlagene Näherungsformel (14) zur Verfügung. Da über mögliche Fehlerquellen bei der Anwendung dieser Rechenmethode keine Aussagen gemacht werden konnten, wurde eine $\chi^2$-Minimierung zur Berechnung der gesuchten Zellparameter $C_m$ und $G_m$ aus der Admittanz $Y_i$ entworfen. Die $\chi^2$-Minimierung stellt ein mathematisches Standardverfahren dar, um Variablen in einem überstimmten Gleichungssystem (13) zu bestimmen. Ihr entscheidender Vorteil ist, daß sowohl weitere Parameter des Zellmodells, wie z.B. die Pipettenkapazität oder möglicherweise der Abdichtwiderstand der Pipette, berechnet werden können als auch Meßdaten bei mehr als zwei Frequenzen in die Berechnung eingebracht werden können. Damit war sie Voraussetzung für die Entwicklung des Vierfrequenz-Synchrondetektors, bei dem eine explizite Auflösung des Gleichungssystems (20) nach den gesuchten Parametern $G_a$, $G_m$, $C_m$ und $C_p$ nicht mehr möglich ist.

Beim Zweifrequenz-Synchrodetektor zeigten sich im Vergleich der Ergebnisse beider Rechenverfahren zwar geringe Unterschiede (einige Prozent) in den Absolutwerten für $G_m$ und $C_m$; die in den Experimenten beobachteten relativen Veränderungen der Leitfähigkeit und Kapazität unterschieden sich dagegen um weniger als 1% (in 5 untersuchten Experimenten).

### 5.1.4 Testmessungen am Zweifrequenz-Synchrondetektor und Vergleich mit den Meßdaten

Um die Unabhängigkeit der Meßgrößen $C_m$ und $G_m$ voneinander zu überprüfen, wurden die in Kapitel 2.3 beschriebenen Testmessungen durchgeführt. Es konnte gezeigt werden, daß der Einfluß einer Leitfähigkeitsänderung auf die Kapazitätsmessung kleiner als 4 fF/nS ist.

Vergleicht man diesen Wert mit den in den Experimenten an Kolonzellen gefundenen Korrelationen zwischen $\Delta G_m$ und $\Delta C_m$ in der Größenordnung von 20-30 fF/nS, so liegen die beobachteten Effekte etwa um den Faktor 5-8 über der Auflösungsgrenze der Meßapparatur. Daher kann eine vorgetäuschte Korrelation, die durch systematische Fehler der Meßapparatur verursacht wird, als Ursache für die beobachteten Kapazitätsänderungen ausgeschlossen werden.

Aus den in den Experimenten gefundenen Leitfähigkeitsänderungen $\Delta G_m$ von etwa 10 nS kann auf einen Fehler der Kapazitätsmessung von weniger als 10 nS x 4 pF/nS $\approx$ 40 fF geschlossen werden. Die beobachteten Kapazitätsänderungen liegen etwa um den Faktor 5 höher. Ein Artefakt der Messung kann also auch hier ausgeschlossen werden.

### 5.1.5 Fehlerquellen bei der Messung mit dem Zweifrequenz-Synchrondetektor

Ein Parameter des Zellmodells, die Pipettenkapazität, konnte mit dem Zweifrequenz-Synchrondetektor nicht bestimmt werden und mußte vor Beginn der Messung durch Abgleich von Hand kompensiert werden. Dies ist jedoch nur in der „Cell-attached“-Konfiguration der Saugelektrode möglich. Tritt spontan bereits bei der Ausbildung des „Seals“ eine Ruptur der Zellmembran im Innern der Pipette auf und bildet sich eine Ganzzell-Konfiguration aus, so ist ein Abgleich nicht mehr möglich.

Im Verlauf des Experiments kann eine Änderung der Pipettenkapazität auftreten. Dies könnte z.B. durch eine Vergrößerung der angelagerten Zellmembranfläche an die Innenseite der Saugelektrode hervorgerufen werden.

Als Verbesserung der bekannten Methode wurde im Laufe unserer Untersuchungen und Messungen die Vierfrequenz-Synchrondetektions-Technik entwickelt, die es erlaubt, auch die Pipettenkapazität zu bestimmen und damit die mögliche Fehlerquelle durch den Pipetten-Abgleich von Hand ausschließt.

### 5.1.6 Kapazitätsmessung mit dem Vierfrequenz-Synchrondetektor

Die Messungen mit dem neuentwickelten Vierfrequenz-Synchrondetektor erforderten keinen Abgleich der Pipettenkapazität mehr, da $C_p$ als vierter Parameter bei der Messung mit bestimmt wurde. Wir verzichteten jedoch nicht ganz auf den Kompensations-Schaltkreis, da ohne Kompensation der Pipettenkapazität der Gesamtstrom über den Saugelektroden-Verstärker um die Hälfte durch den kapazitativen Strom über $C_p$ zunimmt und damit das Signal-Rausch-Verhältnis der Elektronik ungünstiger wird. Eine Kompensations-Einstellung von etwa 3-4 pF bei einer Pipettenkapazität von ca. 5 pF hat sich bei allen Messungen bewährt.

### 5.1.7 Testmessungen am Vierfrequenz-Synchrondetektor

Die Testmessungen am Vierfrequenz-Synchrondetektor zeigten, daß auch die gegenseitige Beeinflussung der Meßgrößen ähnlich gering wie beim Zweifrequenz-Synchrondetektor war. Eine Variation der Pipettenkapazität hatte keinen Einfluß auf die übrigen Meßparameter mehr.

Vergleicht man die Ergebnisse der Testmessungen am $C_m$ oder $G_m$-Testschaltkreis an Zwei- bzw. Vierfrequenz-Synchrondetektor (Abb.12 und Abb.15 bzw. Abb.11 und Abb.16), so erkennt man eine auffällige Übereinstimmung der Streuung der Meßergebnisse. Dies war zunächst nicht zu erwarten, da sowohl die verwendeten Frequenzen als auch das zur Berechnung der Meßgrößen benutzte Gleichungssystem unterschiedlich sind. Eine mögliche Erklärung für die Ähnlichkeit der Meßergebnisse ist, daß die Streuung hauptsächlich durch die Abweichungen der Testschaltkreise selbst hervorgerufen wird, die in beiden Fällen identisch sind.

### 5.1.8 Fehlerquellen bei der Messung mit dem Vierfrequenz-Synchrondetektor und neue Anwendungsmöglichkeiten

Eine mögliche Fehlerquelle könnte auch bei dem neuentwickelten Vierfrequenz-Synchrondetektor in der unzureichenden Beschreibung der Versuchsanordnung durch unserer Ersatzschaltbild liegen. Die Beschreibung der an die Zelle angelegten Saugelektrode durch eine Zugangsleitfähigkeit $G_a$ und eine Pipettenkapazität $C_p$ ist nur eine Näherung für die tatsächliche Verteilung der Kapazitäten und Widerstände in der Saugelektrode.

Der Zugangswiderstand teilt sich auf in einen Längswiderstand der Elektrolytlösung im Innern der Pipette und den Widerstand der Übergangsstelle vom Pipetteninnern in die Zelle. Die Pipettenkapazität ist über die gesamte Länge dieses Widerstandes zwischen Ableitelektrode und der Pipettenspitze verteilt, wobei die Kapazität zur Pipettenspitze hin zunimmt, da die Dicke der als Dielektrikum wirkenden Glaswände sinkt.

Auch der Ohmsche Abdichtwiderstand des „Seals“ selbst ist im Schaltbild nicht berücksichtigt. Zudem kann auch eine kapazitative Kopplung zwischen Pipette und der Zelle über die Zellmembran im Pipetteninnern in Betracht gezogen werden [77].

Die Größen $C_p$ und $G_a$ fassen all diese Eigenschaften der Saugelektrode in nur zwei Parametern zusammen. Der Vierfrequenz-Synchrondetektor bietet die Möglichkeit, weitere der obengenannten Größen in das Zellmodell aufzunehmen und bei der Messung zu berücksichtigen.

Eine neue Einsatzmöglichkeit des Vierfrequenz-Synchrondetektors wäre die Kapazitätsmessung an elektrisch durch *Gap Junctions* gekoppelten Zellen, wenn ein entsprechend angepaßtes Modell für elektrisch gekoppelte Zellen entworfen und neue Parameter wie etwa $G_{Kopplung}$ eingeführt würden.

Theoretisch wären aus den 8 gemessenen Größen $ReY_1$...$ReY_4$ und $ImY_1$...$ImY_4$ bei 4 Frequenzen bis zu 8 Parameter des Zellmodells bestimmbar. Das erforderliche Signal-Rausch-Verhältnis bei der Messung schränkt jedoch die Zahl der zu bestimmenden Parameter auf einer geringere Zahl (etwa 5-6) ein.

### 5.1.9 Die Rolle des Zugangswiderstandes $G_a$, Wahl der Pipettengröße

Je besser die elektrische Verbindung zwischen der Saugelektrode und dem Zellinnern ist, d.h. je höher die Zugangsleitfähigkeit $G_a$ ist, desto günstiger sind die Bedingungen für eine genaue Kapazitätsmessung. Unsere Experimente an den Kolonkrypten zeigten, daß das Rauschen in den gemessenen $C_m$ und $G_m$-Werten mit größerer Zugangsleitfähigkeit abnimmt. Für ein hohes Auflösungsvermögen der Kapazitätsmessung wären also möglichst große Pipetten am günstigsten.

Ein großer Zugang zum Zellinnern hat jedoch andere Nachteile: Durch Dialyse zwischen dem Zellinnern und der Pipette können funktionell wichtige Bestandteile des Zytoplasmas, wie z.B. Enzymsysteme der Transduktionskaskade, aus der Zelle entfernt werden, umgekehrt kann die Pipettenlösung, die ein EGTA-Puffersystem für $Ca^{2+}$ enthält, in die Zelle gelangen und die Konzentration des *second messengers* $[Ca^{2+}]_i$ beeinflussen. In unseren Experimenten zeigte sich, daß insbesondere die Funktion des FSK/IBMX aktivierten cAMP-Weges durch zu große Zugangsleitfähigkeiten beeinträchtigt wurde.

### 5.1.10 Kapazitätsmessung an einem intakten Gewebsverband - Einfluß der Kopplung der Zellen

An Zellen innerhalb eines intakten Gewebsverbandes wie der Kolonkrypte wurde bisher noch keine Kapazitätsmessung beschrieben. Zellen in einem epithelialen Gewebsverband sind häufig durch sogenannte *Gap junctions*, kanalartige Proteine, die zwischen benachbarten Zellen ausgebildet sind, elektrisch miteinander gekoppelt.

Die *Gap junctions* sind für unterschiedliche Ionen aber auch kleinere Proteinmoleküle permeabel. Ihre physiologischen Aufgaben sind der Stoffaustausch und die elektrische Signalübermittlung zwischen den Zellen [30]. Die Leitfähigkeit der *Gap junctions* ist reguliert und kann durch unterschiedliche Einflüsse verändert werden, ein intrazellulärer $Ca^{2+}$-Anstieg führt z.B. zum Verschluß der *Gap junctions.*

Eine Messung der Membrankapazität wird durch eine Kopplung der Zellen untereinander erschwert: Von der Messung wird nicht nur die Kapazität der Membran einer einzelnen Zelle, sondern mehrerer benachbarter Zellen erfaßt. Mit einem einfachen Ersatzschaltbild, wie in Abb.6 dargestellt, können gekoppelte Epithelzellen also nicht beschrieben werden. Zudem kann ein Öffnen oder Schließen der *Gap junctions* [30] die elektrisch zugängliche Membranfläche verändern und damit Kapazitätsänderungen vortäuschen.

Voraussetzung für unsere Kapazitätsmessung war daher, daß an der untersuchten Kolonkrypte keine Kopplung der Einzelzellen untereinander vorhanden ist. Dies konnte von C. Jacobi und Kollegen im Physiologischen Institut der Universität Freiburg gezeigt werden [49]. Mir Hilfe optischer Messungen konnte nachgewiesen werden, daß Farbstoffe wie Lucifer Yellow oder Calcein, für die *Gap junctions* permeabel sind, sich nicht von einer Kryptepithelzelle in ihre Nachbarzellen ausbreiten.

Unsere Kapazitätsmessungen konnten diese Befunde bestätigen: Eine gemessene mittlere Kapazität der Kryptbasiszellen von unter 10 Pikofarad läßt sich nur durch ungekoppelte Einzelzellen erklären. Die „scheinbare“ Zellmembrankapazität in gekoppelten Epithelverbänden, wie z.B. dem Pankreas-Azinus, liegt in der Größenordnung von 100-1000 pF, wie erste von M. Slawik et al. im Physiologischen Institut der Universität Freiburg durchgeführte Messungen ergaben.

### 5.1.11 Die Präparation und Aufbewahrung der Krypten

Ziel der Kryptpräparation war es, einzelne Krypten aus dem Gewebsverband der Kolonschleimhaut abzulösen, ohne die einzelnen Epithelzellen zu schädigen. Dazu wurde ein ursprünglich von Siemer und Gögelein beschriebenes Verfahren [95], das im Physiologischen Institut der Universität Freiburg weiterentwickelt wurde, verwendet [19].

Das Prinzip der Präparationsmethode ist es, durch einen $Ca^{2+}$-Entzug die interzelluläre Verbindung des Epithels zum submukösen Bindegewebe zu lockern und anschließend durch mechanisches Schütteln die Krypten von der Darmoberfläche abzulösen. Mit der verwendeten Präparationsmethode war es möglich, eine große Zahl (100-1000) intakter Einzelkrypten aus einem Darmstück zu gewinnen.

Zur Aufbewahrung mußten die Krypten bei 4° C gelagert werden und erst zum Experiment in die 37° C warme Lösung gegeben werden. Eine Aufbewahrung der Zellen bei 37° C führte, auch in einem Bicarbonat - $CO_2$-Puffersystem, nach 15-30 min zum Absterben der Zellen und zur Auflösung der Kryptstruktur.

Die mögliche Aufbewahrungszeit der Krypten nach der Präparation bis zum Experiment war auch bei 4° C auf etwa 3-5 h eingeschränkt.

## 5.2 Faktoren, die die Membrankapazität beeinflussen können

Die Membrankapazität an Kolonzellen kann durch verschiedene Faktoren beeinflußt werden:

- Die Exozytose von Ionenkanal-enthaltenden Membranvesikeln führt zu einem Kapazitätsanstieg [13,16,73,74].
- Eine von der Ionenkanal-Aktivierung unabhängige Exozytose, z.B. von Mucus, kann ebenfalls die Membrankapazität erhöhen [51].
- Eine Zellschwellung kann zu einem Membrankapazitäts-Anstieg, eine Zellschrumpfung zu einem Abfall der Membrankapazität führen [43].

Der Einfluß des Zellvolumens auf die gemessene Membrankapazität läßt sich wie folgt erklären:

Die Oberfläche eines sezernierenden oder resorbierenden Epithelgewebes ist durch sogenannte *Mikrovilli*, feine Ausstülpungen der Zellmembran (Abb.1) vergrößert [92]. Diese feine Faltung der Zellmembran an der Epitheloberfläche beeinflußt auch die gemessene Membrankapazität: Ist die Zelle geschrumpft, liegen die Faltungen der Zellmembran eng aneinander, der Längswiderstand innerhalb eines *Mikrovillus* ist sehr hoch. Damit ist nur ein Teil der tatsächlichen Membranoberfläche innerhalb der Falten für die Kapazitätsmessung zugänglich, die Gesamtkapazität der Zellmembran ist niedriger als ihrer Gesamtfläche entspricht.

Bringt man eine Zelle z.B. durch hypotone Lösung zum Anschwellen, so können sich die Membraneinstülpungen teilweise entfalten. Dadurch sinkt der Längswiderstand innerhalb der Membranfalten, die elektrisch zugängliche Membranfläche wird größer. Die Membrankapazität, die der Fläche der Zellmembran proportional ist, steigt an. Setzt man dagegen die Zelle einer hypertonen Schrumpfung aus, so verkleinert sich durch stärkere Faltung der Membran ihre Oberfläche, die Kapazität $C_m$ sinkt.

Um zwischen den obengenannten Prozessen, die die Membrankapazität verändern, unterscheiden zu können, wurden folgende Untersuchungen durchgeführt:

1) Korrelationen zwischen den Leitfähigkeits- und Kapazitätsänderungen:

   Bei einer Aktivierung von Ionenkanälen durch Exozytose erwartet man einen kausalen Zusammenhang zwischen dem Kapazitätsanstieg, der Zeichen der Exozytose ist, und der gemessenen Leitfähigkeitsänderung, die durch die neu eingebauten Ionenkanäle entsteht. Dieser Zusammenhang zeigt sich mathematisch in einer starken Korrelation der Parameter [84,100].

2) Berechnung der Kanaldichte der möglichen Exozytosevesikel:

   Die Zahl der Ionenkanäle pro Fläche in einem möglicherweise vorhandenen Exozytosevesikel kann abgeschätzt werden [48,102]. Aus der Kapazitätsänderung $\Delta C_m$ kann mit Hilfe von Gleichung (4) die exozytierte Membranfläche, aus der Leitfähigkeitsänderung $\Delta G_m$ und der abgeschätzten Einzelkanalleitfähigkeit $g_i$ kann die Zahl der Kanäle N nach Gleichung (27) bestimmt werden.

$$\Delta G = N \cdot g_i \qquad (27)$$

3) Zellschwellung und Zellschrumpfung
   a) Um den Einfluß des Zellvolumens auf die Membrankapazität zu untersuchen, können die Zellen durch hypotone Lösung zum Schwellen sowie durch hypertone Lösung zum Schrumpfen gebracht werden [43,105].
   b) Eine Hemmung des Ionentransportes durch den $Na^+2Cl^-K^+$-Kotrtansporter führt ebenfalls zum Schrumpfen der Zellen [43].

4) Veränderung des Zytoskeletts durch Toxine

Mit Hilfe von Toxinen, die das Zytoskelett zerstören, kann damit auch der Aktin- oder Tubulin-abhängige Einbau von Membranvesikeln in die Zellmembran verhindert werden [83,101,109].

## 5.3 Agonisteninduzierte Veränderungen

### 5.3.1 Erhöhung des $[Ca^{2+}]_i$ durch CCH und ATP

Untersuchungen an verschiedenen Zellen (CHO-Zellen [102], $HT_{29}$-Zellen [35]) haben gezeigt, daß eine Erhöhung der intrazellulären $Ca^{2+}$-Aktivität zu einem Membrankapazitäts-Anstieg führt. Auch in unseren Untersuchungen zeigte sich nach Stimulation mit den Agonisten CCH und ATP eine geringe Kapazitätszunahme. Parallel zur Kapazitätszunahme bewirkt der $Ca^{2+}$-Anstieg an Kolonzellen die Öffnung $Ca^{2+}$-regulierter basolateraler Kalium-Kanäle [19,36]. Die Zellen hyperpolarisieren dadurch und die Triebkraft für die luminale $Cl^-$-Sekretion steigt [37].

Es stellt sich daher die Frage, ob zwischen dem Kapazitätsanstieg und dem Leitfähigkeitsanstieg ein kausaler Zusammenhang besteht. Vorangegangene Untersuchungen an $HT_{29}$-Kolonkarzinomzellen ergaben Hinweise auf einen solchen Zusammenhang: Kapazitäts- und Leitfähigkeitsänderungen sind miteinander korreliert und haben einen ähnlichen Zeitverlauf [35].

Die vorliegenden Untersuchungen machen jedoch einen solchen Zusammenhang an der Kolonkrypte unwahrscheinlich: Zwischen den beobachteten Kapazitäts- und Leitfähigkeitsänderungen besteht nur eine schwache Korrelation, in etwa einem Viertel der Experimente mit CCH zeigte sich trotz einer Leitfähigkeitszunahme kein Kapazitätsanstieg oder sogar ein Kapazitätsabfall (Abb.33). Auch die Steigung der Regressionsgeraden ist mit 18.4 fF/nS sehr flach: Schätzt man aus der mittleren Leitfähigkeitszunahme die Zahl der neu eingebauten Kanäle ab und aus dem Kapazitätsanstieg die neu hinzugekommene Membranfläche, so ergibt sich eine Kanaldichte der Exozytosevesikel von etwa 40 Kanälen/µm². Diese Anzahl erscheint im Vergleich zur $K^+$-Kanalzahl in Einzelkanaluntersuchungen an Kolonkrypten [7] (etwa 4 Kanäle pro Membranfleck von einigen µm² Größe) sehr hoch.

Eine kausale Verknüpfung zwischen der Kapazitätserhöhung und dem Leitfähigkeitsanstieg durch Agonisten, die $[Ca^{2+}]_i$ erhöhen, ist am Kolon also sehr unwahrscheinlich. Die beobachteten Kapazitätsänderungen lassen sich vielmehr durch eine von der Aktivierung der Ionenleitfähigkeit unabhängige Exozytose z.B. von Mucus [51] oder durch eine Vergrößerung des Zellvolumens bei der Sekretion (Abschnitt 5.5) erklären [43].

### 5.3.2 Erhöhung der zytosolischen Konzentration von cAMP durch FSK/IBMX

Die cAMP-Erhöhung öffnet an der Kolonkrypte einerseits (sehr kleine) basolaterale Kalium-Kanäle (mit einer Leitfähigkeit < 3 pS) und andererseits eine luminale Chlorid-Leitfähigkeit, die vermutlich mit dem CFTR (*cystic fibrosis transmembrane regulator*) assoziiert ist [9,28,36]. Der Mechanismus, wie durch die Aktivierung von CFTR eine Chlorid-Leitfähigkeit gebildet wird, ist in der Literatur umstritten [23,40,74]. Einerseits wird vorgeschlagen, daß dies durch eine (Proteinkinase A abhängige) Phosphorylierung regulatorischer Domänen dieses Proteins geschieht [23,25,88,99], andererseits wird angenommen, daß CFTR die Exozytose und Endozytose anderer Ionenkanäle reguliert, d.h. den sogenannte „membrane traffic“ beeinflußt oder aber selbst exozytiert wird [3,8,74].

Während an $HT_{29}$-Zellen eine Aktivierung von CFTR eine parallele Erhöhung der Membranleitfähigkeit und der Membrankapazität bewirkte [35], konnte an CFTR-exprimierenden CHO-Zellen kein Zusammenhang zwischen der Leitfähigkeitserhöhung und dem Kapazitätsanstieg beobachtet werden [47].

Ein Ergebnis der vorliegenden Untersuchungen an der Kolonkrypte ist, daß cAMP die Membranleitfähigkeit stark erhöht, während die Effekte auf die Membrankapazität nur gering sind. Die Korrelation zwischen den beobachteten Kapazitäts- und Leitfähigkeitseffekten ist zudem mit 33 fF/nS schwach ausgeprägt.

Die mittlere Leitfähigkeitserhöhung betrug 7.3 nS, was 1000-2000 neuen $Cl^-$-Kanälen entspricht (bei einer angenommenen Einzelkanalleitfähigkeit von 4-8 pS). Wenn also alle aktivierten $Cl^-$-Kanäle, die die Membranleitfähigkeit erhöhen, durch Exozytose in die Membran eingebaut würden, müßten sie in einer Fläche von 24 $\mu m^2$ komprimiert sein. Die berechnete Kanaldichte von 50-100 Kanälen/$\mu m^2$ erscheint sehr hoch. In Einzelkanaluntersuchungen an $HT_{29}$-Zellen wurde eine Kanaldichte von wenigen Kanälen pro Membranfläche in einer Saugelektrode (einige µm) gefunden [45,59].

Daher machen es die Messungen sehr unwahrscheinlich, daß die cAMP-vermittelte Aktivierung von $Cl^-$-Kanälen über Exozytose abläuft. Der dennoch beobachtete Kapazitätsanstieg kann wiederum durch eine Volumenzunahme der Zellen erklärt werden: Der luminale Chlorid-Ausstrom führt zum Absinken der intrazellulären Chlorid-Konzentration, diese ist wiederum zusammen mit dem cAMP-Anstieg Stimulus für den basolateralen $Na^+2Cl^-K^+$-Kotransporter [105]. Die erhöhte $Na^+$, $Cl^-$ und $K^+$-Aufnahme läßt die Zellen anschwellen, $C_m$ steigt durch eine Entfaltung der Membran.

## 5.4 Osmotische Veränderungen des Zellvolumens

### 5.4.1 Hypotone Zellschwellung

Eine hypotone Zellschwellung führt in unterschiedlichen Geweben zum Öffnen von Ionenleitfähigkeiten als Zeichen der osmotischen Gegenregulation der Zellen [14,60]. Die Hyperpolarisation, die in unseren Experimenten an der Kolonkrypte beobachtet werden konnte, ist dabei Zeichen für eine vorwiegend aktivierte Kalium-Leitfähigkeit [36]. Häufig wird die Schwellungs-induzierte Leitfähigkeit durch kleine und räumlich in der Zelle begrenzte $Ca^{2+}$-Transienten hervorgerufen [80].

Der Kapazitätsanstieg, der gleichzeitig mit der Leitfähigkeitserhöhung beobachtet wurde, war jedoch ausgeprägter als bei den anderen $Ca^{2+}$-freisetzenden Agonisten, während die Leitfähigkeitsänderungen in der gleichen Größenordnung lagen. Dies spricht wiederum für eine volumeninduzierte Kapazitätsänderung. Auch der Zeitverlauf der Kapazitäts- und Leitfähigkeitsänderungen läßt sich nur hierdurch erklären: Während zunächst (nach 30 Sekunden) noch eine signifikante Korrelation zwischen der Volumenzunahme, die sich im Kapazitätsanstieg widerspiegelt, und der gegenregulatorischen Reaktion (Leitfähigkeits-Anstieg) besteht, ist nach einer Minute keine Korrelation mehr zu erkennen, die Zellen mit hoher Membranleitfähigkeit weisen sogar die niedrigsten Membrankapazitäten auf. Dies kann nur durch eine Volumenabnahme bei erfolgreicher Gegenregulation erklärt werden.

### 5.4.2 Hypertone Zellschrumpfung

Während für viele Zellen die Aktivierung nichtselektiver Kationenkanäle durch hypertone Zellschwellung in der Literatur gefordert wurde [56,104], konnten B. Weyand und Kollegen im Physiologischen Institut der Universität Freiburg zeigen [107], daß die beobachtete Leitfähigkeitsabnahme und die Depolarisation durch einen intrazellulären $Ca^{2+}$-Abfall und das nachfolgende Schließen $Ca^{2+}$-abhängiger $K^{+}$-Kanäle hervorgerufen wird. Die beobachtete Kapazitätsabnahme spricht wiederum für eine volumeninduzierte Einfaltung der Zellmembran, eine aktive Endozytose von $K^{+}$-Kanälen ist bisher nicht beschrieben.

## 5.5 Die Hemmung des Kotransporters durch Azosemid

Neuere Untersuchungen am $Na^{+}2Cl^{-}K^{+}$-Kotransporter der Rektaldrüse des Dornhaies (*Squalus acanthies*), die in vielen Eigenschaften als Modell für ein NaCl-sezernierendes Epithel dient [34,42], konnten zeigen, daß bei der Stimulation der Epithelzellen zur Sekretion auch der Kotransporter aktiviert wird. Durch die erhöhte $Na^{+}$, $Cl^{-}$ und $K^{+}$-Aufnahme kommt eine nachfolgende osmotische Zellschwellung zustande, die (über eine Entfaltung der Zellmembran) für einen Kapazitätsanstieg verantwortlich sein kann.

Auch in unseren Untersuchungen an Kolonkrypten konnten wir nach der Stimulation der Zellen zur Sekretion einen Kapazitätsanstieg beobachten. Um die Rolle des Kotransporters dabei zu untersuchten, hemmten wir nach der Stimulation der luminalen $Cl^{-}$-Leitfähigkeit den Kotransporter durch Azosemid. Dies führte zu einer intrazellulären $Cl^{-}$-Verarmung, das Nernst-Potential für $Cl^{-}$ erhöhte sich und die Zellen hyperpolarisierten [18].

Gleichzeitig konnten wir eine Abnahme der Membrankapazität beobachten. Dies bestätigt, daß eine Zellschrumpfung über eine verstärkte Einfaltung der Zellmembran zum Absinken der Membrankapazität führt.

## 5.6 Veränderung des Zytoskeletts mit Hilfe von Toxinen

Um den Einfluß des Zytoskeletts bei der Aktivierung der Ionenleitfähigkeiten in der Kolonkrypte zu untersuchen, wurden die Kolonkrypten verschiedenen Toxinen für das Zytoskelett ausgesetzt. Die Exozytose von Membranvesikeln benötigt intakte zelluläre Strukturen wie einen Golgi-Apparat, Mikrotubuli und Aktin-Filamente [48,74,91]. Während die ersten beiden besondere Bedeutung bei der Bildung und dem Transport sekretorischer Vesikel haben, scheinen Aktin-Filamente wichtig für den letzten Schritt der Exozytose, die Fusion mit der Zellmembran zu sein [12]. Im Augenblick ist jedoch noch wenig darüber bekannt, welche Rolle diese Prozesse bei der Aktivierung von Ionenkanälen in Säugetier-Zellen besitzen.

### 5.6.1 Toxin A und B aus *Clostridium difficile*

Die Toxine A und B aus dem Bakterium *Clostridium difficile* sind Proteine mit einem Molekulargewicht von 308 kD bzw. 269 kD, die im Jahre 1991 sequenziert wurden und deren molekularer Wirkmechanismus im Jahre 1995 aufgeklärt werden konnte [101]. Sie sind Auslöser der pseudomembranösen Kolitis und wirken bereits in extrem niedrigen Konzentrationen. Die letale Dosis für eine 25 Gramm schwere Maus beträgt für Toxin A und B nur etwa 10 ng [101].

Beide Toxine gehören zu den sogenannten Glucosyltransferasen. Sie entfalten ihre Wirkung durch Glykosylierung kleiner regulatorischer Enzyme der Zelle, den sogenannten Rho, Rac und Cdc42-GTPasen. Während Toxin A nur die Proteine Rho A,B und C glykosyliert [109], kann das Toxin B zusätzlich auch die GTPasen Rac1,2 und Cdc42 hemmen [53]. Die Rho-Proteine sind unter anderem am Aufbau des Aktin-Zytoskeletts, der Kontraktion der glatten Muskulatur und an der Ausbildung von Zell-zu-Zell-Kontakten beteiligt. Die Rac-Proteine haben zusätzlich Einfluß auf Faltungsprozesse der Zellmembran und die Zellmigration bei Zellen des Immunsystems [1].

Die Anwendung der Toxine führt an verschiedenen Geweben zu sichtbaren morphologischen Veränderungen der Zellen in Form eines Verlustes der Zellform und einer Abrundung des Zellkörpers, die Ausdruck der Zerstörung des Zytoskeletts sind. An verschiedenen Zellen, wie z.B. Fibroblasten, $T_{84}$-Kolonkarzinomzellen oder CHO-Zellen konnte der spezifische Abbau der Aktin-Filamente durch fluoreszenzoptische Untersuchungen nachgewiesen werden [47,101].

Es gibt zwei Möglichkeiten, die bakteriellen Toxine A und B aus *Clostridium difficile* an Kolonkrypten einzusetzen: Zum einen können die isolierten Krypten in einer Toxin-haltigen Lösung inkubiert werden, wobei das Toxin aktiv über die Zellmembran in die Zellen aufgenommen wird [101], zum anderen kann das Toxin in die Pipettenlösung gegeben werden und durch Diffusion direkt ins Zellinnere gelangen.

Aus vorangegangenen Untersuchungen an Kulturzellen des Chinesischen Hamster-Ovars (*Chinese Hamster Ovary Cells, CHO*) ist bekannt, daß das Toxin A aus *Clostridium difficile* nach einer einstündigen Inkubation bei 37° C zu deutlichen morphologischen Veränderungen der Zellen führt [47,102]: Die Zellen runden sich ab und passen, als Ausdruck des fehlenden Zytoskeletts, ihre Form der Badströmung an. Um die Wirksamkeit der verwendeten Toxine A und B sicherzustellen, wurden sie zu Kontrolle in parallelen Ansätzen an CHO-Zellen getestet.

Bei den isolierten Kolonzellen war eine Inkubation bei 37° C nicht möglich, da die frisch isolierten Kolonkrypten bei dieser Temperatur innerhalb von 15-30 Minuten absterben. Daher wurden die Kolonzellen bei 4° C inkubiert. Da über die Transportrate der Toxine durch die Zellmembran und damit die notwendige Inkubationsdauer bei niedrigen Temperaturen nichts bekannt ist, haben wir die Zellen über eine Zeitspanne von einer bis zu fünf Stunden inkubiert. Danach war weder eine Veränderung ihrer Morphologie zu beobachten noch zeigte sich eine Abweichung ihrer elektrischen Antwort auf Agonisten gegenüber den Kontrollzellen.

Daher applizierten wir alternativ die Toxine durch die Pipettenlösung. Untersuchungen und theoretische Betrachtungen von E. Neher und A. Marty legen nahe, daß Makromoleküle durch Diffusion aus der Pipette innerhalb weniger Minuten ins Zellinnere gelangen können [86]. Im Gegensatz zur Inkubation, bei der alle Zellen der Krypte dem Toxin ausgesetzt werden, kann bei der Dialyse über die Saugelektrode keine Aussage über morphologische Änderungen gemacht werden, da nur eine Einzelzelle betroffen ist.

Ein Ergebnis der in der vorliegenden Arbeit durchgeführten Untersuchungen war, daß weder nach Inkubation mit den Toxinen A und B noch nach Dialyse mit Toxin über die Saugelektrode die Antwort der Kolonzellen auf eine Stimulation mit den Agonisten ATP, CCH oder FSK/IBMX beeinflußt wurde. Auch die Effekte einer osmotischen Zellschwellung oder -schrumpfung auf die Membrankapazität und -leitfähigkeit blieben unverändert.

Daher erscheint es unwahrscheinlich, daß das Aktin-Zytoskelett an der Aktivierung der Ionenleitfähigkeiten der Kolonzelle beteiligt ist. Auch der Kapazitätsanstieg nach der Stimulation der Zellen zur Sekretion konnte unverändert beobachtet werden. Dies spricht ebenfalls dafür, daß keine aktiver Exozytoseprozess, sondern eine Entfaltung der Zellmembran bei der Sekretion die Ursache der beobachteten Kapazitätserhöhung darstellt.

Eine andere Erklärungsmöglichkeit für die beobachteten Befunde könnten ein Aktin-unabhängiger Exozytoseprozess oder die Unwirksamkeit der Toxine an nativen Kolonzellen sein. Gegen eine Aktin-unabhängige Exozytose spricht, daß an anderen Zellen, wie z.B. an 16HBE-Zellen [48] oder Zellen des proximalen Tubulus [94,98], Aktin-Filamente eine wichtige Rolle bei der Exozytose von Membranvesikeln spielen. Eine Hemmung anderer Strukturen, wie z.B. von Tubulin-Filamenten hat dagegen keinen Einfluß auf die Reaktion der Zellen [48].

Ein vollkommen neuartiger Exozytosemechanismus für Ionenkanäle, der weder Aktin- noch Tubulin-vermittelt ist, erscheint zwar ebenfalls als unwahrscheinlich, kann aber anhand der vorliegenden Daten nicht ausgeschlossen werden. Gegen eine Unwirksamkeit der Toxine an nativen Kolonzellen spricht, daß das Kolon den natürliche Angriffspunkt der beiden Toxine beim Menschen und im Tierversuch darstellt. So konnte in Tierversuchen eine hämorrhagische Flüssigkeitssekretion und ein Verlust der Barrierefunktion des Kolons nach der Anwendung von Toxin A nachgewiesen werden [101].

### 5.6.2 Cytochalasin B

Auch die Cytochalasine, Mykotoxine aus dem Schimmelpilz *Helminthosporum dematioideum*, sind eine häufig zur Untersuchung des Zytoskeletts verwendete Substanzklasse [20,48]. Die mehr als 20 bekannten Cytochalasine hemmen alle spezifisch und reversibel die durch Aktin-Filamente vermittelte Teilung des Zytoplasmas (Cytokinese), haben aber keinen Einfluß auf die Kernteilung (Mitose). Sie können daher zur Erzeugung vielkerniger Riesenzellen verwendet werden [20].

In Säugetierzellen hemmen sie alle Bewegungsprozesse, die mit der Aktin-Filamentbildung zusammenhängen, wie z.B. die Aggregation von Blutplättchen, die Ausbildung von *Mikrovilli* oder die Phagozytose bei Makrophagen [20]. Der molekulare Me-chanismus ihrer Wirkung ist eine Hemmung des Anbaus neuer Aktin-Filamente am sogenannten „Stachel-Ende“ der Aktin-Filamente, von dem aus die Aktin-Fäden verlängert werden. Der allmähliche Abbau der Aktin-Polymere von ihrem entgegengesetzten Ende her führt dann zu deren Depolymerisierung und schließlich dem Verlust des gesamten Aktin-Zytoskeletts.

An 16HBE-Bronchialepithelzellen konnte der Abbau des Aktin-Zytoskeletts nach Behandlung mit Cytochalasin D in fluoreszenzmikroskopischen Untersuchungen gezeigt werden [48]. Gleichzeitig veränderte sich die Morphologie der Zellen: Sie zeigten irreguläre Zellformen und blasenförmige Ausstülpungen. Die Dialyse von Cytochalasin D über die Pipettenlösung konnte die $Ca^{2+}$ und cAMP-vermittelte Leitfähigkeit um mehr als 90% unterdrücken.

Die vorliegenden Untersuchungen an Kolonkrypten dagegen zeigten, daß Cytochalasin B weder Veränderungen der Leitfähigkeitsantwort auf die Agonisten FSK/IBMX noch auf CCH hervorruft. Die Beteiligung des Aktin-Zytoskeletts an der Aktivierung der luminalen $Cl^-$ oder basolateralen $K^+$-Leitfähigkeit der Kolonkrypte erscheint daher unwahrscheinlich.

Eine gleichzeitige Regulation der Ionenleitfähigkeit über zwei unterschiedliche Wege, sowohl die Öffnung membranständiger Kanäle als auch die Exozytose neuer Kanäle, kann aufgrund der vorliegenden Daten jedoch nicht ausgeschlossen werden. Der Anteil der Exozytose an der Aktivierung der Leitfähigkeit müßte jedoch im Vergleich zu dem durch Regulation membranständiger Kanäle sehr gering sein, da die Hemmung der Exozytose keine signifikante Abnahme des Gesamt-Leitfähigkeitsanstieges bewirkte.

Die Kapazitätszunahme bei der Sekretion auch nach der Hemmung der Exozytose spricht wiederum für eine volumeninduzierte Zellmembran-Entfaltung bei der Sekretion, wie sie auch an andern Geweben beobachtet werden konnte [43].

### 5.6.3 Phalloidin

Phalloidin, ein zyklisches Heptapeptid aus dem grünen Knollenblätterpilz, ruft seine zytotoxische Wirkung durch irreversible Bindung an F-Aktin hervor [78]. Mit dem Fluoreszenz-Farbstoff FITS markiertes Phalloidin wird daher häufig als optischer

Marker für das Aktin-Zytoskelett eingesetzt [48]. An 16HBE-Bronchialepithelzellen konnte nach der Dialyse von Phalloidin über die Saugelektrode eine Abnahme des Leitfähigkeitsanstieges durch die *second messenger* $Ca^{2+}$ und cAMP um 60% beobachtet werden [48].

Ein Ergebnis der vorliegenden Experimente an Kolonzellen war, daß die Phalloidin-Behandlung keinen Einfluß auf die Aktivierbarkeit der Ionenleitfähigkeiten der Kolonzelle hatte. Da weder die cAMP vermittelte $Cl^-$-Leitfähigkeit noch die $Ca^{2+}$ aktivierte $K^+$-Leitfähigkeit signifikant beeinflußt wurden, kann geschlossen werden, daß eine Aktin-vermittelte Exozytose keine oder nur eine untergeordnete Rolle bei der Aktivierung der Leitfähigkeit spielt.

Auch die Kapazitätszunahme bei der Sekretion konnte unverändert beobachtet werden, was wiederum für eine volumeninduzierte Membranentfaltung spricht. Auch der Kapazitäts- und gegenregulatorische Leitfähigkeitsanstieg bei der osmotischen Zellschwellung wurde nicht signifikant beeinflußt. Eine Erklärung dafür könnte sein, daß der Einfluß des Aktin-Zytoskeletts bei der Volumenregulation der Zelle eher gering ist. Eine nur partielle Inhibition des Aktin-Zytoskeletts durch das eingesetzte Phalloidin, wie sie auch in 16HBE-Zellen beobachtet wurde [48], kann jedoch nicht ausgeschlossen werden.

### 5.6.4 Colchicin

Colchicin, ein Alkaloid aus der Herbstzeitlosen, ist ein hochwirksames Mitosegift. Sein molekularer Wirkmechanismus ist eine irreversible Zerstörung der Mikrotubuli, wodurch eine Fixierung der Mitose in der Metaphase eintritt [20,103].

In Untersuchungen an $T_{84}$-Zellen konnte nach der Inkubation mit Colchicin eine Abnahme der cAMP-abhängigen Aktivierung von $Cl^-$-Kanälen gefunden werden, während $Ca^{2+}$-aktivierte $Cl^-$-Ströme nicht beeinflußt wurden [24,73]. In Experimenten an 16HBE-Zellen dagegen konnte keine Abnahme der cAMP-abhängigen $Cl^-$-Ströme nach Anwendung verschiedener Toxine, die Mikrotubuli depolymerisieren, gefunden werden [48].

In unseren Experimenten an Kolonkrypten konnte ebenfalls keine Beeinflussung der Leitfähigkeitszunahme durch $Ca^{2+}$ oder cAMP freisetzende Agonisten beobachtet werden. Tubulin-Filamente scheinen daher keine oder nur eine untergeordnete Rolle bei der Aktivierung der $Cl^-$-Leitfähigkeit der Kolonkrypte zu spielen.

Ein zweifacher Aktivierungs-Mechanismus sowohl über Exozytose als auch über das Öffnen membranständiger Kanäle, bei dem die Exozytose im Vergleich zur Kanalöffnung nur einen geringen Anteil zur Gesamtleitfähigkeit beiträgt, kann jedoch nicht ausgeschlossen werden.

## 5.7 Die Rolle der Exozytose bei Aktivierung von Ionenleitfähigkeiten am Kolon

Epitheliale Ionenkanäle werden durch eine Reihe unterschiedlicher Mechanismen reguliert. So können die *second messenger* $Ca^{2+}$ und cAMP direkt oder indirekt über Proteinkinasen Kanalproteine, die in der Membran vorhanden sind, regulieren. Es gibt jedoch auch zahlreiche Hinweise, daß auch das Zytoskelett, z.B. durch die Exozytose von Membranvesikeln, die Ionenkanäle enthalten, eine wichtige Rolle bei der Regulation epithelialer Ionenkanäle spielt [24,35,63,94,98]:

Aus den vorliegenden Untersuchungen ergibt sich, daß eine Exozytose von Membranvesikeln oder eine Beteiligung des Aktin- oder Tubulin-Zytoskeletts bei der Aktivierung der Ionenleitfähigkeiten des Kolons keine oder nur eine untergeordnete Bedeutung besitzen. Dies wird durch folgende Befunde belegt:

- Bei der Stimulation zur Sekretion zeigen Kryptbasiszellen des Kolons nur einen geringen Membrankapazitäts-Anstieg, der zudem nur schwach mit der Leitfähigkeitszunahme korreliert.
- Hypotone Zellschwellung und hypertone Zellschrumpfung bewirken die größten Änderungen der Membrankapazität. Der bei der Sekretion der Zellen beobachteten Kapazitätsanstieg läßt sich daher als eine Zellschwellung und daraus resultierende Vergrößerung der „effektiven“ Membranfläche bei der Entfaltung der Zellmembran deuten.
- Diese Interpretation konnte durch die Hemmung des $Na^+2Cl^-K^+$-Kotransporters bestätigt werden: Eine Hemmung des basolateralen Ioneneinstroms bei der Sekretion führt zu einem Kapazitätsabfall der Zellen.
- In unseren Experimenten an Kolonkrypten fanden wir keine Reduktion der cAMP vermittelten $Cl^-$ oder $Ca^{2+}$-vermittelten $K^+$-Leitfähigkeit der Zellen nach der An-wendung verschiedener Toxine, die mit dem Aufbau des Aktin oder Tubulin-Zytoskeletts interferieren.

Die für andere Zellen geforderte Regulation der CFTR-assoziierten $Cl^-$-Leitfähigkeit oder der $K^+$-Leitfähigkeit durch Aktin- oder Tubulin-vermittelte Exozytoseprozesse läßt sich anhand der Untersuchungsergebnisse für Kolonkrypten nicht bestätigen.

Mit den vorliegenden Ergebnissen kann jedoch nicht ausgeschlossen werden, daß cAMP und $Ca^{2+}$ auf zwei unabhängigen Wegen die Ionenleitfähigkeit regulieren: Zum einen öffnen die *second messenger* in der Membran vorhandene Ionenkanäle, zum anderen werden, wenn auch nur in weitaus geringerer Zahl, neue Kanäle in die Membran eingebaut.

Weiterhin kann auch ein gänzlich neuartiger Exozytosemechanismus für Ionenkanäle, der in einer balancierten Exozytose von kanalhaltigen und einer Endozytose von kanalfreien Membranvesikeln besteht oder von anderen Proteinen des Zytoskeletts wie z.B. Annexinen oder Profilin vermittelt wird [15,68], anhand der vorliegenden Daten nicht ausgeschlossen werden.

# 6. Zusammenfassung

Die vorliegende Arbeit befaßte sich mit der Fragestellung, ob die Regulation der NaCl-Sekretion in Kryptbasiszellen des Kolons über die Exo- und Endozytose von Ionenkanälen erfolgt. Zunächst wurde mit Hilfe von Kapazitätsmessungen an der Kolonepithelzelle untersucht, ob eine Aktivierung ihrer Ionenleitfähigkeiten durch die *second messenger* Calcium ($Ca^{2+}$) und cyclisches Adenosinmonophosphat (cAMP) zu einem Membrankapazitätsanstieg führt und damit von Exozytoseereignissen begleitet ist. Hierzu wurde die erstmals 1993 von R. und V. Rohliçek vorgeschlagene Zweifrequenz-Synchrondetektionsmethode [89] eingesetzt und weiterentwickelt:

Mit zwei neu gebauten Modellschaltkreisen für die elektrischen Eigenschaften einer Einzelzelle wurde die Meßgenauigkeit der verwendeten Apparatur getestet. Eine neue Meßtechnik wurde entwickelt, um bei der Kapazitätsmessung gleichzeitig auch das Zellmembranpotential bestimmen zu können. Als Weiterentwicklung der Zweifrequenz-Synchrondetektionstechnik wurde erstmals ein Vierfrequenz-Synchrondetektor zur Kapazitätsmessung eingesetzt, der es erlaubt, neue elektrische Parameter in das Zellmodell einzuführen und die Störgröße der Pipettenkapazität ($C_p$) zu eliminieren.

Bei der Stimulation der Kolonkrypten zur Sekretion über die *second messenger* $Ca^{2+}$ und cAMP konnte ein starker Anstieg der Membranleitfähigkeit ($G_m$), aber nur eine geringe Zunahme der Membrankapazität ($C_m$) beobachtet werden. Leitfähigkeits- und Kapazitätsänderungen zeigten nur eine schwache Korrelation. Die größten Veränderungen der Membrankapazität konnten durch eine osmotische Zellschwellung ($C_m$ steigt) mit hypotoner Badlösung und Zellschrumpfung ($C_m$ fällt) mit hypertoner Lösung hervorgerufen werden.

Die Zerstörung des Aktin- und Tubulin-Zytoskeletts der Krypten mit den Toxinen A und B aus *Clostridium difficile* sowie *Cytochalasin B*, *Phalloidin* und *Colchicin* hatte keine Auswirkungen auf die Aktivierbarkeit der Ionenleitfähigkeiten und den Membrankapazitätsanstieg bei der Sekretion.

Daher erscheint es sehr unwahrscheinlich, daß Exozytoseprozesse an der Regulation der cAMP aktivierbaren luminalen Chloridleitfähigkeit oder der $Ca^{2+}$ aktivierten basolateralen $K^+$-Leitfähigkeit beteiligt sind. Der beobachtete Membrankapazitätsanstieg kann vielmehr als eine Entfaltung der Zellmembran bei der Sekretion gedeutet werden.

Dies konnte durch die Hemmung des $Na^+2Cl^-K^+$-Kotransporters mit Azosemid bestätigt werden: Wenn die basolaterale Chloridaufnahme in die Krypten über den Kotransporter verhindert wird, verarmen die Zellen bei der Sekretion an Chlorid und verlieren nachfolgend osmotisch Wasser, die Membrankapazität sinkt.

# Literatur

# Danksagung

An dieser Stelle möchte ich allen danken, die mich bei der Anfertigung dieser Arbeit unterstützt haben.

Meinen beiden „Diplomeltern“, Prof. Dr. Greger und Prof. Dr. Königsmann, möchte ich für die Betreuung dieser Arbeit danken. Prof. Greger gilt mein besonderer Dank für die Überlassung des Themas, das meine beiden Studienfächer Physik und Medizin berücksichtigt und die Ermöglichung dieser Arbeit am Physiologischen Institut. Seine enthusiastische Art zu experimentieren und seine vielen engagierten Diskussionen haben mir gezeigt, was wissenschaftliche Forschung bedeutet.

Prof. Königsmann danke ich für seine Bereitschaft, meine Diplomarbeit im Fach Physik fakultätsübergreifend zu betreuen. Seine wertvollen Vorschläge und die aktive Mitwirkung bei der Lösung aller physikalischen und mathematischen Probleme waren mir eine große Hilfe.

Dr. Martin Hug sei ganz herzlich gedankt für die experimentelle Einführung in die Arbeitsmethodik. Ohne seine Geduld und seinen Humor wäre mir die Kunst des „Patchens“ nicht so schnell gelungen, und auch seine kompetente Durchsicht dieser Arbeit war mir eine große Hilfe. Meinen beiden Mitstreitern beim „Patchen“, Malte Hübner und Uli Beschorner, sei für die gute Zusammenarbeit bei den oft schwierigen Experimenten und Karsten Hamm, Dirk Heizmann und Richard Warth für die bei der Präparation der Kolonkrypten gedankt. Mit Annette Höfer, Michael Köttgen und Dirk Kerstan hat die Arbeit im gleichen Labor viel Freude gemacht.

Dr. Ulrich Fröbe und Robert Busche aus der Elektronik-Werkstatt gilt mein besonderer Dank für ihre Mithilfe bei allen technischen Problemen und für die Konstruktion und den Bau des Vierfrequenz-Synchrondetektors. Ohne ihre Mitarbeit wäre ein Teil dieser Arbeit nicht zustandegekommen.

Ganz herzlich danke ich allen technischen Assistenten, insbesondere Viktoria Müller, Helga Schauer, Jutta Siebert, Alexander Bausch, Tobias Fehrenbach und Wolfgang Rohm, für das Ansetzen der Experimentierlösungen.

# Veröffentlichung

*Kongreßbeitrag zum Joint Congress of the German Physiological Society and the Scandinavian Physiological Society, Hamburg, März 1998 (Poster):*

Hübner M, Schill C, Greger R, Hug MJ (1998) Agonist induced changes in membrane conductance and capacitance of rat colonic crypt cells.
Pflügers Arch 435 Supp 6:R182 (Abstr.)

Ich versichere, daß die vorliegende Arbeit selbständig verfaßt wurde und keine anderen als die angegebenen Quellen und Hilfsmittel benutzt wurden.